\documentclass[%
 aip,
 amsmath,amssymb,
 reprint,%
]{revtex4-1}

\usepackage{graphicx}
\usepackage{dcolumn}
\usepackage{bm}
\usepackage[dvipsnames]{xcolor}
\usepackage[utf8]{inputenc}
\usepackage[T1]{fontenc}
\usepackage{mathptmx}
\usepackage{etoolbox}
\usepackage{comment}
\usepackage{xcolor}
\usepackage{caption}
\usepackage{subcaption}
\usepackage[newcommands]{ragged2e}
\usepackage{caption}
\makeatletter
\def\@email#1#2{%
 \endgroup
 \patchcmd{\titleblock@produce}
  {\frontmatter@RRAPformat}
  {\frontmatter@RRAPformat{\produce@RRAP{*#1\href{mailto:#2}{#2}}}\frontmatter@RRAPformat}
  {}{}
}%
\makeatother
\begin{document}


\title[Finite Size Effects in CO$_{2}$ hydrates]{Three-phase equilibria of hydrates from computer simulation. II.~Finite-size effects in the carbon dioxide hydrate}

\author{J. Algaba}
\affiliation{%
Laboratorio de Simulación Molecular y Química Computacional, CIQSO-Centro de Investigación en Química Sostenible and Departamento de Ciencias Integradas, Universidad de Huelva, 21007 Huelva, Spain
}%
\author{S. Blazquez}
\affiliation{ Departamento de Química Física, Facultad de Ciencias Químicas, Universidad Complutense de Madrid, 28040 Madrid, Spain
}%
\author{E. Feria}
\affiliation{%
Laboratorio de Simulación Molecular y Química Computacional, CIQSO-Centro de Investigación en Química Sostenible and Departamento de Ciencias Integradas, Universidad de Huelva, 21007 Huelva, Spain
}%
\author{J. M. Míguez}
\affiliation{%
Laboratorio de Simulación Molecular y Química Computacional, CIQSO-Centro de Investigación en Química Sostenible and Departamento de Ciencias Integradas, Universidad de Huelva, 21007 Huelva, Spain
}%
\author{M. M. Conde$^*$}%
\affiliation{ 
Departamento de Ingeniería Química Industrial y del Medio Ambiente, Escuela Técnica Superior de Ingenieros Industriales, Universidad Politécnica de Madrid, 28006, Madrid, Spain
}%

\author{F. J. Blas$^*$}
\email{felipe@uhu.es}
\affiliation{%
Laboratorio de Simulación Molecular y Química Computacional, CIQSO-Centro de Investigación en Química Sostenible and Departamento de Ciencias Integradas, Universidad de Huelva, 21007 Huelva, Spain
}%

\date{\today}

\begin{abstract}
In this work, the effects of finite size on the determination of the three-phase coexistence temperature ($T_3$) of carbon dioxide (CO$_2$) hydrate have been studied by molecular dynamic simulations and using the direct coexistence technique. According to this technique, the three phases involved (hydrate-aqueous solution-liquid CO$_2$) are placed together in the same simulation box. By varying the number of molecules of each phase it is possible to analyze the effect of simulation size and stoichiometry on the $T_3$ determination. In this work, we have determined the $T_3$ value at 8 different pressures (from 100 to 6000 bar) and using 6 different simulation boxes with different numbers of molecules and sizes. In 2 of these configurations, the ratio of the number of water and CO$_2$ molecules in the aqueous solution and the liquid CO$_2$ phase is the same as in the hydrate (stoichiometric configuration). In both stoichiometric configurations, the formation of a liquid drop of CO$_2$ in the aqueous phase is observed. This drop, which has a cylindrical geometry, increases the amount of CO$_2$ available in the aqueous solution and can in some cases lead to the crystallization of the hydrate at temperatures above $T_3$, overestimating the $T_3$ value obtained from direct coexistence simulations. The simulation results obtained for the CO$_{2}$ hydrate
confirm the sensitivity of $T_{3}$ depending on the size and
composition of the system, explaining the discrepancies
observed in the original work by Míguez \emph{et al.} [\emph{J. Chem Phys.} \textbf{142}, 124505 (2015)]. Non-stoichiometric configurations with larger unit cells show convergence of $T_{3}$ values, suggesting that finite-size effects for these system sizes, regardless of drop formation, can be
safely neglected. The results obtained in this work highlight that the choice of a correct initial configuration is essential to accurately estimate the three-phase coexistence temperature of hydrates by direct coexistence simulations. 
\end{abstract}

\maketitle
$^*$Corresponding authors: felipe@uhu.es and maria.mconde@upm.es

\section{Introduction}

Gas hydrates, crystalline structures formed by the encapsulation of gas molecules within water cages, have garnered significant attention due to their implications in energy storage, gas separation, and environmental processes. \cite{Sloan2003a,Koh2012a,Sloan2008a,Ripmeester2022a} 
Gas hydrates are a subject of considerable interest due to their diverse applications depending on the guest molecule they encapsulate. Hydrogen hydrates, for instance, have been proposed as an alternative method for storing hydrogen, presenting potential applications in energy storage. \cite{barthelemy2017hydrogen, chen2023rapid, zhang2022hydrogen} On the other hand, methane hydrates represent a natural and alternative source of energy, particularly when found in seabed deposits. \cite{Sloan2003a, CG_1988_71_41, ARE_1990_15_53, GRL_34_L22303_2007, N_2007_445_303} However, in the context of natural gas transport, hydrate formation within pipelines poses a significant challenge, prompting research into hydrate inhibition strategies. Ionic liquids, for instance, have garnered attention as inhibitors in mitigating hydrate formation during gas transport. \cite{additives_book_hydrates, GHIASI2021114804}
In addition to their occurrence on Earth, clathrate hydrates play a crucial role in the context of icy satellites within our solar system, particularly under conditions involving salty water. \cite{JPCC_1_80_2020,PCCP_2017_19_9566,D1CP02638K,PRIETOBALLESTEROS2005491,KARGEL2000226,10.1130/G22311.1,https://doi.org/10.1002/2014RG000463}

Carbon dioxide (CO$_2$) hydrates have become a pivotal area of research due to their profound implications in addressing global environmental challenges, particularly those associated with the increasing levels of carbon dioxide in the Earth's atmosphere.\cite{peter2018reduction,yoro2020co2} 
Thus, CO$_2$ hydrates are of particular interest in the context of carbon capture and sequestration (CCS) strategies, which aim to mitigate the adverse impacts of anthropogenic CO$_2$ emissions.\cite{d2010carbon,gao2020industrial,wang2020research,nguyen2022technical,zheng2020carbon} 

In 2015, some of us studied the three-phase equilibrium line of CO$_2$ hydrate employing the direct coexistence technique. \cite{Miguez2015a} We showed that to accurately replicate the experimental behavior of T$_3$ at various pressures, it was necessary to introduce positive deviations (specifically, $\xi$=1.13) to the energetic Lorentz-Berthelot (LB) rules. By incorporating these deviations, we successfully reproduced the three-phase line of CO$_2$ hydrate. It is noteworthy that our focus was primarily on the coexistence phase when CO$_2$ was in its liquid state (i.e., above 4 MPa).

When delving into the realm of simulations concerning CO$_2$ hydrates, it is noteworthy to highlight several notable contributions in the field. 
In a pioneering work, Tung \emph{et al.}~\cite{Tung2011a} computed the three-phase line of the carbon dioxide hydrate using the TIP4P-Ew\cite{horn2004development} model for water and the EPM2\cite{harris1995carbon} for CO$_2$. They obtained a $T_3$ of $285(2)\,\text{K}$ at $400\,\text{bar}$, in good agreement with our work of 2015.\cite{Miguez2015a}
Another such significant work was conducted by Costandy \emph{et al.} in 2015.\cite{Costandy2015a} Similar to the work of Miguez \emph{et al.},~\cite{Miguez2015a} they used the direct coexistence technique and utilized the same force fields (TIP4P/Ice for water and TraPPe for CO$_2$). However, they employed different cutoff values and altering water-CO$_2$ interactions due to a distinct deviation from the LB rules. Their findings, particularly at 400 bar, yielded a $T_3$ of $283.5(1.7)\,\text{K}$, differing slightly from the results obtained by Miguez~\emph{et al.} of $287(2)\,\text{K}$ at the same pressure.\cite{Miguez2015a}
In a separate endeavor, Waage and coworkers,\cite{Waage2017a} employing a different technique,  evaluated the dissociation temperature of CO$_2$ hydrate using interfacial energy calculations. Remarkably, they employed the same force fields as Costandy ~\emph{et al.}~\cite{Costandy2015a} and Miguez \emph{et al.}~\cite{Miguez2015a} However, the dispersive interactions and cutoff values implemented by Waage and the team were distinct from the other works. Their results showed a $T_3$ of $284.5\,\text{K}$ at $400\,\text{bar}$. 

There exists other significant works in which the $T_3$ of CO$_2$ hydrates is computed using different force fields for water and CO$_2$\cite{Jiao2021a,Hao2023a} or using Monte Carlos simulations in the Grand Canonical ensemble.\cite{Qiu2018a} Even there are theoretical works in which the three-phase equilibrium of this hydrate is evaluated by using  PC-SAFT and  Peng-Robinson equation\cite{El2016a} or by using semiempirical equations.\cite{jager2013a} Theoretical calculations using the van der Waals and Platteeuw approach have also been employed to evaluate the occupation of the CO$_2$ hydrates.\cite{Sun2005b} Also, the kinetics of these hydrates have been studied.\cite{Tung2011a,blazquez2023growth}
Even the addition of salt to the hydrates (due to they can be found at the seabed) has been computationally studied for CO$_2$ hydrates\cite{Wang2023a} (and recently for methane hydrates in a wide range of concentrations).~\cite{Fernandez-Fernandez2019a,Blazquez2023b}
Collectively, these works offer insights into the thermodynamics and stability of hydrates under diverse conditions.

Importantly, in 2022 we developed a new methodology to evaluate the $T_3$ of gas hydrates baptized as the \textit{solubility method}.\cite{Grabowska2022b,Algaba2023a} In short, by calculating on the one hand the solubility of gas in a gas-liquid system (which decreases with temperature) and on the other hand solubility of gas in a hydrate-liquid system (which increases with temperature). The two solubility curves intersect at a certain temperature that is the $T_3$ at a fixed pressure. We employed this approach for the methane hydrate\cite{Grabowska2022b} and for the carbon dioxide hydrate\cite{Algaba2023a} with larger system sizes than in the direct coexistence simulations of previous works but using the same force fields (TIP4P/Ice + TraPPe) and the same cutoff. By using this methodology we estimated a $T_3$ of $290(2)\,\text{K}$ at $400\,\text{bar}$ which slightly differs from previous calculations.

The discussed previous works suggest that system sizes or cutoff values can have an effect on the determination of the $T_3$. 
This observation aligns with earlier research that has delved into finite-size effects across various systems. Previous studies, for Lennard-Jones potentials\cite{panagiotopoulos1994molecular} or for the evaluation of surface tensions, \cite{orea2005oscillatory, binder2000computer} have investigated the influence of finite-size effects in different contexts. Additionally, the impact of finite sizes has been explored in square well fluids.\cite{vortler2008simulation} Notably, a recent study conducted a comprehensive analysis of how finite sizes affect the determination of the melting temperature of ice.\cite{Conde2017a} This body of research collectively underscores the significance of considering finite-size effects in simulations across various systems and provides valuable insights into their implications on observed phenomena.

In this work, we have conducted a thorough examination of the finite-size effects influencing the determination of the three-phase equilibrium temperature for CO$_2$ hydrates. Similar to our parallel work on methane hydrate finite-size effects, we have employed the TIP4P/Ice force field for water and the TraPPe for CO$_2$, and we systematically have varied system sizes by manipulating the number of molecules in the hydrate, liquid, and gas phases.

The structure of our study is organized as follows: Section II outlines the methodology and simulation details, providing a comprehensive overview of our approach. In Section III, we present the results, encompassing the three-phase equilibrium temperatures for different system sizes of carbon dioxide hydrate at different pressures. Finally, Section IV encapsulates the main conclusions drawn from this extensive study, summarizing key findings and insights into the impact of finite-size effects on the determination of carbon dioxide hydrate equilibrium temperatures.

\begin{table*}
\begin{tabular}{c c c c c c c c c c c c c c c c c c cc c c c c}
\hline
\hline
Configuration & &   \multicolumn{5}{c}{Hydrate phase} & &
{H$_{2}$O-rich liquid phase} & &
{CO$_{2}$-rich liquid phase} & & Stoichiometric & & \multicolumn{3}{c}{Simulation box size}\\
     \cline{3-7}\cline{15-17}
     & \, & Unit Cell & \, & Water & \, & CO$_2$ & \, & Water &  \,  & CO$_2$ & & & & $L_x=L_y$ (nm) & &
$L_z$ (nm)\\
\hline
0 & & 2x2x2 && 368 & &  64  && 368  & & 192  & & No &  & 2.4 & & 7.2\\
1 & & 2x2x2 && 368 & &  64  && 368  & & 64  & & Yes & & 2.4 & & 5.2\\
2 & & 3x3x2 & & 828 & &  144  & & 828  & & 432 & & No & & 3.6 & & 6.7\\
3 & & 3x3x3 & & 1242 & &  216  & & 1242  & & 648 & & No &  & 3.6 & & 11.3\\
4 & & 3x3x3 & & 1242 & &  216  & & 1242  & & 216 & & Yes &  & 3.6 & & 7.4\\
5 & & 4x4x2 & & 1472 & &  256  & & 1472  & & 768 & & No & & 4.8 & & 6.7\\
6 & & 4x4x4 & & 2944 & &  512 & & 2944  & & 1536 & & No & & 4.8 & & 13.3\\
\hline
\hline
\end{tabular}
\caption{\label{tabla-moleculas} \justifying{Initial number of molecules for each phase (hydrate phase, H$_{2}$O-rich liquid phase, and CO$_{2}$-rich liquid phase) in the different configurations studied in this work. For the hydrate phase, the replication factor of the unit cell is indicated in each case. Configuration 0 corresponds to the original system studied by M\'{\i}guez \emph{et al.}~\cite{Miguez2015a} but it is not studied in this paper. The last columns indicate whether the system composition is stoichiometric as well as the simulation box sizes.}}
\end{table*}

\section{Methodology and Simulation Details}

In order to study finite-size effects on the three-phase coexistence line of the CO$_2$ hydrate, we have employed the same methodology introduced by Conde and Vega.~\cite{Conde2010a} Following this methodology, the three phases involved (CO$_2$ hydrate, aqueous, and pure CO$_2$ phases) are put together in the same simulation box. Due to the use of periodic boundary conditions, this arrangement ensures that the three phases are in contact, with one of the phases surrounded by the other two. In this work, all the simulation boxes have been generated using the same arrangement explained previously regardless of the size and number of molecules of the system.

The CO$_2$ hydrate exhibits a sI crystallographic structure. The unit cell of the sI hydrate structure is formed by six tetradecahedrons, 5$^{12}$6$^2$, and two pentagonal dodecahedrons, 5$^{12}$ cages. Hence, one sI hydrate unit cell is formed by 46 molecules of water and 8 cages. The initial sI CO$_2$ hydrate unit cell has been built taking explicitly into account that hydrates are proton-disordered structures. To obtain this arrangement, hydrogen atoms are placed using the algorithm proposed by Buch \emph{et al.}~\cite{Buch1998a} This allows the generation of solid configurations satisfying the Bernal-Fowler rules,~\cite{Bernal1933a} with zero (or at least negligible) dipole moment. Also, in this work, we have assumed single occupancy of all the CO$_2$ hydrate cages, which means that each cage of the hydrate is occupied by a molecule of CO$_2$. In this work, 6 different initial configurations have been built by varying the replication factor of the hydrate unit cell and the number of initial molecules of water and CO$_2$ in the aqueous and CO$_2$ phases respectively (see Table \ref{tabla-moleculas}),

Once the initial configuration is generated, $NPT$ simulations are carried out by fixing the pressure at 8 different values (see Table \ref{t3_results}) and varying the temperature. The pressure range extends from $100$ to $6000\,\text{bar}$. Under these pressure conditions, CO$_{2}$ is in liquid phase for all studied configurations. If the selected temperature is above the temperature at which the three phases coexist, $T_3$, the hydrate phase becomes unstable and melts. If the selected temperature is below the $T_3$, the hydrate phase will grow until extinguishing the CO$_2$ and/or aqueous phases depending on the initial amount of molecules of CO$_2$ and water in both phases respectively. If the amount of water and CO$_2$ in the aqueous and CO$_2$ phases is the same as that in the hydrate phase (stoichiometric composition), at $T<T_3$, the hydrate will grow until extinguishing both phases, and the final configuration box will be formed by a unique bulk hydrate phase (configurations 1 and 3 from Table \ref{tabla-moleculas}). Contrary, if the amount of CO$_2$ molecules in the CO$_2$ phase is larger than the number of molecules of water in the aqueous phase (using the stoichiometric composition of the hydrate as a reference), the hydrate will grow until extinguishing the aqueous phase. In this case, the final configuration box will be formed by a CO$_2$ hydrate phase in equilibrium with a pure CO$_2$ phase (configurations 2, 4, 5, and 6 from Table \ref{tabla-moleculas}). The evolution of the system can be determined from the analysis of the potential energy as a function of the simulation time. If $T<T_3$, the potential energy will decrease as a consequence of the increment of hydrogen bonds formed when the hydrate phase grows. Contrary, if $T>T_3$, the potential energy will increase as a consequence of the destruction of the hydrogen bonds of the CO$_2$ hydrate phase. Finally, the dissociation temperature, $T_3$, is estimated by analyzing if the hydrate phase grows or melts, at a fixed value of pressure, as a function of the temperature. The $T_3$ value is estimated as the intermediate value between the lowest temperature at which the CO$_2$ hydrate melts and the highest temperature at which the CO$_2$ hydrate grows at a given value of pressure. Some of us have previously used this methodology to study the three-phase coexistence line of the CO$_2$ hydrate.~\cite{Miguez2015a} However, in this previous work, only one size system was used and the finite size effects were not taken into account.

\begin{table*}
\begin{ruledtabular}
\begin{tabular}{ccccccccc}
& \multicolumn{7}{c}{Configuration} & Experiment \\
\hline
& 0 & 1 & 2 & 3 & 4 & 5 & 6 & \\
$P$ (bar) & 2x2x2 & 2x2x2 & 3x3x2 & 3x3x3 & 3x3x3 & 4x4x2 & 4x4x4 &  \\
\hline
100 & 284(2) & 301(2) & 283(2) & 287(2)  &  289(2) & 282(2) & 289(2) & 283.6\\ 
400 &  287(2) & 299(2) & 286(2) & 289(2)  & 291(2) & 284(2) & 289(2) & 286.2\\
1000 &  289(2) & 299(2) & 287(2) & 294(2)  & 299(2) & 286(2) & 294(2) & 289.7\\
2000 &  292(2) & 301(2) & 291(2) & 296(2) & 303(2) & 289(2) & 296(2) & 292.6\\ 
3000 &  287(2)  & 301(2) & 291(2) & 296(2) &  301(2)& 291(2) & 296(2)& 293.7\\  
4000 &  284(2) & 296(2) & 289(2) & 293(2) &  299(2)& 289(2) & 293(2)& 293.4\\
5000 &  -- & 289(2) & 284(2) & 289(2) &  296(2) & 281(2)& 291(2)& 292.1\\
6000 &  -- & 279(2) & 276(2) & 286(2) &  291(2) & 276(2) & 289(2)& -- \\
\end{tabular}
\caption{\justifying{Three-phase coexistence temperatures ($T_3$), at different pressures, for the CO$_2$ hydrate obtained for the 6 size-dependent configurations as obtained from MD-$NPT$ computer simulations. Simulations for configuration 0  correspond to the original simulation obtained the M\'{\i}guez \emph{et al.}~\cite{Miguez2015a} Experimental data taken from the literature is also included.~\cite{Sloan2008a}}} 
\label{t3_results}
\end{ruledtabular}
\end{table*}

In this work, all the molecular dynamic simulations have been carried out using GROMACS (version 4.6, double precision).~\cite{VanDerSpoel2005a} CO$_{2}$ molecules are modeled through the TraPPE (Transferable Potentials for Phase Equilibria) force field~\cite{Potoff2001a} and water molecules are described using the widely-known TIP4P/Ice model.~\cite{Abascal2005b} As in our previous work,~\cite{Miguez2015a} Lorentz-Berthelot combining rules have been applied, but the Berthelot combining rule for the unlike dispersive interactions between water and CO$_2$ molecules has been modified by a  $\xi=1.13$ factor. This factor was used in order to match the results obtained in our previous work\cite{Miguez2015a} with experimental results taken from the literature. However, must be taken into account that this factor is rectifying the discrepancies with the experimental results, not only for the weakness of the unlike interaction descriptions of the molecular models but also for the finite-size effects presented in our previous study. We have used a Verlet leapfrog algorithm\cite{Cuendet2007a} to numerically solve Newton's motion equations. The time step used is $2\,\text{fs}$ since all the models employed in this work are rigid. We have also used the Nosé-Hoover thermostat\cite{Nose1984a} and the Parrinello-Rahman barostat,\cite{Parrinello1981a} to ensure that simulations are performed at constant temperate and pressure. The time constant used for both, thermostat and borastat, is $2\,\text{ps}$. It is important to remark that the barostat has been applied anisotropically to avoid any stress from the solid hydrate structure. We have used a cutoff distance of $1.0\,\text{nm}$ for dispersive and Coulombic interactions. No long-range corrections have been applied for the dispersive interactions. For the case of the coulombic interactions, long-range particle mesh Ewald (PME) corrections~\cite{Essmann1995a} have been applied with a width of the mesh is $0.1\,\text{nm}$ and a relative tolerance of $10^{-5}$.

\section{Results}

In this section, we show the results obtained to understand the role of finite-size effects in determining the dissociation temperature, as a function of pressure, of the CO$_{2}$ hydrate. As in paper I~\cite{paperI}, we estimate the $T_{3}$ in different systems of varying sizes.  Table~\ref{tabla-moleculas} shows the initial number of molecules forming each phase and the unit cell replication factor for the 6 size-dependent configurations. We have also included, for comparison reasons, the configuration 0 corresponding to the system studied by M\'{\i}guez \emph{et al.}~\cite{Miguez2015a} In this work, we extend the study of paper I and consider 8 different pressures in each configuration, from $100$ to $6000\,\text{bar}$. This allows to analyze how the $T_{3}$ depends on the size of the system simulated in a wide range of pressures.

As in our previous work, we consider two different scenarios: systems in which the H$_{2}$O- and CO$_{2}$-rich liquid phases are formed from stoichiometric configurations, and systems in which the liquid phases are formed from configurations with a higher number of CO$_{2}$ molecules in the CO$_{2}$-rich liquid phase. In the first case, when $T<T_3$, the simulations evolve into a single phase of CO$_{2}$ hydrate (configurations 1 and 4). In the second case, the systems evolve into two phases, CO$_{2}$ hydrate and CO$_{2}$-rich liquid phase (configurations 2, 3, 5, and 6). Note that in the first estimation of the three-phase coexistence temperature of the CO$_{2}$ hydrate, the configuration used was non-stoichiometric (configuration 0).

 It is worth mentioning that the dissociation line of the CO$_{2}$ hydrate is not entirely analogous to that of the CH$_{4}$ hydrate considered in paper I.~\cite{paperI} In this case, the three-phase involves the corresponding solid phase and two liquid phases, a water-rich liquid phase and a CO$_{2}$-rich liquid phase; in the CH$_{4}$ case, one of the fluid phases is not liquid since the conditions of temperature and pressure are supercritical and the CH$_{4}$-rich phase behaves as a fluid of low density that can be considered as a gas phase.

\begin{figure*}
     \centering
     \begin{subfigure}[hbt]{0.315\textwidth}
         \centering
         \includegraphics[width=1.10\textwidth]
        {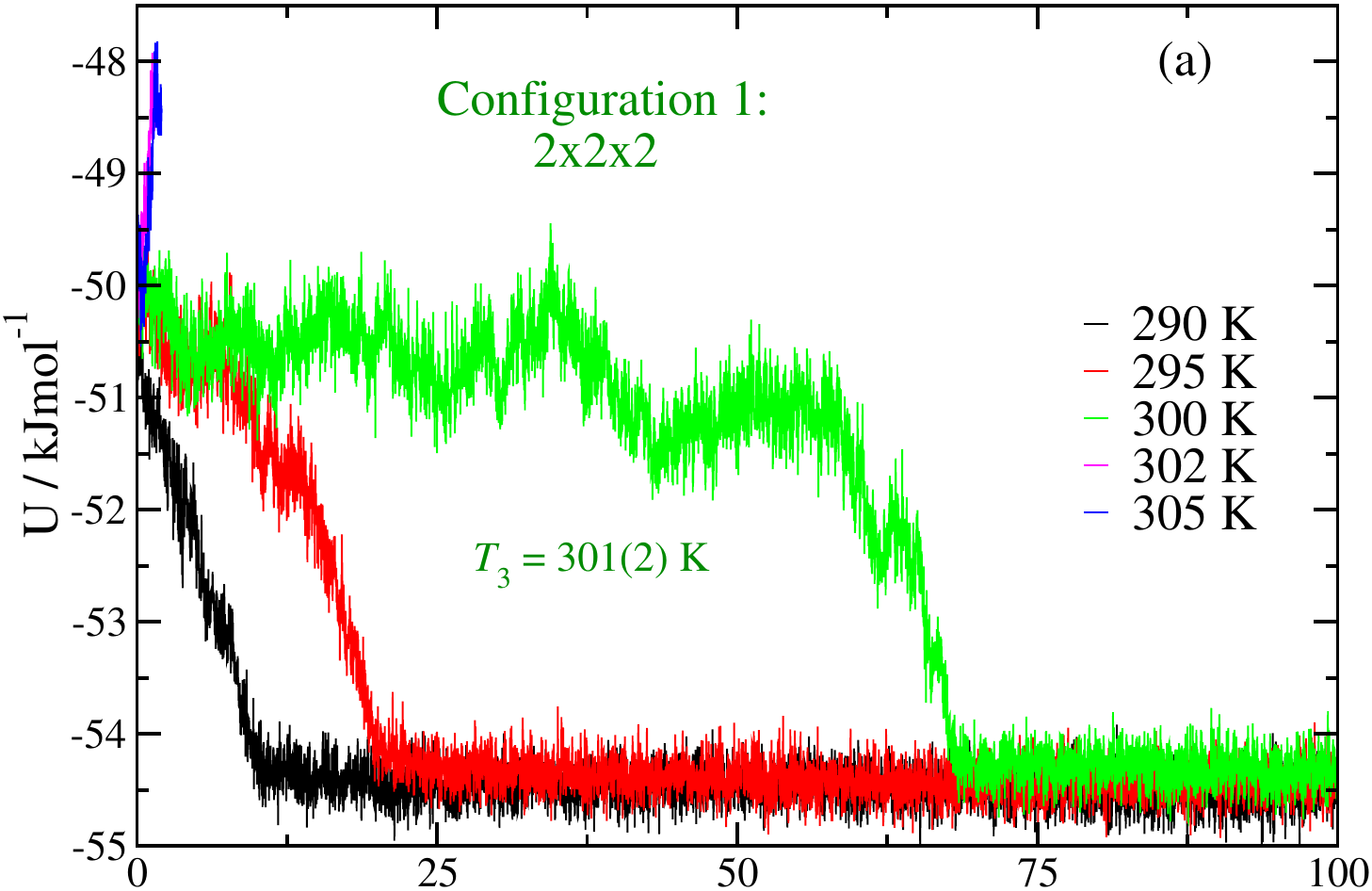}
         \caption{Conf. 1}
         \label{conf1}
     \end{subfigure}
     \hfill
     \begin{subfigure}[hbt]{0.30\textwidth}
         \centering
         \includegraphics[width=1.10\textwidth]{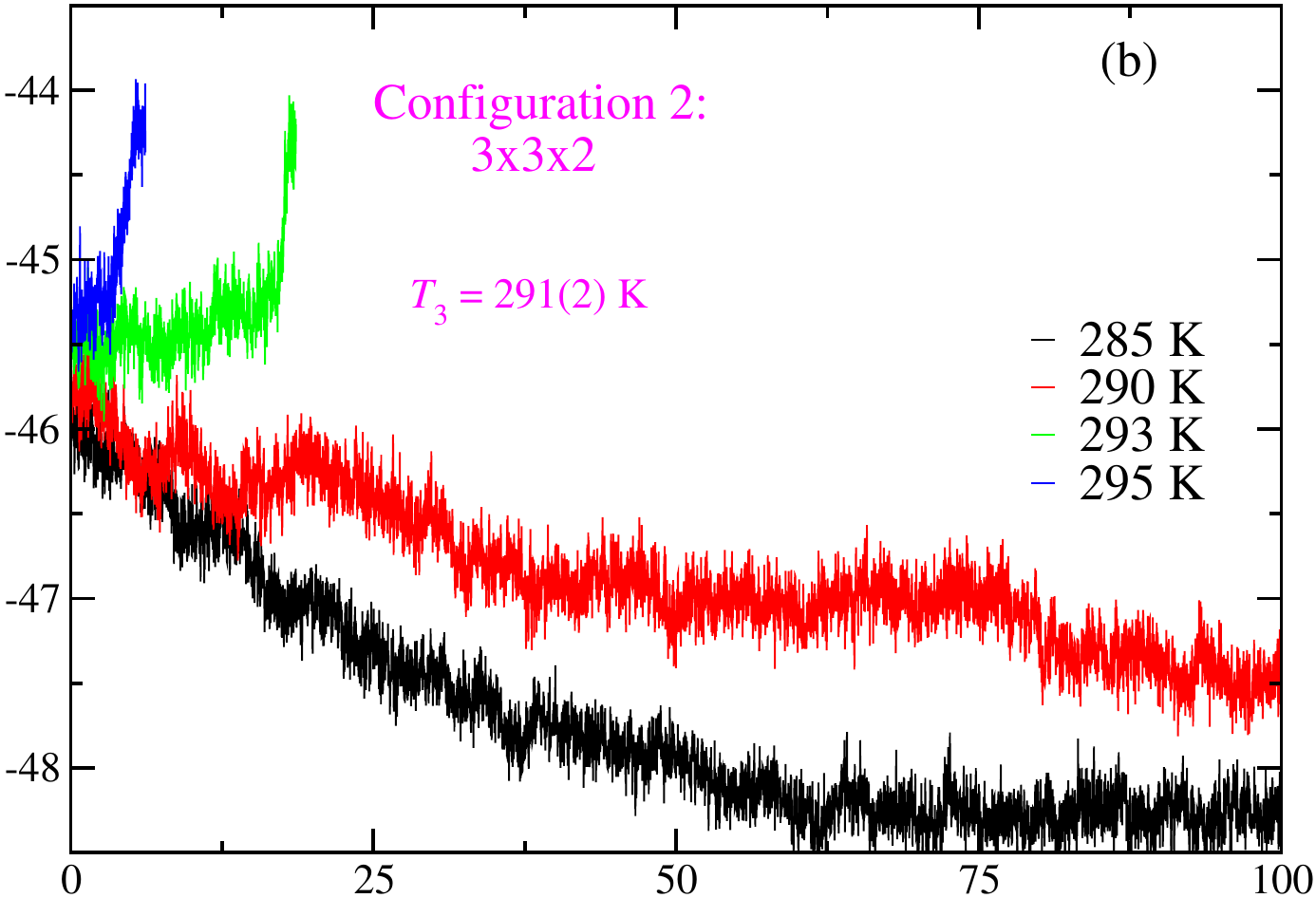}
         \caption{Conf. 2}
         \label{conf2}
     \end{subfigure}
     \hfill
     \begin{subfigure}[hbt]{0.30\textwidth}
         \centering
         \includegraphics[width=1.10\textwidth]{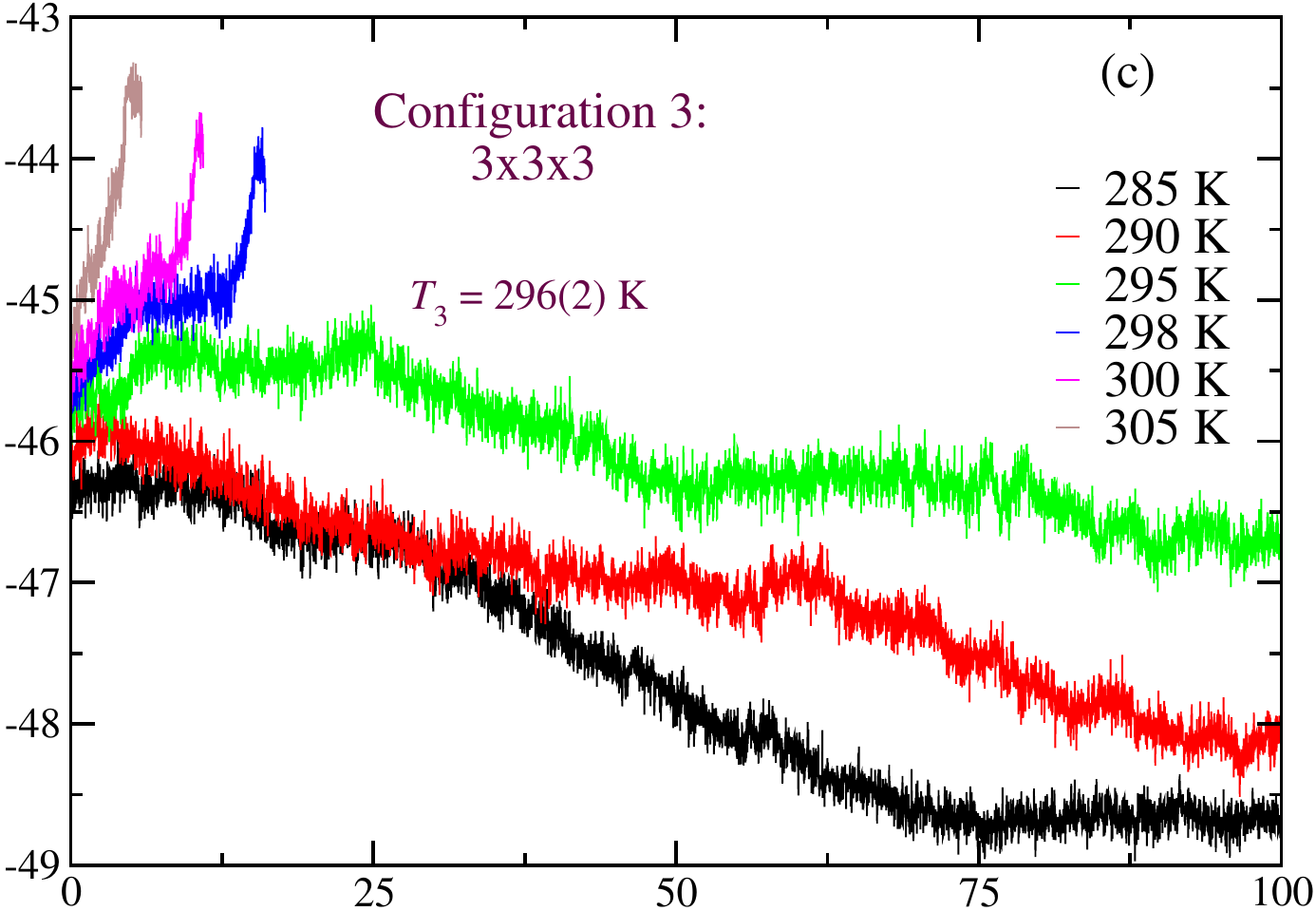}
         \caption{Conf. 3}
         \label{conf3}
     \end{subfigure}
     \hfill
     \begin{subfigure}[hbt]{0.315\textwidth}
         \centering
         \includegraphics[width=1.10\textwidth]{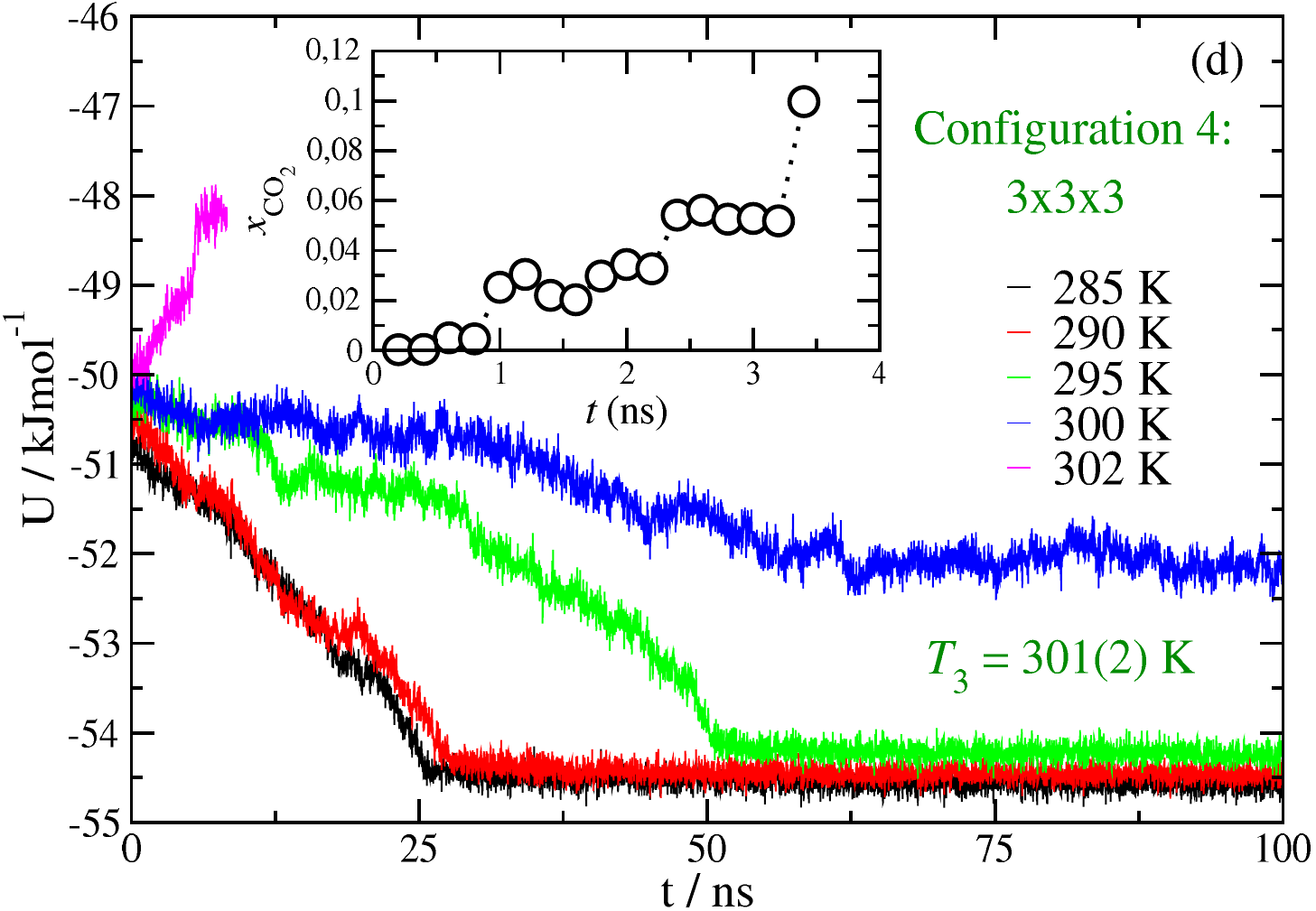}
         \caption{Conf. 4}
         \label{conf4}
     \end{subfigure}
     \hfill
     \begin{subfigure}[hbt]{0.30\textwidth}
         \centering
         \includegraphics[width=1.10\textwidth]{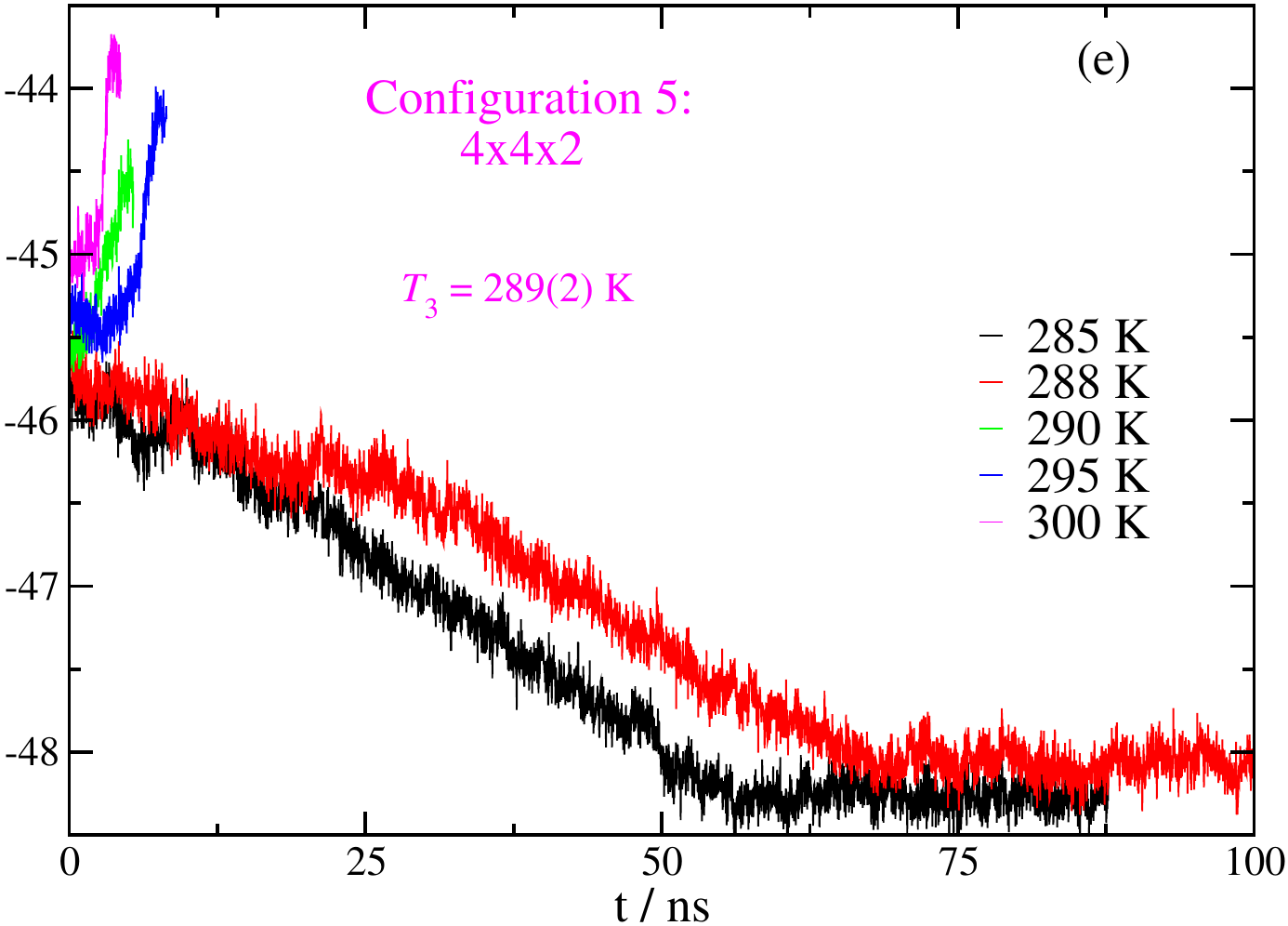}
         \caption{Conf. 5}
         \label{conf5}
     \end{subfigure}
     \hfill
     \begin{subfigure}[hbt]{0.30\textwidth}
         \centering
         \includegraphics[width=1.10\textwidth]{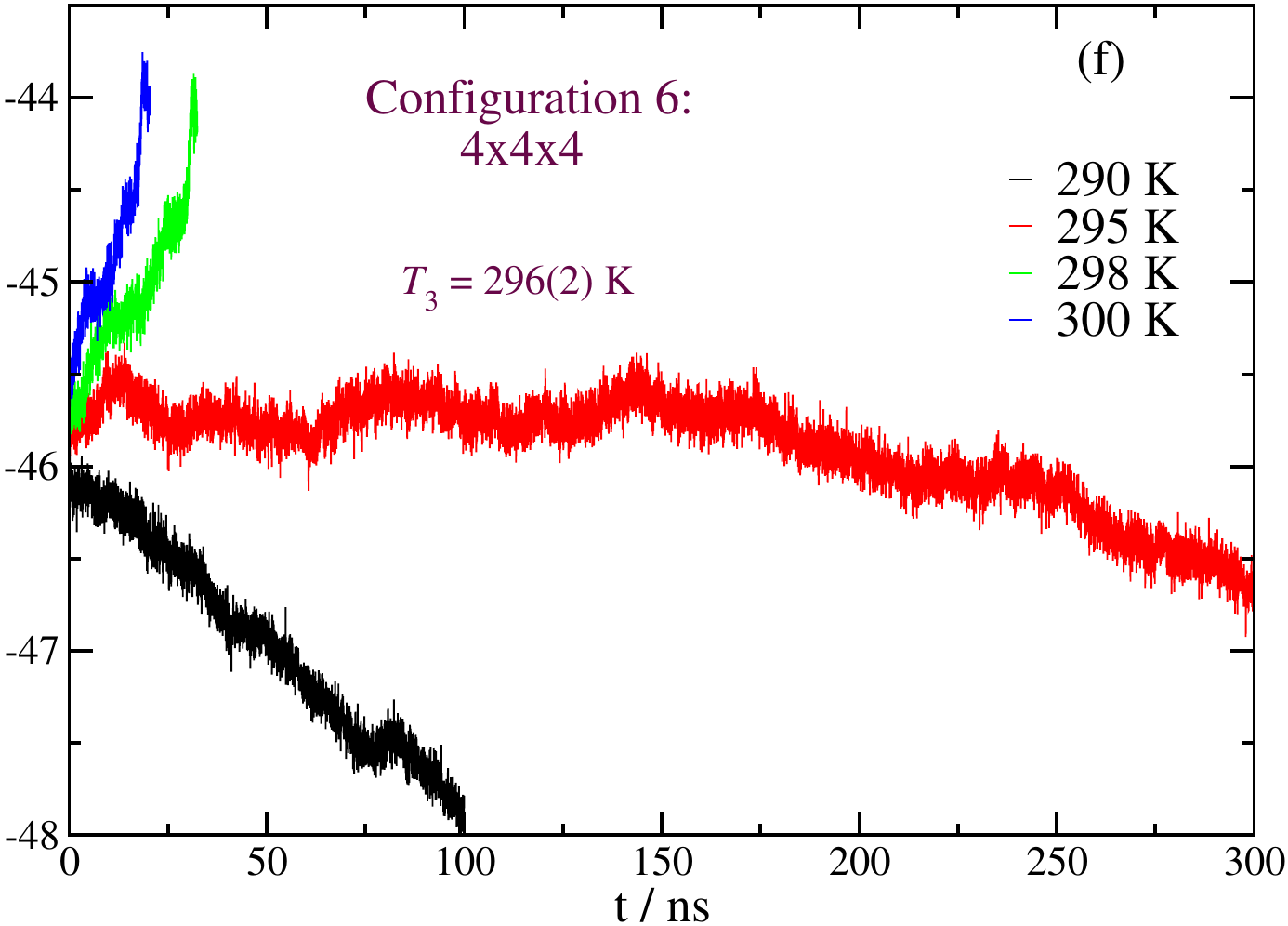}
         \caption{Conf. 6}
         \label{conf6}
     \end{subfigure}
\caption{
\justifying{Evolution of the potential energy, as a function of time, for the 6 size-dependent configuration analyzed in the work. The results are obtained from MD-$NPT$ simulations at $2000\,\text{bar}$. Notice that panel (d) includes an inset with the molar fraction of CO$_{2}$ in the aqueous phase as a function of time at $285\,\text{K}.$}}
\label{u_evolution}
\end{figure*}

\subsection{Effect of the stoichiometric configuration}

We first consider two stoichiometric configurations, i.e., configurations in which the number of molecules of CO$_{2}$ and H$_{2}$O in the hydrate phases is the same as those in both liquid phases. The first configuration we analyze (configuration 1) is a system formed from 368 and 64 water and CO$_{2}$ molecules in the hydrate phase (all the cages of the hydrates are fully occupied, i.e., 8 CO$_{2}$ molecules per 46 water molecules in each unit cell in a $2\times 2\times 2$ sI structure) and the same number of molecules of each species in the liquid phases. According to the nomenclature used previously in paper I,~\cite{paperI} this is a stoichiometric configuration. As we have already seen, it is important to compare the results obtained from simulations using stoichiometric systems with simulation data obtained using non-stoichiometric configurations. In this work, we compare the results with simulation data obtained by M\'{\i}guez \emph{et al.}~\cite{Miguez2015a} in which some of us estimated the three-phase coexistence temperature of the CO$_{2}$ hydrate using the same solid configuration ($2\times 2\times 2$) surrounded by liquid water (with the same number of water molecules) and a second liquid phase formed from $192$ CO$_{2}$ molecules, i.e., the triple number than that of the hydrate phase. This system is the configuration 0. According to the nomenclature used in this work, configuration 0 is not stoichiometric (see Table~\ref{tabla-moleculas}).

We have determined the three-phase coexistence temperature of configuration 1, $T_{3}$, in a wide range of pressures, from $100$ to $6000\,\text{bar}$, following the methodology already explained in paper I~\cite{paperI} and in Section II. We first concentrate at a given pressure, $2000\,\text{bar}$, and then analyze the behavior of the system at different pressures. Fig.~\ref{conf1} shows the evolution of the potential energy of the system, $U$, as a function of time, at $2000\,\text{bar}$ and temperatures from $290$ to $305\,\text{K}$. As can be seen, two clear behaviors are observed. At the highest temperatures, $302$ and $305\,\text{K}$, the potential energy increases very quickly over time, indicating the melting of the CO$_{2}$ hydrate. Note that potential energy curves obtained at $302$ and
$305\,\text{K}$ require very careful examination since the hydrate melts very
quickly, in less than $3\,\text{ns}$, approximately. However, at low temperatures, $300$, $302$, and $305\,\text{K}$, the potential energy shows a sharp decrease, which is more pronounced as the temperature is lower, indicating that the hydrate solid phase grows. Since at $302\,\text{K}$ the potential energy increases and at $300\,\text{K}$, the potential energy decreases, the three-phase coexistence temperature at $2000\,\text{bar}$ is estimated at $T_{3}=301(2)\,\text{K}$. It is interesting to compare this prediction with the $T_{3}$ corresponding to the configuration 0 at the same pressure. According to Table~\ref{t3_results}, the $T_{3}$ obtained by M\'{\i}guez \emph{et al.}~\cite{Miguez2015a} was $292(2)\,\text{K}$. Based on this, the dissociation temperature of the stoichiometric configuration is $9\,\text{K}$ above that of the non-stoichiometric configuration of the liquid phases. This result is in agreement with our findings in paper I,~\cite{paperI} indicating that the use of stoichiometric configurations produces an overestimation of the $T_{3}$.

Note that characteristic times at which the CO$_2$ hydrate crystallizes are shorter than those compared with the CH$_{4}$ hydrate. This is clearly seen comparing Figs.~\ref{conf1} of this work and Fig.~2a of paper I.~\cite{paperI} In both cases, the system has exactly the same number of molecules in each phase. As can be seen, in the case of the CO$_{2}$, crystallization of the hydrate occurs relatively soon, between $10$ and $75\,\text{ns}$. However, the CH$_{4}$ hydrate needs even $200\,\text{ns}$ to crystallize if temperature is close of the $T_{3}$. This is an expected result since hydrate growth rate is controlled by mass transfer processes, which depends on the solubility and diffusivity of the guest molecules in water. Note that not only solubility but also diffusivity of CO$_{2}$ in water are one order of magnitude higher than those of CH$_{4}$ in water.~\cite{Grabowska2022a,Algaba2023a}

As we have previously mentioned, stoichiometric configurations simulated at temperatures below the $T_{3}$ of the hydrate, at the corresponding pressure, evolve into singular phases of hydrate. This is confirmed by the sharp decrease in the potential energy already shown in Fig.~\ref{conf1}. As we will see later, this behavior indicates that the growing mechanism of the hydrate from these configurations is not only due to a layer-by-layer formation of the hydrate. In fact, as it happens with the CH$_{4}$ hydrate,~\cite{paperI} this behavior confirms the presence, not of a bubble of CO$_{2}$ but a liquid drop of CO$_{2}$ within the liquid phase. Recall that CO$_{2}$ and water exhibit liquid-liquid immiscibility at the conditions at which the hydrate forms.~\cite{Sloan2008a} This liquid drop collapses, at a certain time, producing a very quick formation of the hydrate phase corroborated by the abrupt decrease in the potential energy at this time. We have checked that the liquid drop mechanism is observed in the whole range of pressures considered in configuration 1. It is also important to remark on the quantitative difference found between the decrease of the potential energy of the CH$_{4}$ and CO$_{2}$ hydrates. In the first case, as can be seen in Fig.~2a of paper I,~\cite{paperI} the sharp decrease is more pronounced than that compared with the CO$_{2}$ hydrate, as shown in Fig.~\ref{conf1} of this work. This indicates a quicker dissolution of the CH$_{4}$ bubbles than the CO$_{2}$ liquid drop. A more detailed account of the formation of the liquid drops in the CO$_{2}$ hydrate is presented below.

As we have already mentioned in paper I,~\cite{paperI} this bubble formation has been previously observed by several authors. Walsh \emph{et al.}~\cite{Walsh2009a,Walsh2011a} calculated
nucleation rates after observing spontaneous nucleation
of methane hydrate preceded by the formation of a bubble. In 2010, Conde and
Vega~\cite{Conde2010a} also observed bubble formation before
hydrate growth when determining the $T_3$ of the methane hydrate. This bubble formation was also
shown by Liang and Kusalik~\cite{Liang2011a} for H$_2$S systems. After
these pioneering studies, other works have studied the effect of bubble formation in the dissociation temperature
of methane hydrates.~\cite{Yagasaki2014a,Grabowska2022a,Fang2023a,Bagherzadeh2015a}

In paper I, we have concentrated on the estimation of the $T_{3}$ at a fixed pressure, $400\,\text{bar}$. In this work, we extend the study and consider the dissociation temperature of the CO$_{2}$ hydrate in a wide range of pressures. We have used the same procedure to evaluate the $T_{3}$ of the hydrate in a wide range of pressures, from $100$ to $6000\,\text{bar}$. At each pressure, we have simulated several temperatures, separated $5\,\text{K}$, from $280$ up to $310\,\text{K}$, approximately. The results obtained are presented in Table~\ref{t3_results} (predictions obtained for the configuration 0 by M\'{\i}guez \emph{et al.}~\cite{Miguez2015a} are also included for comparison reasons).

\begin{figure*}
\centering
\hspace*{-0.2cm}
\includegraphics[width=0.75\columnwidth]{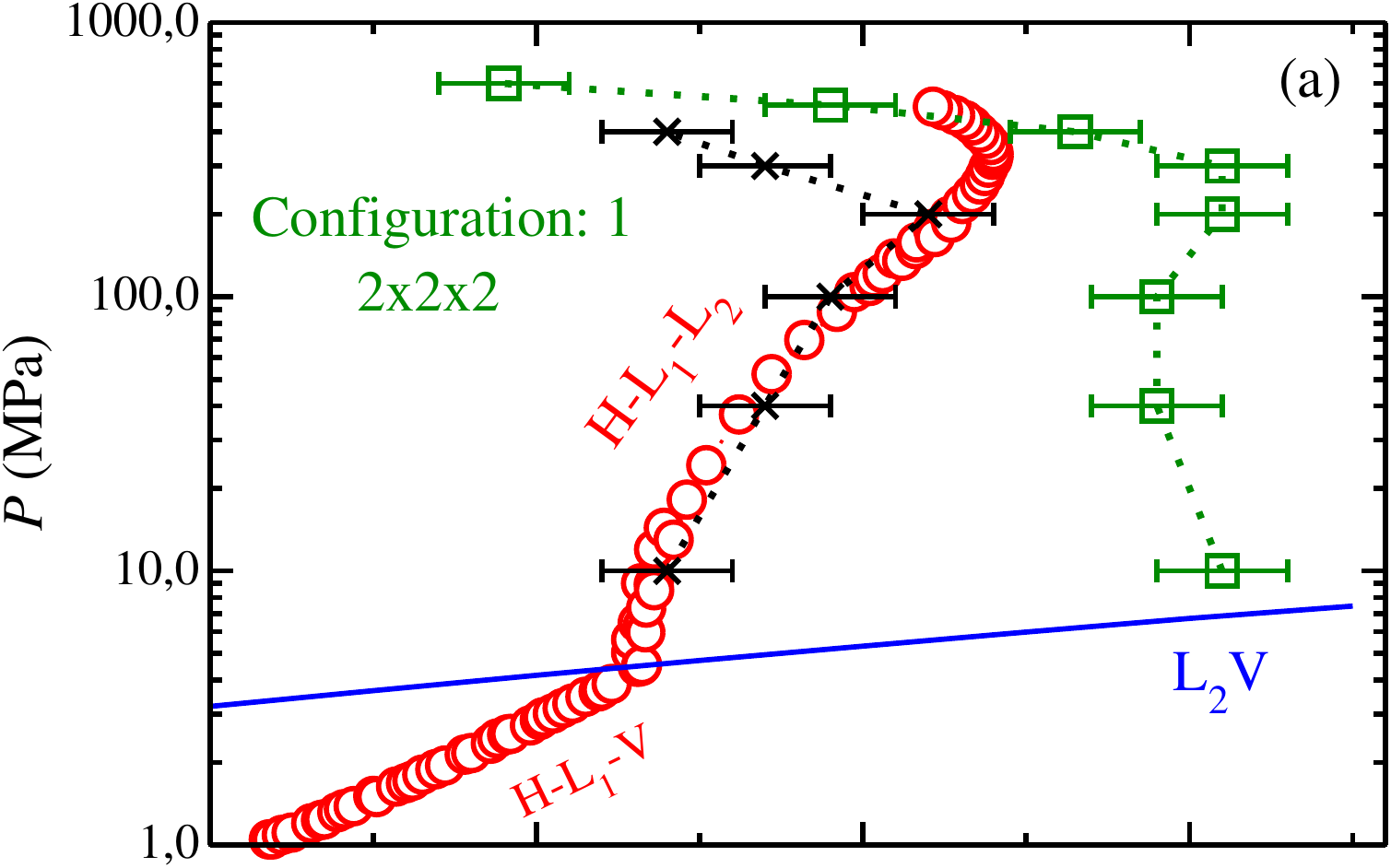}
\includegraphics[width=0.65\columnwidth]{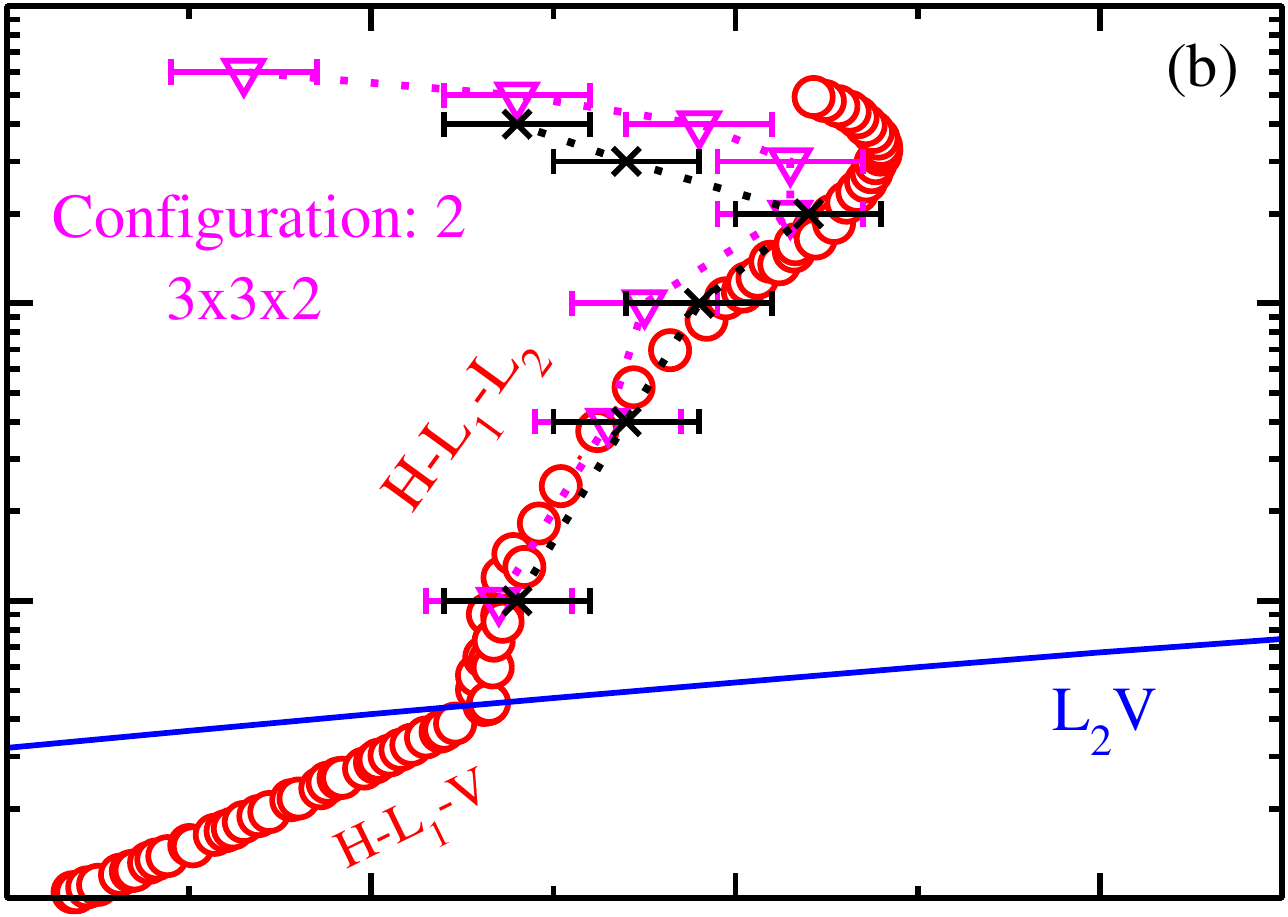}
\includegraphics[width=0.65\columnwidth]{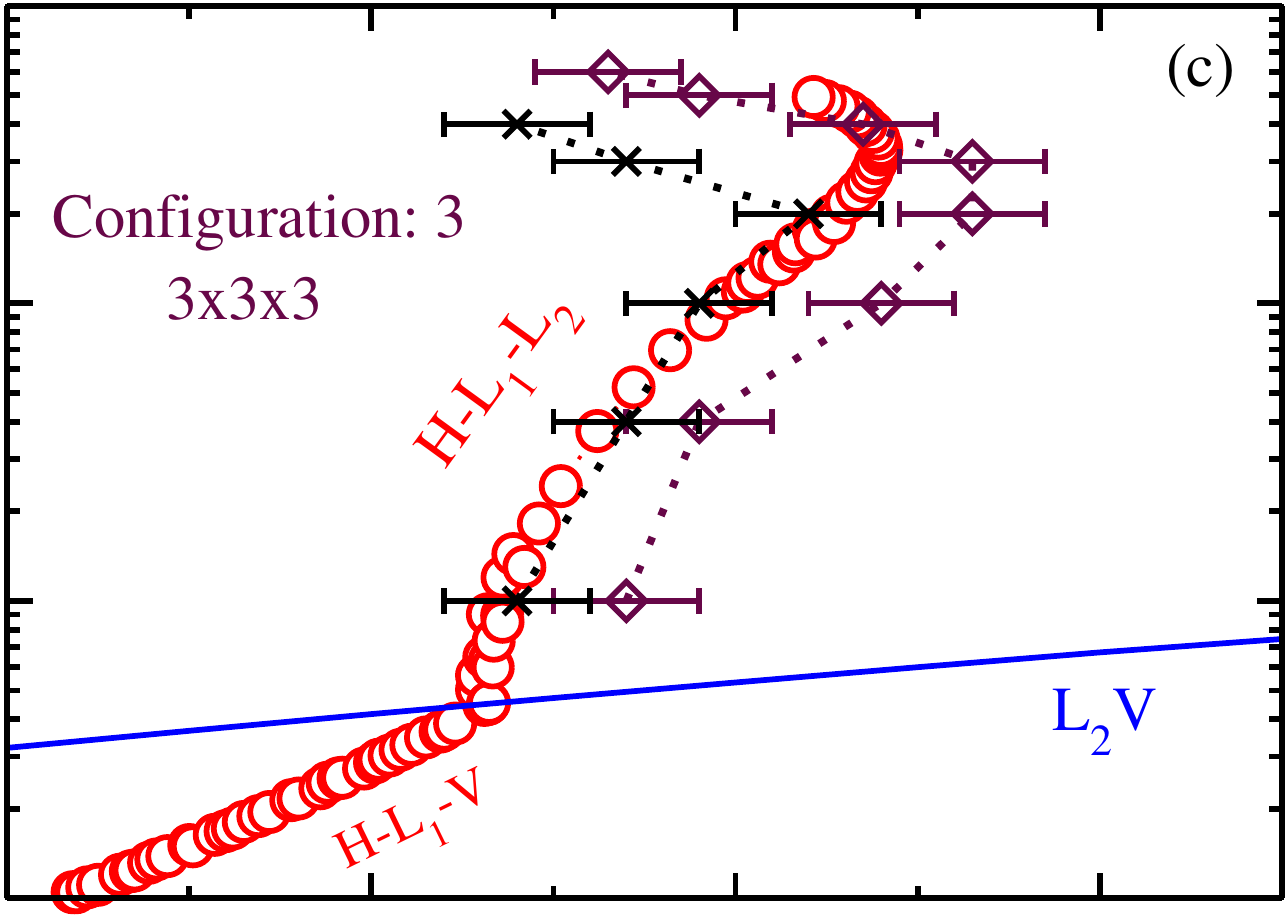}
\includegraphics[width=0.73\columnwidth]{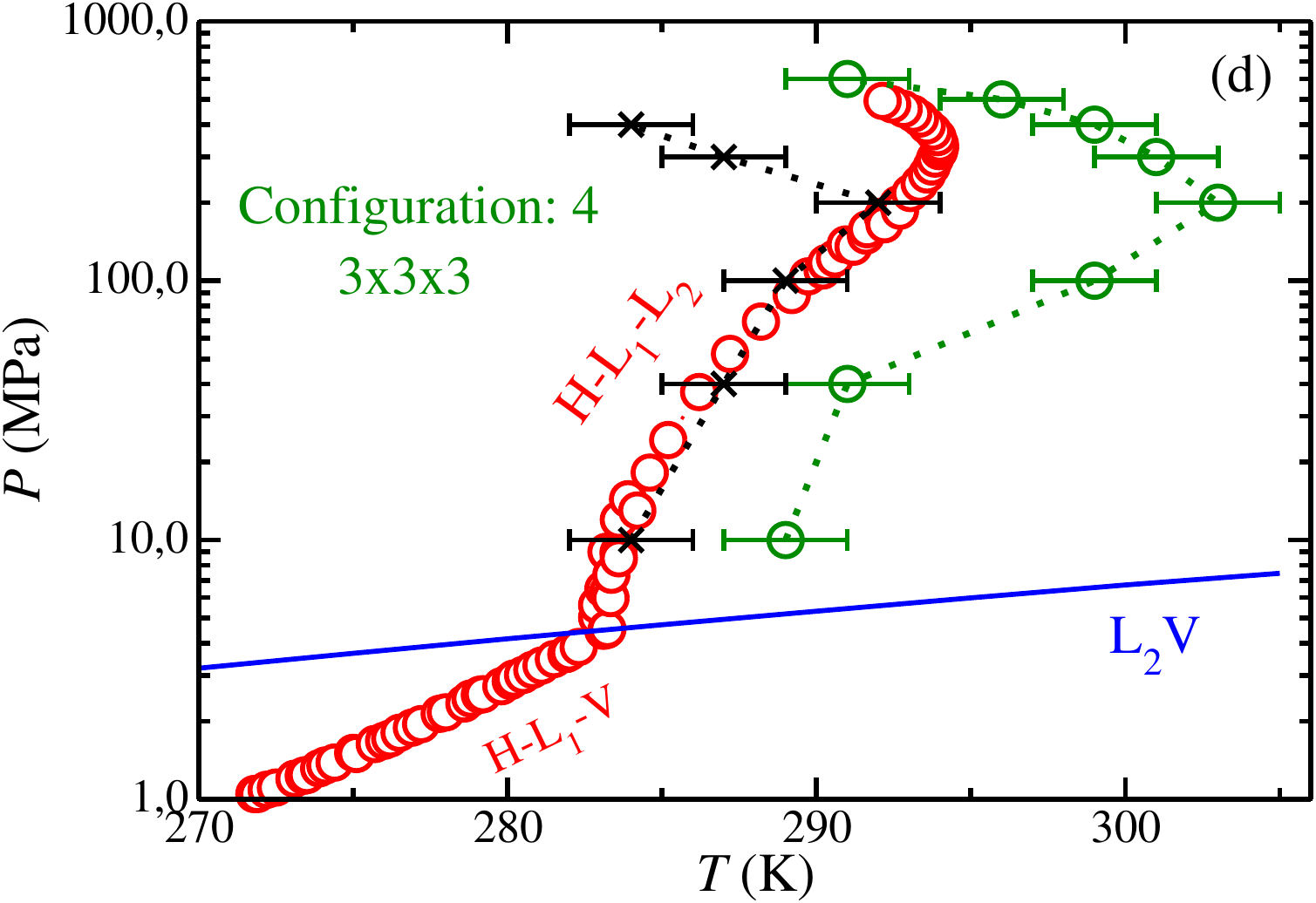}
\includegraphics[width=0.65\columnwidth]{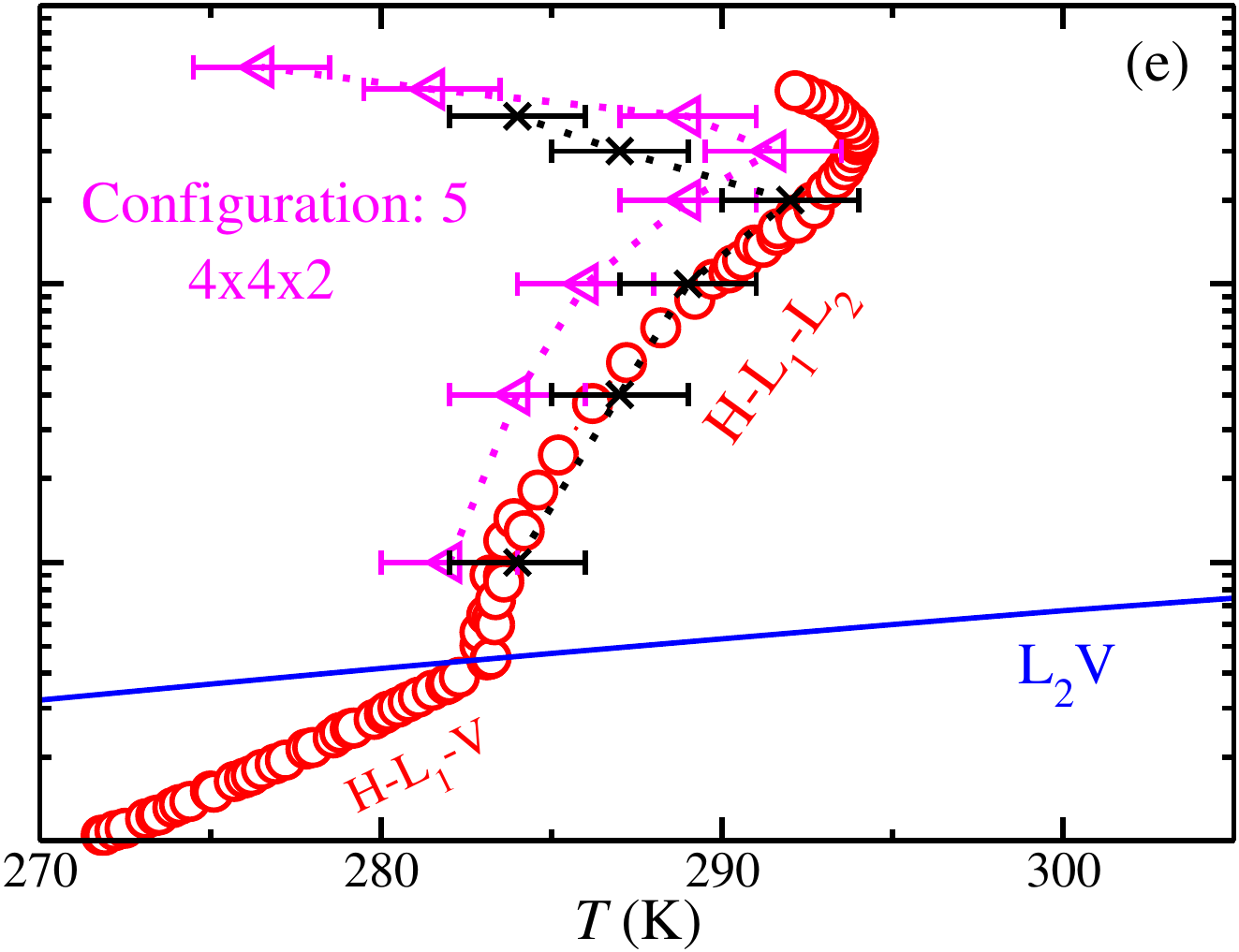}
\includegraphics[width=0.65\columnwidth]{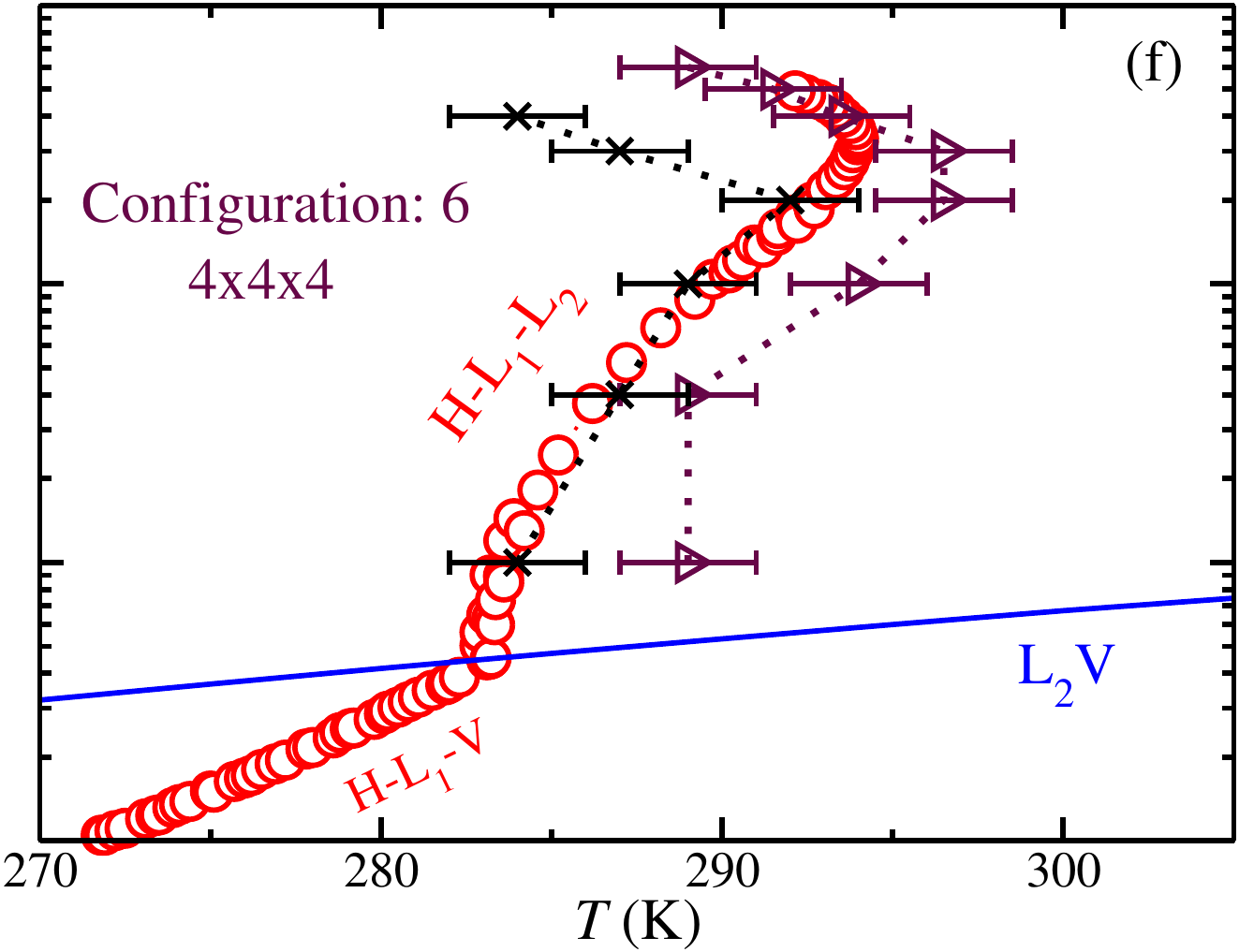}
\caption{\justifying{{Pressure–temperature or $PT$ projection of the dissociation line of the CO$_{2}$ hydrate. The green squares (a), violet down triangles (b), maroon diamonds (c), green circles (d), violet left triangles (e), and maroon right triangles (f) represent the predictions obtained from simulation using the corresponding configuration. The red circles represent the experimental data taken from the literature~\cite{Sloan2008a} and the blue curve is the experimental vapor pressure of the CO$_{2}$. Black crosses stand for the results obtained in the work of M\'{\i}guez \emph{et al.}~\cite{Miguez2015a}}}}
\label{pt_results}
\end{figure*}

To have an overall vision of the results obtained along the whole range of pressures, we have represented all the $T_{3}$ estimated in this work in a $PT$ or pressure-temperature projection, as shown in Fig.~\ref{pt_results}. Particularly, Fig.~\ref{pt_results}a shows the diagram corresponding to configuration 1. We have also included the simulation results of the configuration 0, studied by M\'{\i}guez \emph{et al.}~\cite{Miguez2015a} using a non-stoichiometric configuration (see Table~\ref{tabla-moleculas} for further details). As can be seen, the main effect of using a stoichiometric instead of a non-stoichiometric configuration is to displace the whole dissociation line towards high temperatures. Particularly, the dissociation line of configuration 1 is located $9-14\,\text{K}$, depending on the pressure, above the dissociation line of configuration 0 (non-stoichiometric). Note that at $100\,\text{bar}$, differences between both $T_{3}$ values are $14\,\text{K}$. See the Table~\ref{t3_results} for further details. Differences between the $T_{3}$ values of stoichiometric and non-stoichiometric ($2\times 2\times 2$) are much larger than in the case of the CH$_{4}$ hydrate.~\cite{paperI}\\

We now study a second stoichiometric configuration but considering a larger system. In this case, we replicate the unit cell of the hydrate three times ($3\times 3\times 3$), instead of the two times of configuration 1 ($2\times 2\times 2$). According to this, the initial hydrate phase is now formed by $1242$ and $216$ water and CO$_{2}$ molecules, respectively. In order to have the same stoichiometry in the liquid phase, we surround the initial hydrate phase by a slab with $1242$ water molecules and a second slab with $216$ CO$_{2}$ molecules. This system corresponds to configuration 4 presented in Table~\ref{tabla-moleculas}, which is similar to configuration 6 presented in paper I for the CH$_{4}$ hydrate.~\cite{paperI} It is interesting to compare the results obtained from this new stoichiometric configuration and check if the CO$_{2}$ hydrate shows the same general behavior.

Fig.~\ref{conf4} shows the evolution of the potential energy of configuration 4, as a function of time, at $2000\,\text{bar}$ and several temperatures, from $285$ to $302\,\text{K}$. As can be seen, there is an increase in the time needed to observe the crystallization or melting of the hydrate. It is worthy to mention that the time required to observe one of the behaviors is not so large as in the case of the CH$_{4}$ hydrate. This indicates again that CO$_{2}$ hydrates grow much quicker than CH$_{4}$ hydrates due to the high solubility of CO$_{2}$ in water.~\cite{blazquez2023growth} The dissociation temperature of the hydrate at $2000\,\text{bar}$ ranges between $302$ (the lowest temperature showing an increase in potential energy) and $300\,\text{K}$ (the highest temperature showing a decrease in potential energy). According to this, $T_{3}=301(2)\,\text{K}$. The corresponding dissociation temperature of configuration 0 (non-stoichiometric), at the same pressure ($2000\,\text{bar}$), is $292(2)\,\text{K}$. This is the same value obtained in the case of configuration 1. The potential energy curves, as functions of time for temperatures below the $T_{3}$, exhibit a similar behavior to those observed in configuration 1 (see Figs.~\ref{conf1} and \ref{conf4}), with two minor differences: (1) the slope of the sharp decrease is now lower; and (2) the time required to observe the decrease is higher. The reason is due to the large size of configuration 4 compared with that of configuration 1 ($3.375$ times in terms of number of molecules). As we will see later, this also confirms the quick formation of liquid drops before the layer-by-layer growth of the hydrate at temperatures below the $T_{3}$.

\begin{figure*}[htp]
    \centering
\vspace*{0.5cm}
\includegraphics[width=1.1\columnwidth]{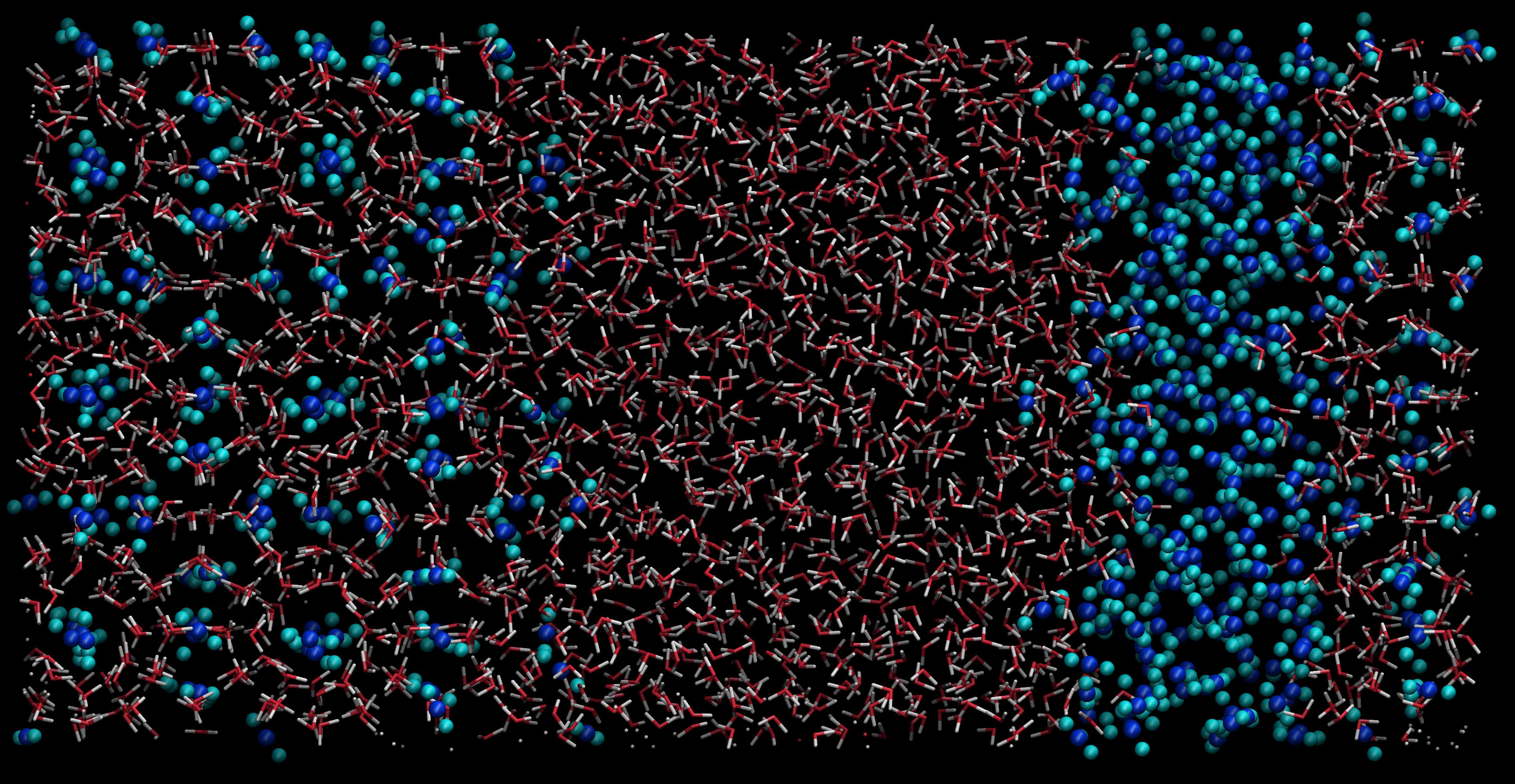}
\vspace*{0.5cm}

\includegraphics[width=1.1\columnwidth]{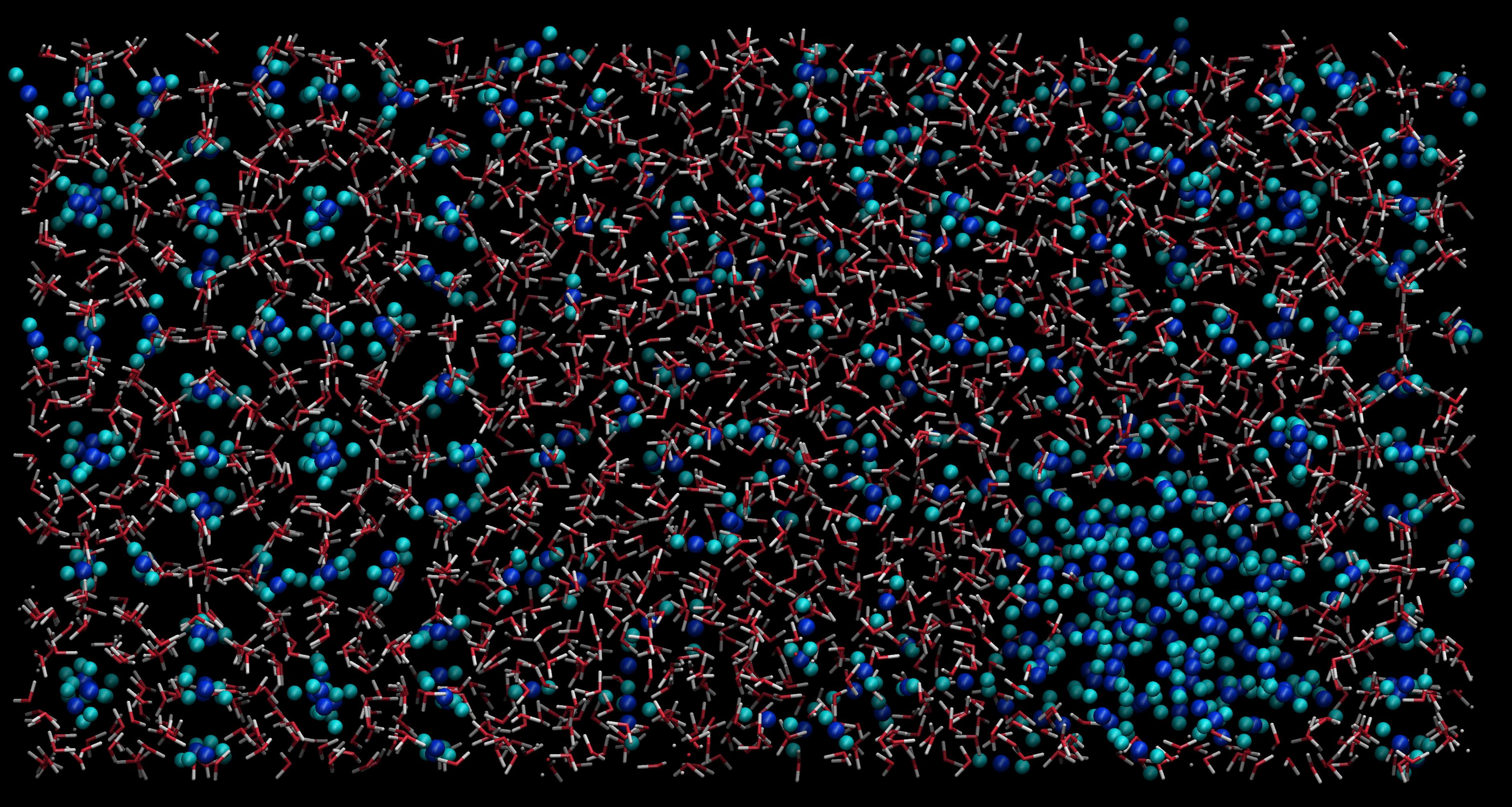}
\vspace*{0.5cm}

\includegraphics[width=1.1\columnwidth]{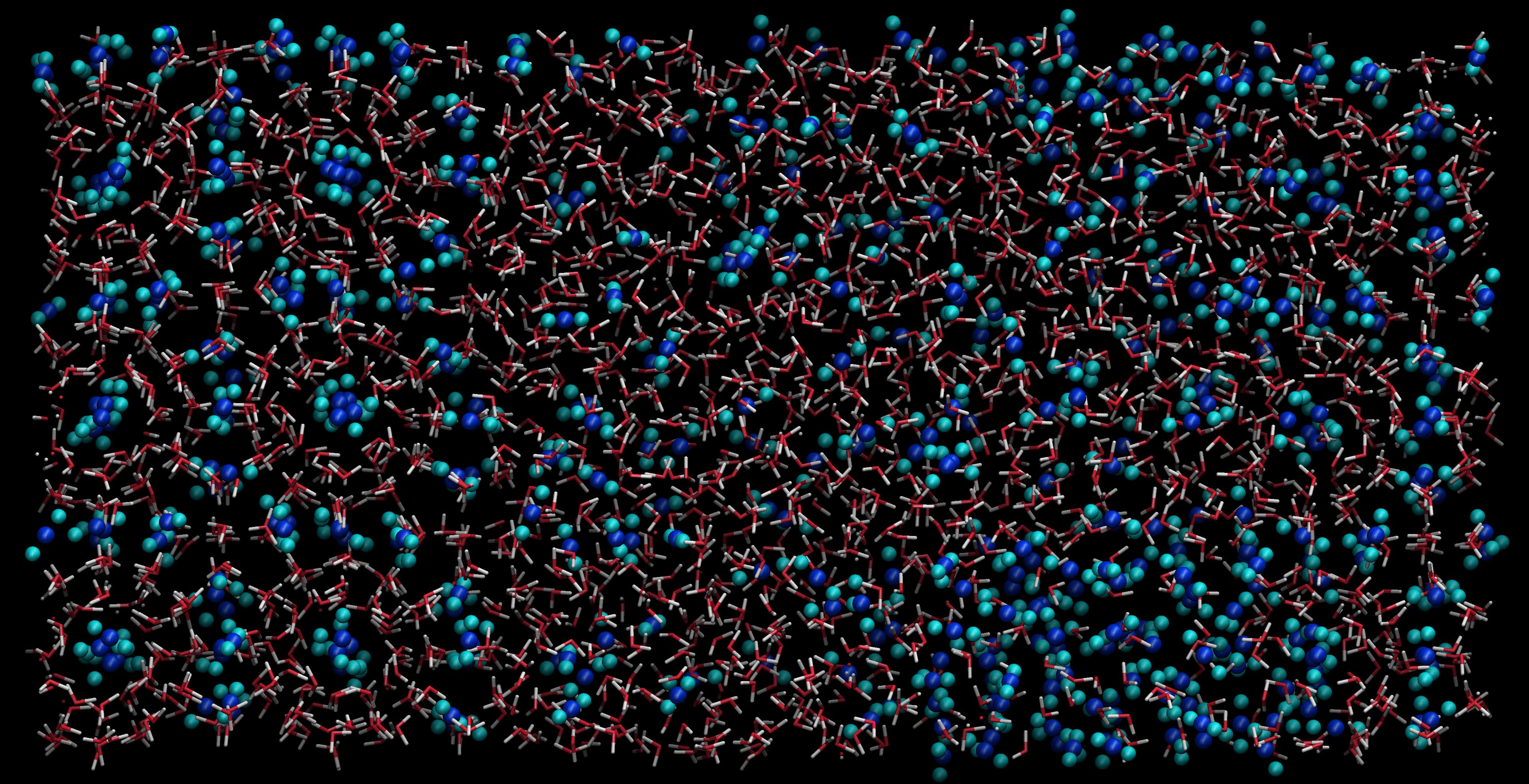}
\vspace*{0.5cm}

\includegraphics[width=1.1\columnwidth]{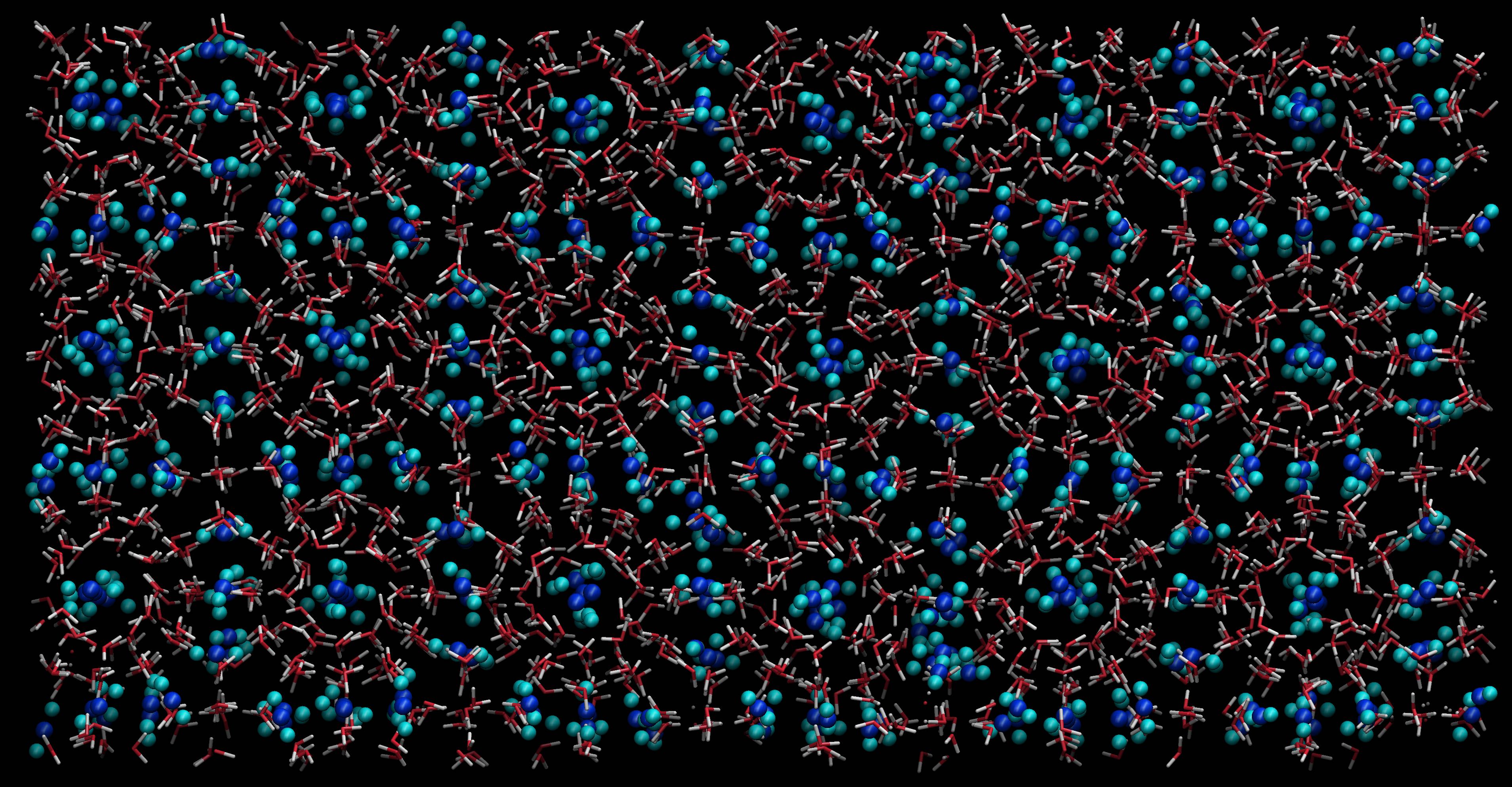}
    \caption{\justifying{
    Snapshots taken at different times during $100\,\text{ns}$ simulation run for configuration 4 at 285 K and 2000 bar. The formation of a liquid drop and the growth of carbon dioxide hydrate are shown. From top to bottom: $t=0$ ns, three-phase initial configuration. $t=3$ ns, formation of a carbon dioxide drop within the liquid phase. $t=4$ ns, rupture of the carbon dioxide drop and the formation of an oversaturated solution. $t=25\,\text{ns}$, complete growth of the CO$_{2}$ hydrate. Water molecules are represented as white and red sticks and CO$_{2}$ molecules as cyan (carbon atom) and blue spheres (oxygen atoms).
    }}
    \label{drop}
\end{figure*}

The complete dissociation line of configuration 4 is also presented in Fig.~\ref{pt_results}d in a pressure-temperature projection of the phase diagram as in the previous case. We have also included the results corresponding to configuration 0 obtained by M\'{\i}guez \emph{et al.}~\cite{Miguez2015a}. As it happens for configuration 1 (Fig.~\ref{pt_results}a),
the dissociation line of configuration 4 is shifted towards higher temperatures at all pressures. However, the displacement is pressure-dependent if we compare it with the displacement observed in the case of configuration 1. Now, the shift is smaller at low pressures, similar at intermediate pressures, and larger at high pressures. This plot, with the results discussed in the previous paragraph, confirms that the liquid drop formation during the CO$_{2}$ hydrate growth is caused by the stoichiometric composition of the system, regardless of its size and the guest.

As in the case of the CH$_{4}$ hydrate study in our paper I,~\cite{paperI} it is possible to observe the complete growth sequence of a liquid drop for configurations 1 and 4. Here we only show the formation of the CO$_{2}$ liquid drop in system 4, which corresponds to the initial $3\times 3\times 3$ configuration of the hydrate. It is important to mention that we have observed the same mechanism of formation of liquid drops, not only at all the pressures considered in this work ($100-6000\,\text{bar}$) for this configuration but also at all of the pressures for configuration 1 ($2\times 2\times 2$ system). The main difference between drops observed in configuration 1 is that they exhibit a smaller size compared with those formed in configuration $4$. As an example, Figure \ref{drop} shows four snapshots that illustrate the evolution of the configuration 4, at 2000 bar and 285 K. In the first snapshot (top), the initial hydrate phase is located on the left, with the pure water phase in the middle, and the CO$_{2}$ pure phase on the right ($0\,\text{ns}$). In the second snapshot (second panel from top), a liquid drop of CO$_{2}$ is formed within the water-rich liquid phase that can be clearly observed next to the water-CO$_{2}$ interface ($3\,\text{ns}$). It is interesting to compare this drop with the CH$_{4}$ bubble presented in paper I.~\cite{paperI} In this case, the drop is more difficult to distinguish than in the methane case due to the high solubility of CO$_{2}$ in the water-rich liquid phase. Then, the liquid drop ruptures and supersaturated aqueous solution of CO$_{2}$ appears in the central region of the simulation box, as can be seen in the third snapshot from the top shown in Fig.~\ref{conf4} ($4\,\text{ns}$). Finally, the last snapshot (bottom) shows the complete formation of the hydrate phase ($25\,\text{ns}$). Notice that this time corresponds to the end of the slope shown in Fig.~\ref{conf4} when the complete growth of the hydrate occurs. As we have already mentioned, this is the expected behavior since configuration 4 corresponds to a stoichiometric system and the final state corresponds to a single hydrate phase.

In the supplementary material, we provide a movie of the simulation trajectory at $285\,\text{}K$ and $2000\,\text{K}$ for configuration 4. The movie illustrates the diffusion of CO$_{2}$ molecules from the CO$_{2}$-rich liquid phase to the aqueous phase and the formation of the droplet. It shows how the droplet gradually reduces in size (thus, growing the hydrate at the same time) until it ruptures, resulting in a supersaturated solution and the complete growth of the hydrate phase. The visualization of the trajectory reveals a curved interface between the CO$_{2}$ droplet and aqueous solution, contrasting with the planar interfaces observed in the rest of the phase coexistence in the system.

The mechanism of the formation of the bubble, in the case of the CH$_{4}$ hydrate, and the liquid drop, in the case of the CO$_{2}$ hydrate, is completely analogous. However, the time scales are very different. In the first case, the formation of the bubble corresponds to the exact moment when the potential energy starts to drop abruptly (see Fig.~2f of paper I). Contrary, in the second case (CO$_{2}$), the formation of the drop starts at $0.8\,\text{ns}$. Note that potential energy in this case starts to drop at the beginning of the simulation. In addition to this, the temporal sequence at which the bubble forms, it ruptures, and the solid phase occupies the whole simulation box in the case of the CH$_{4}$ hydrate is $415\,\text{ns}\rightarrow 440\,\text{ns}\rightarrow 500\,\text{ns}$, approximately. However, in the case of the CO$_{2}$ hydrate is really different: $3\,\text{ns}\rightarrow 4\,\text{ns}\rightarrow 25\,\text{ns}$, approximately. The reason for these differences has to be found in the high difference between the solubility of CO$_{2}$ and CH$_{4}$ in water: in this case, the high concentration of CO$_{2}$ in water acts as a driving force that enhances the diffusion of CO$_{2}$ molecules in the water-rich liquid phase, the formation of the liquid drop, and finally, the formation of the complete hydrate phase. To corroborate this point, we have also calculated the solubility of CO$_{2}$ in the aqueous phase as a function of time. As it happens in paper I,~\cite{paperI} we observe a large increment of CO$_{2}$ concentration in the aqueous solution just at the beginning of the droplet formation. This can be clearly seen in the inset of Fig.~\ref{conf4}. The same behavior has been previously observed by
Kusalik and coworkers for other hydrates.~\cite{Hall2016a}

\begin{figure}[htp]
    \centering
\vspace*{0.5cm}
\includegraphics[width=0.9\columnwidth]{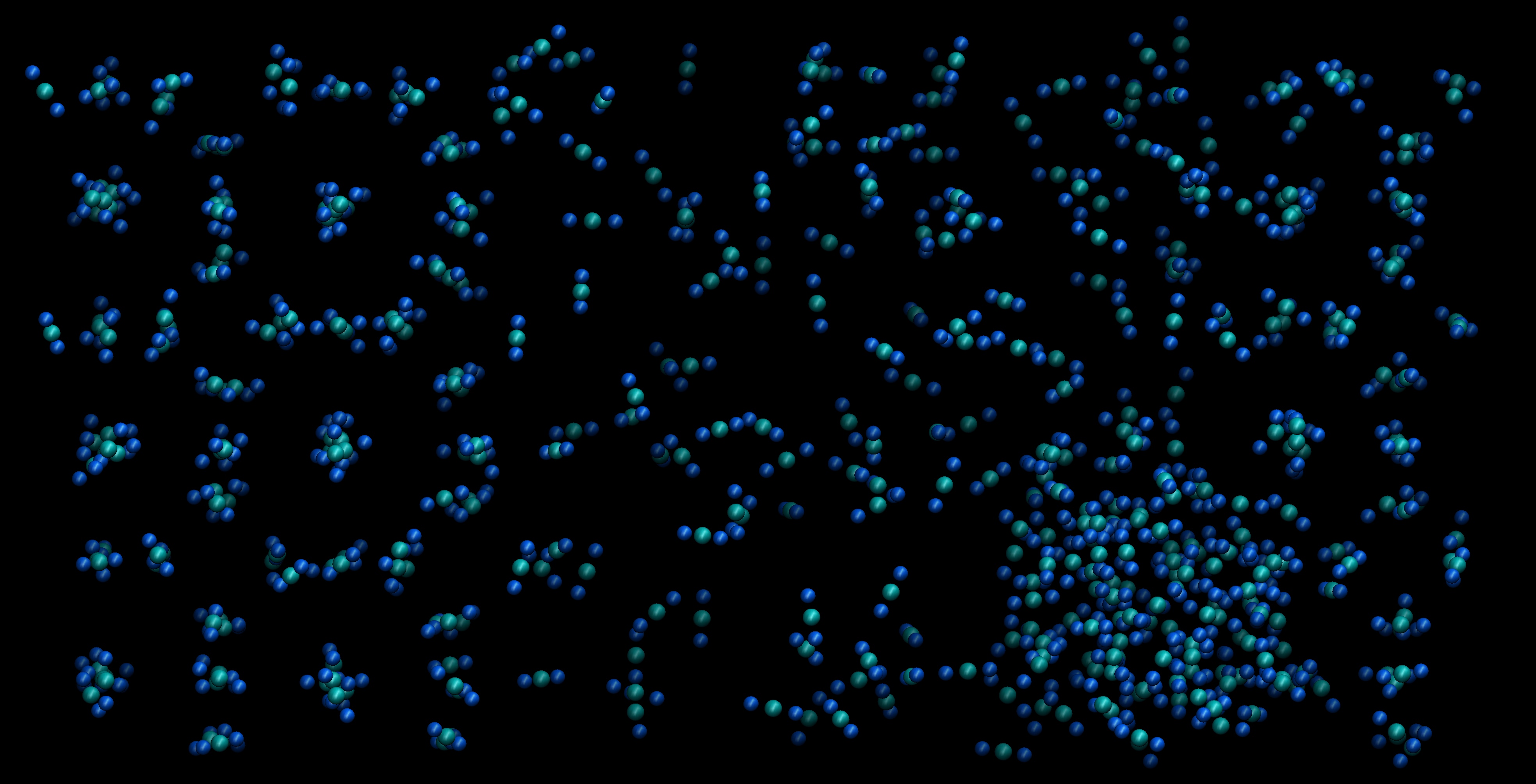}
\vspace*{0.5cm}

\includegraphics[width=0.9\columnwidth]{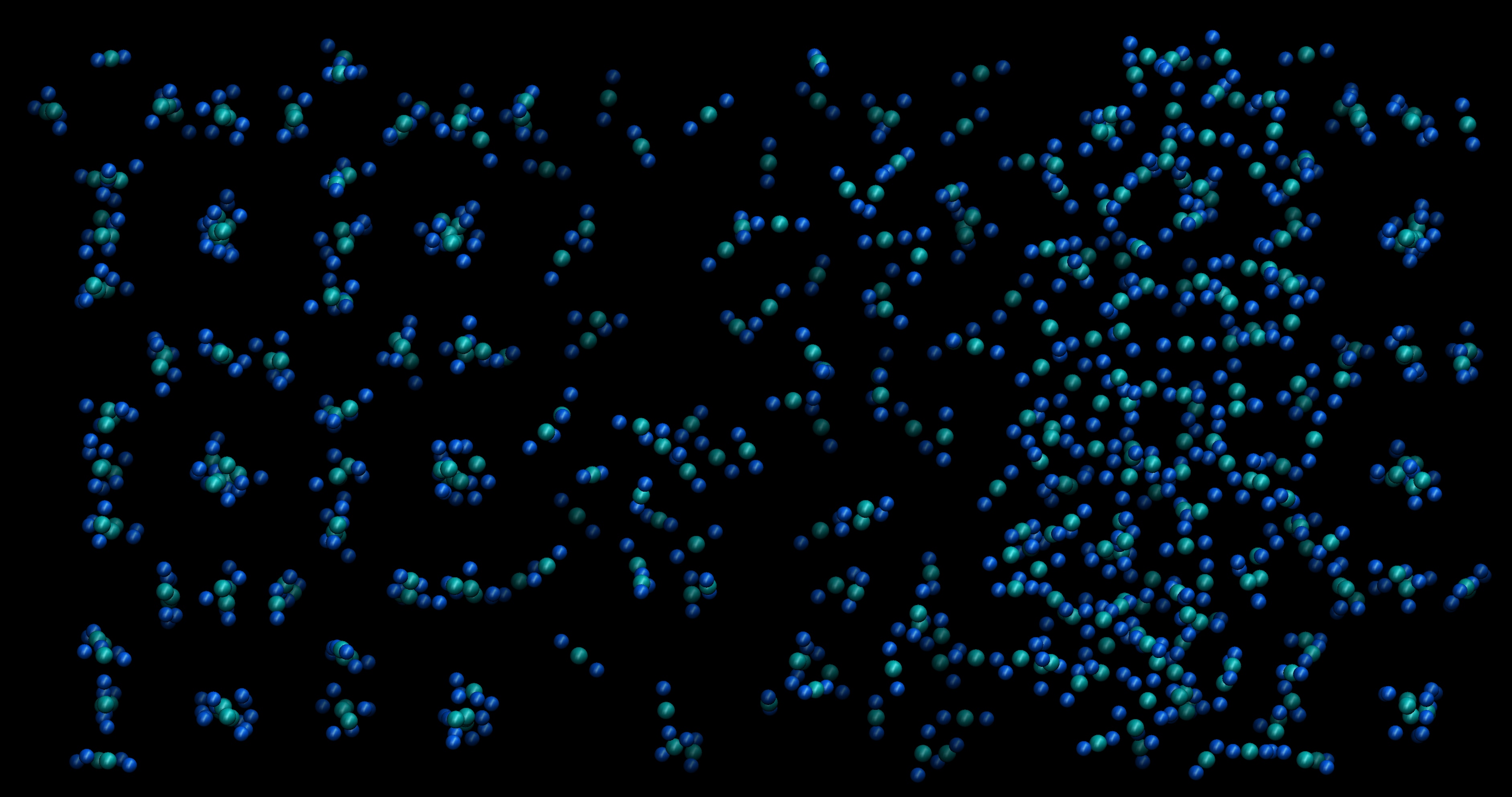}
\vspace*{0.5cm}

    \caption{\justifying{
    Snapshots of the $xz$ (top) and $yz$ (bottom) planes taken for configuration 4 at $285\,\text{K}$ and $2000\,\text{bar}$, showcasing the emergence of a CO$_{2}$ liquid drop within the liquid phase at $t=3\,\text{ns}$. For clarity, water molecules are omitted, while CO$_{2}$ molecules are represented as cyan (carbon atom) and blue spheres (oxygen atoms). CO$_{2}$ molecules are visible within the hydrates on the left and right sides of the figure. In the central region, CO$_{2}$ molecules are shown forming a cylindrical liquid drop.  
    }}
    \label{drop2}
\end{figure}  

As we have already mentioned, in paper I we have observed the formation of CH$_{4}$ bubbles for stoichiometric configurations of the hydrate.~\cite{paperI} Particularly, we have discussed in detail not only the formation and stability of the CH$_{4}$ bubble, but also its shape. Simulation results of paper I demonstrate that the CH$_{4}$ bubbles exhibit a cylindrical shape instead of a spherical one. An obvious question arises in this context: are the CO$_{2}$ liquid drops observed in the current simulations spherical or cylindrical, as in the case of the CH$_{4}$ hydrate? To answer this question, we analyze the snapshot presented in Fig.~\ref{drop}b ($t=3\,\text{ns}$) corresponding to the simulation of configuration 4 at $285\,\text{K}$ and $2000\,\text{bar}$. It is important to recall again that we have observed the formation of liquid drops of CO$_{2}$, not only at this particular pressure but in the whole range of pressures along the dissociation line of the hydrate. Particularly, we have also observed the drops in configuration 1 ($2\times2\times2$). However, since the solubility of CO$_{2}$ in water is high ($10$ times higher than that of CH$_{4}$ in water, approximately), it is more difficult to distinguish the drops. For this reason, we concentrate here on liquid drops of configuration 4.

Fig.~\ref{drop2} shows the projection of the $xz$ plan (top), as well as the $yz$ projection (bottom) of the same configuration. To help the visualization of the drop we have omitted the water molecules. As can be seen, the second projection ($yz$) clearly shows that the liquid drop extends along the whole $y$-axis of the simulation box. As it discussed in paper I,~\cite{paperI} and according to the Laplace equation, cylindrical interfaces result in lower solubility than spherical ones but higher than planar interfaces. If the liquid drops were stable and sufficiently large, the dissociation temperature would be shifted towards higher temperatures compared with the planar interface. In this work, as in paper I, drops of CO$_{2}$ are only stable for $1-20\,\text{ns}$, even less than in the case of the CH$_{4}$ bubbles ($10-25\,\text{ns}$).

As we have mentioned in the previous paragraph, the enhanced solubility of CO$_{2}$ in the aqueous phase in the presence of droplets compared to a planar interface at constant temperature can be understood in the context of the Laplace equation. According to this, the pressure inside the cylindrical droplets exceeds $2000\,\text{bar}$ (the outside pressure of the system). This higher internal pressure yields a higher chemical potential, leading to a higher molar fraction assuming ideal behavior for CO$_{2}$ in water. This excess of pressure can be estimated following our previous work.~\cite{Grabowska2022a} Here, we use the Laplace equation for droplets with cylindrical geometry ($\Delta P =\gamma/R$) to estimate the pressure inside the droplet. Here $\Delta P$ is the difference of pressure inside and outside the cylindrical drop, $\gamma$ is the water-CO$_{2}$ interfacial tension, and $R$ is the radius of the cylinder. To this end, we have calculated the aqueous solution - CO$_{2}$-rich liquid-liquid interfacial free energy using the same procedure as in our previous work.~\cite{Algaba2023a} The interfacial tension at $2000\,\text{bar}$ and $285\,\text{K}$ is $\gamma=30.2(3)\,\text{mJ/m}^2$. We have also calculated the radius for the observed droplet at the same thermodynamic conditions. In this case, $R=0.76(10)\,\text{nm}$, approximately. Using the Laplace equation, the internal pressure is about $2400(50)\,\text{bar}$. As expected, this pressure is higher than the $2000\,\text{bar}$ of the global system, leading to higher solubilities of CO$_{2}$.

In summary, as it happens in the case of the CH$_{4}$ hydrate, the presence of liquid drops of CO$_{2}$ modifies the prediction of the $T_{3}$ of CO$_{2}$ hydrates. This is only observed in liquid configurations that have stoichiometric configurations, such as configurations 1 and 4 studied in this work. Hence, we do not recommend the use of stoichiometric configurations in order to get reliable predictions of the $T_{3}$ of CO$_{2}$ hydrates. Particularly, the recommendation for future studies is to use the configuration $3$ ($3\times3\times3$), which is non-stoichiometric, formed from an appropriate number of water and CO$_{2}$ molecules that allow to simulate the system in a reasonable time. It is worth mentioning, as it happens with bubbles in the case of the methane hydrate studied in paper I,~\cite{paperI} that the formation of the droplets is expected not only in systems with stoichiometric compositions (i.e., when the ratio of molecules of CO$_{2}$ in the CO$_{2}$-rich liquid phase to that of water in the liquid phase is 8/46 , i.e., 0.174) but also in systems with lower values of this ratio. In fact, in these cases, the formation of the droplet is expected to occur at shorter times.

The formation of droplets should be directly related with the thickness of the CO$_{2}$-rich liquid phase in contact with the aqueous solution. To identify if there exists a critical thickness below which droplet formation is expected, we have simulated a water - CO$_{2}$ planar interface, without hydrate phase, using the direct coexistence simulation technique. Particularly, we use 1242 water molecules with a CO$_{2}$ liquid phase with different numbers of CO$_{2}$ molecules (i.e., $648$, $500$, $450$, $400$, and $350$). Simulations are performed using the anisotropic isothermal-isobaric or $NP_{z}\mathcal{A}T$ ensemble, with a fixed interfacial area of the systems of  $3.6\times3.6=12.96\,\text{nm}^2$, at $2000\,\text{bar}$ and $290\,\text{K}$. In all cases, simulations are run during at least $60\,\text{ns}$. We observe the formation of the droplets only in the systems formed from $400$ CO$_{2}$ molecules or less. According to this, our estimation of the critical thickness of the CO$_{2}$ slab is about 1.53-1.58 nm, approximately. In other words, when the thickness of the CO$_{2}$ phase is larger than $1.58\,\text{nm}$ no droplet is formed in $60\,\text{ns}$.

In any case, further work is needed to determine precisely under which conditions the planar water-CO$_{2}$ interface is not stable with respect to the formation of a cylindrical or spherical droplets as has been done for one component systems in other studies.~\cite{MacDowell2006a,Singh2019a,Montero2022a} In fact, in the future, it would also be interesting to study the droplet shape in bigger systems, as we have also stated in paper I.~\cite{paperI}

Before finishing this section, it is important to briefly mention how simulation results obtained from stoichiometric configurations 1 and 4 compare with experimental data taken from the literature. As can be seen in Figs.~\ref{pt_results}a and \ref{pt_results}d, simulation predictions exhibit large deviations with respect to experimental data. This is expected since, according to our previous discussion, the use of these kinds of configurations produces an overestimation of the $T_{3}$ at all the pressures along the dissociation line of the hydrate. We shall discuss in detail this issue in the last section, where we present the finite-size effect of non-stoichiometric configuration on the T$_{3}$ along the dissociation line of the CO$_{2}$ hydrate.

\subsection{Effect of the overall size}

Once we have analyzed in detail the effect of using two different stoichiometric configurations to determine the dissociation temperature of the CO$_{2}$ hydrate in a wide range of pressures, we now turn on to study the finite-size effects on $T_{3}$ for non-stoichiometric configurations. In this first section, we concentrate on finite-size effects due to the same increase in the number of molecules in each of the phases involved. 

The control configuration used in this section is again configuration 0, with $992$ total number of molecules, and an initial configuration formed from $432$ ($368+64$) water and CO$_{2}$ molecules in the hydrate phase, $368$ water molecules in the water phase, and $192$ CO$_{2}$ molecules in the CO$_{2}$-rich liquid phase. Note that the non-stoichiometric composition of the liquid phases is verified since the number of CO$_{2}$ molecules in that phase is three times the number corresponding to the stoichiometric composition ($192=3\times 64$). Note that the size of the interfacial area and the length of the simulation box perpendicularly to the interface also increase. We first determine the dissociation temperatures, in the whole range of pressures already considered in the previous section, of configuration 3. According to Table~\ref{tabla-moleculas}, this configuration contains a total number of molecules equal to $3348$. This means that the system size, in terms of the number of molecules, is multiplied by a factor of $3.375$, keeping the non-stoichiometric composition of the liquid phases. As can be seen in Table~\ref{tabla-moleculas}, the number of molecules of each species is multiplied by this factor in each of the phases forming the initial simulation box of configuration 3.

We have determined the three-phase coexistence temperature of configuration 3, in the same range of pressures considered previously (from $100$ to $6000\,\text{bar}$). Fig.~\ref{conf3} shows the evolution of the potential energy of the system, $U$, as a function of time, at $2000\,\text{bar}$ and temperatures from $285$ to $305\,\text{K}$. As in the case of the stoichiometric configurations, we observe the same two behaviors: at the highest temperatures, $298$, $300$, and $305\,\text{K}$, the potential energy increases very quickly over time, indicating the melting of the CO$_{2}$ hydrate. However, at low temperatures, $295$, $290$, and $285\,\text{K}$, the potential energy shows a decrease, which is more pronounced as the temperature is lower, indicating that the hydrate solid phase is growing.  The three-phase coexistence temperature at $2000\,\text{bar}$ is estimated at $T_{3}=296(2)\,\text{K}$ (at $298\,\text{K}$ the potential energy increases and at $295\,\text{K}$, it decreases).

It is interesting to compare the evolution of the potential energy obtained in this configuration and those corresponding to the stoichiometric configurations, 1 and 4. Although system sizes are different, it is clear that the characteristics sharp decreases observed in Figs.~\ref{conf1} and \ref{conf4} are not seen in Fig.~\ref{conf3}. The reason, as clearly stated in paper I,~\cite{paperI} is that configuration 3 is not stoichiometric. According to this, the $T_{3}$ value corresponding to this configuration is reliable and can be compared with confidence with that obtained by M\'{\i}guez \emph{et al.}~\cite{Miguez2015a} for configuration 0. According to Table~\ref{t3_results}, the T$_{3}$ at $2000\,\text{bar}$ was $292(2)\,\text{K}$ (configuration 0). In other words, the dissociation temperature of the configuration 3 is $4\,\text{K}$ above that of the configuration 0. This result suggests that there is a finite-size effect on $T_{3}$ that displaces the dissociation temperature towards higher temperatures, making the hydrate phase more stable at this pressure.

To confirm the stabilization of the hydrate when the number of molecules considered in the simulation box is increased (by a factor of $3.375$), we have determined the $T_{3}$ at the whole range of pressures, from $100$ to $6000\,\text{bar}$. Results are presented in Fig.~\ref{pt_results}c. As can be seen, the dissociation line corresponding to configuration 3 (maroon diamonds) is systematically displaced with respect to the results obtained from configuration 0 (black crosses). Simulation data obtained from MD-$NPT$ simulations is also included in Table~\ref{t3_results}. Displacement is not homogeneous. At low pressures (below $400\,\text{bar}$), the T$_{3}$ increases $\sim2\,\text{K}$, resulting in good agreement with previous results within the error bars. However, as the pressure is increased, displacement increases $~\sim4-5\,\text{K}$ at intermediate pressures ($1000$ and $2000\,\text{bar}$. Finally, at the highest pressures ($\ge 3000\,\text{bar}$), the T$_{3}$ is shifted $7-9\,\text{K}$.

To better understand the finite-size effect on the location of the dissociation line of the CO$_{2}$ hydrate, we go further and consider a new larger system. Configuration 6 is formed by a total number of molecules equal to $7936$, $3456$ corresponding to the (stoichiometric) hydrate phase, and $4480$ molecules to the non-stoichiometric liquid phases. Note that now, the total number of molecules in configuration 6 is $8$ times larger than in configuration 0. Details of the particular number of water and CO$_{2}$ molecules can be inspected in Table~\ref{tabla-moleculas}. As in the previous cases, we first concentrate on the determination of the T$_{3}$ at an intermediate pressure, $2000\,\text{bar}$. As can be seen in Fig.~\ref{conf6}, the system melts at the two highest temperatures considered, $298$ and $300\,\text{K}$, and freezes at $295$ and $290\,\text{K}$. According to this, the $T_{3}$, at $2000\,\text{bar}$, is $296(2)\,\text{K}$. Similarly to what happens with the previous non-stoichiometric configuration 3 (Fig.~\ref{conf3}), the evolution of the potential energy, as a function of time, behaves without the characteristic sharp drop associated with configuration 1 and 4 shown in Figs.~\ref{conf1} and \ref{conf4}. Note that now, the time required to observe crystallization is much larger than in smaller systems: in configuration 3, we observe crystallization before $100\,\text{ns}$. However, in configuration 6 we see that at temperatures close to the $T_{3}$, i.e., $295\,\text{K}$, crystallization occurs for times higher than $300\,\text{ns}$.

According to the results discussed in the previous paragraph, the $T_{3}$ predicted in configurations 3 and 6 are equal, $296(2)\,\text{K}$. Taking into account the results obtained for the CH$_{4}$ hydrate in paper I,~\cite{paperI}, an obvious question arises: is it necessary to simulate systems as large as the configuration 6 to
predict $T_{3}$ values without finite-size effects? To answer this question, we have obtained the rest of the dissociation temperatures at different pressures. The results are presented in Fig.~\ref{pt_results}f. As can be seen, there is a similar shift of the dissociation line of configuration 6 (maroon right triangles) with respect to that of configuration 0 (black crosses) in Figs.~\ref{pt_results}c and \ref{pt_results}f, indicating that we have achieved an asymptotic limit. A careful inspection of Table~\ref{t3_results} confirms this hypothesis: except for the $T_{3}$ value at $100\,\text{bar}$, the $T_{3}$ values of configurations 3 and 6 are identical (within the error bars). In summary, configurations 3 and 6, formed from the larger unit cells (e.g., $3\times 3 \times 3$ and $4\times 4\times 4$) show convergence of the $T_{3}$ values. This suggests that no finite-size effects exist for these system sizes.

\subsection{Effect of size of the interfacial area}

In the previous section, we have analyzed the finite-size effects of non-stoichiometric configurations on the dissociation line of the CO$_{2}$ hydrate varying the number of molecules in the system isotropically, i.e., increasing the initial hydrate phase in the three directions ($2\times 2\times 2$, $3\times 3\times 3$, and $4\times 4\times 4$). In addition to this, we consider the same number of water molecules in the aqueous phase and with three times more molecules of CO$_{2}$ in the CO$_{2}$-rich liquid phase to avoid stoichiometric configurations in the liquid phases. We have demonstrated that there exist finite-size effects associated with these changes.

We now investigate if the size of the interfacial area of the simulation box affects the $T_{3}$ of the CO$_{2}$ hydrate. To this end, we proceed similarly but considering two configurations in which the initial hydrate phases are formed replicating the unit cell as $3\times 3\times 2$ (configuration 2) and $4\times 4\times 2$ (configuration 5). The corresponding liquid phases are formed from the same number of water molecules in the aqueous phase as in the hydrate and with the number of CO$_{2}$ molecules equal to three times those existing in the hydrate. The particular number of molecules in each configuration can be seen in Table~\ref{tabla-moleculas}. Note that the total number of molecules in configurations 2 and 5 are $2232$ and $3969$, respectively.

Figs.~\ref{conf2} and \ref{conf5} show the evolution of the potential energy of configurations 2 and 5, as functions of time, at $2000\,\text{bar}$, respectively. The same general trend is observed in both figures, similar to those presented by configurations 3 and 6 (Fig.~\ref{conf3} and \ref{conf6}), i.e., since both configurations are non-stoichiometric they do not show the sharp decrease in potential energy. Following the same procedure used in previous sections, $T_{3}$ of both configurations can be easily determined by inspecting the behavior of potential energy at different temperatures. The $T_{3}$ values predicted by the molecular models are $291(2)$ and $289(2)\,\text{K}$ for configurations 2 and 5, respectively. These results suggest that the $T_{3}$ values of both configurations are equal (within the error bars) and also equal to the $T_{3}$ value corresponding to the configuration 0 ($2\times 2\times 2$) at this pressure. This is a non-expected result according to the behavior observed in the previous section, especially if we take into account that configuration 5 is formed from $3968$ molecules, a $20\%$ more molecules than in configuration 3 ($3\times 3\times 3$). For this former configuration, $T_{3}$ is equal to $296(2)\,\text{K}$ at the same pressure.

To clarify this point, we have also obtained the $T_{3}$ values of configurations 2 and 5 at lower and higher pressures, as we have already done with the configurations previously studied. Fig.~\ref{pt_results}c and \ref{pt_results}f show the pressure-temperature projection of the dissociation line of the CO$_{2}$ hydrate using both configurations. The numerical data obtained from MD-$NPT$ simulations are also presented in Table~\ref{t3_results}. As can be seen in the plots, both configurations exhibit very similar $T_{3}$ values in the whole range of pressures. Particularly, at low pressures, between $100$ and $2000\,\text{bar}$, configurations 0 (black crosses) and 2 (violet down triangles) present the same $T_{3}$ values, while configuration 5 (violet left triangles) seem to exhibit slightly lower values of $T_{3}$. However, both results are within the error bars. At higher pressure, above $3000\,\text{bar}$, the dissociation temperatures of configurations 2 and 5 are above those of configuration 0. Nevertheless, the results are again within the error bars and no significant differences are observed between temperatures at the corresponding pressure (see Table~\ref{t3_results} for further details).

In summary, we have analyzed the finite-size effects on the $T_{3}$ for configurations 2 ($3\times 3\times 2$) and 5 ($4\times 4\times 2$). The main difference between these two configurations is that configuration 5 has a larger interfacial area in contact with the aqueous and CO$_{2}$-rich liquid phases ($4\times 4$ unit cells) than that of configuration 2 ($3\times 3$ unit cells). Results indicate that finite-size effects on $T_{3}$ values are negligible, at least within error bars.

\begin{figure*}
     \centering
     \begin{subfigure}[hbt]{0.25\textwidth}
         \centering
    \includegraphics[width=1.0\textwidth]
        {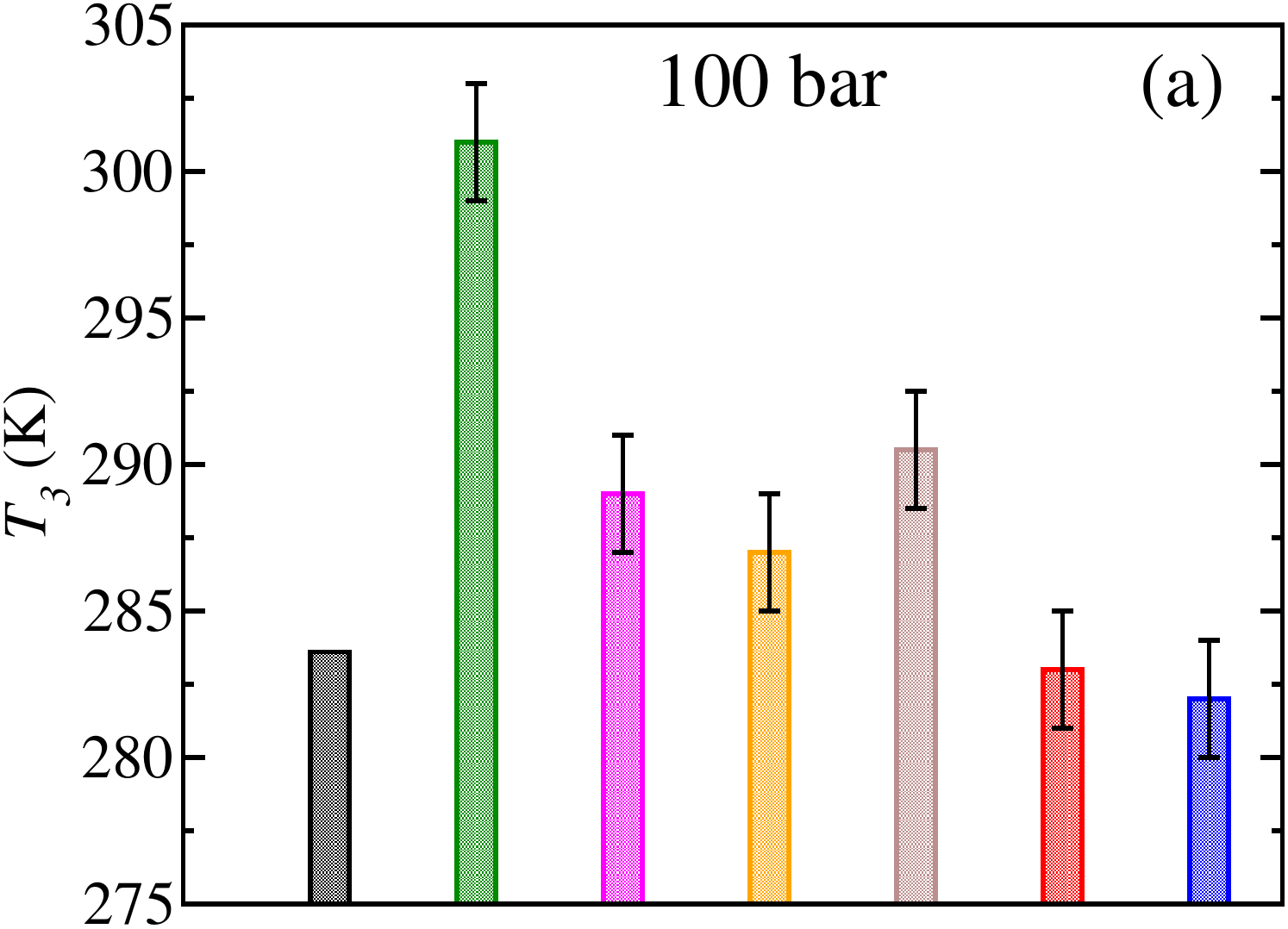}
     \end{subfigure}
     \hfill
     \begin{subfigure}[hbt]{0.23\textwidth}
         \centering         \includegraphics[width=1.0\textwidth]{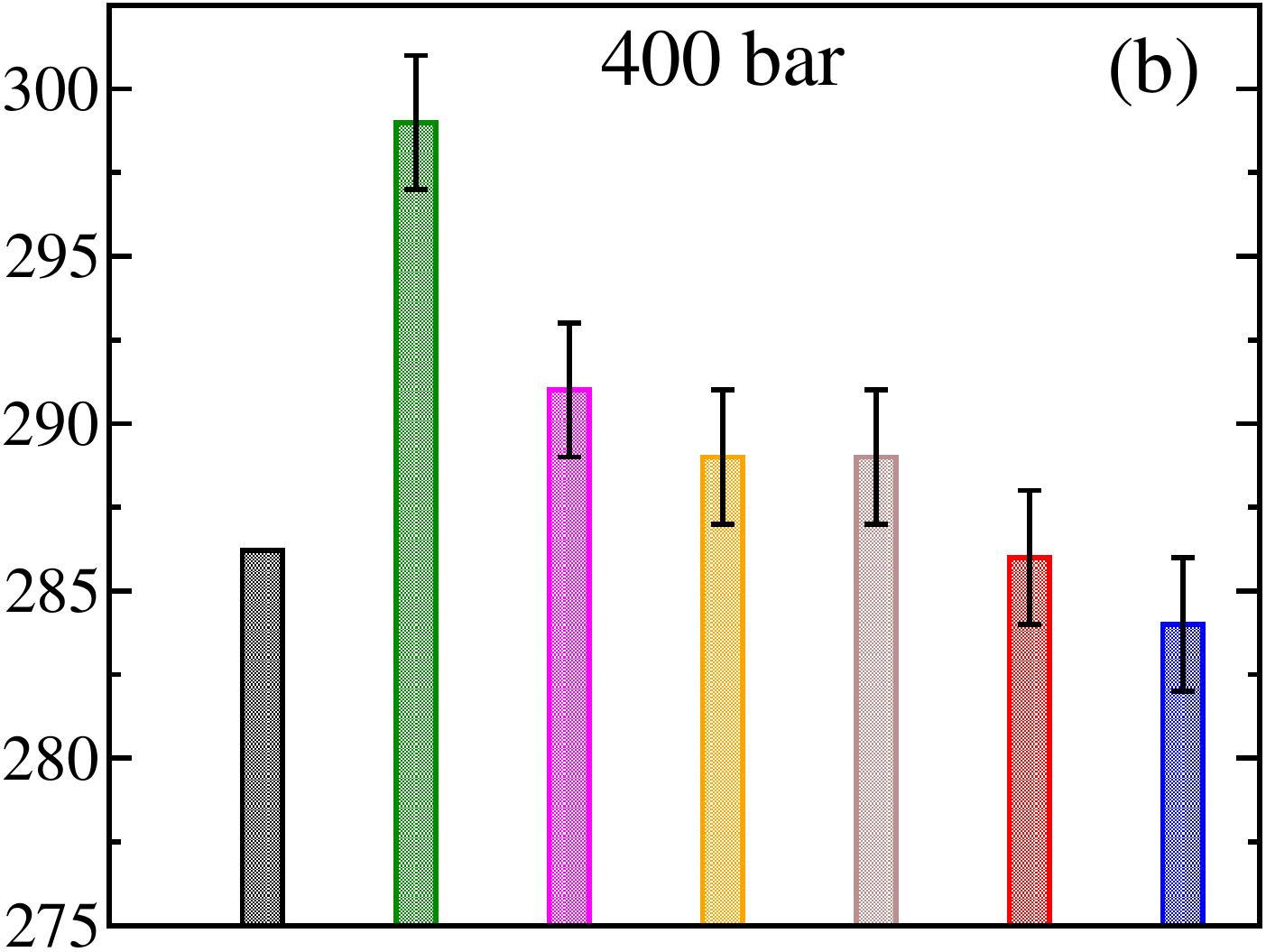}
     \end{subfigure}
     \hfill
     \begin{subfigure}[hbt]{0.23\textwidth}
         \centering         \includegraphics[width=1.0\textwidth]{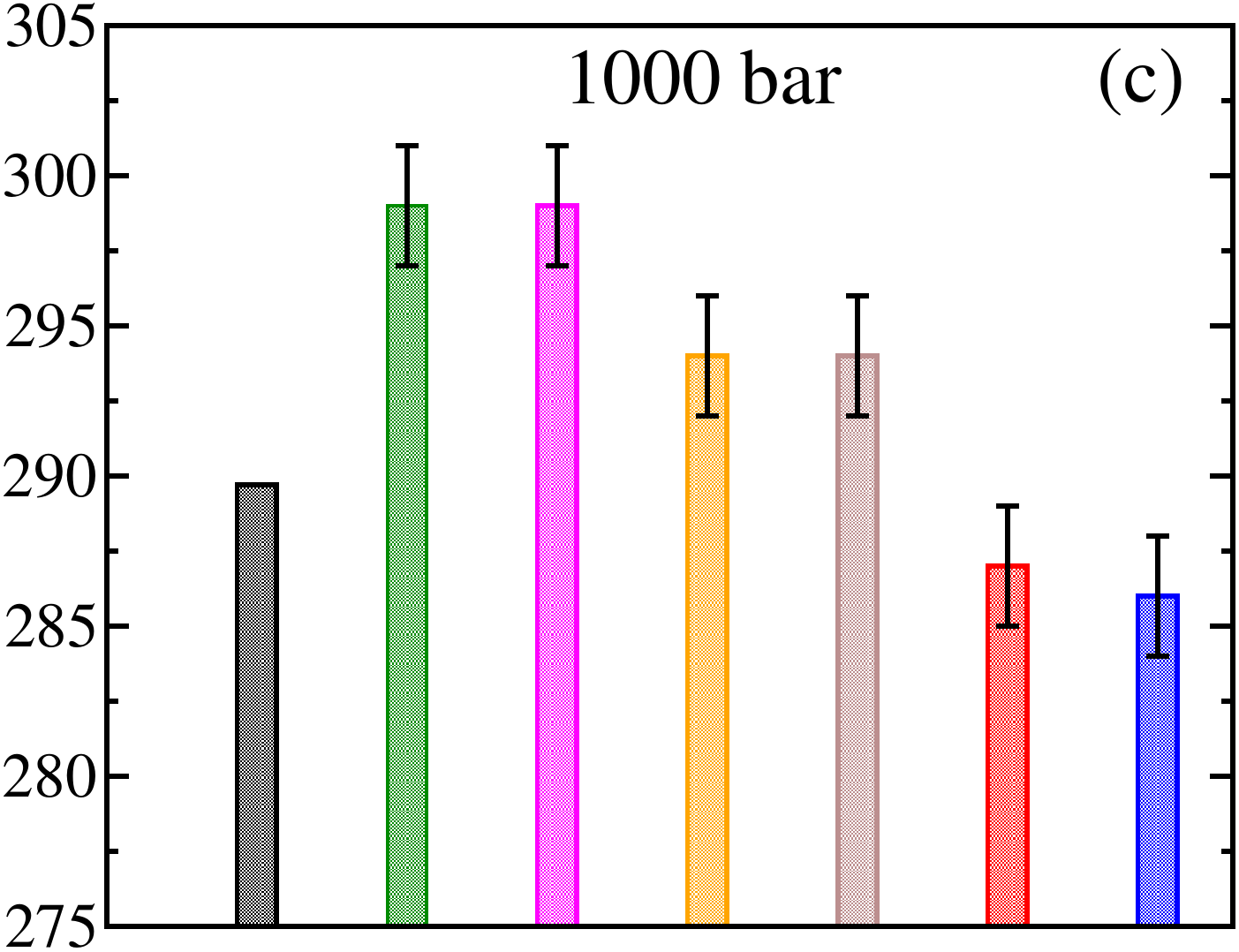}
     \end{subfigure}
     \hfill
     \begin{subfigure}[hbt]{0.23\textwidth}
         \centering         \includegraphics[width=1.0\textwidth]{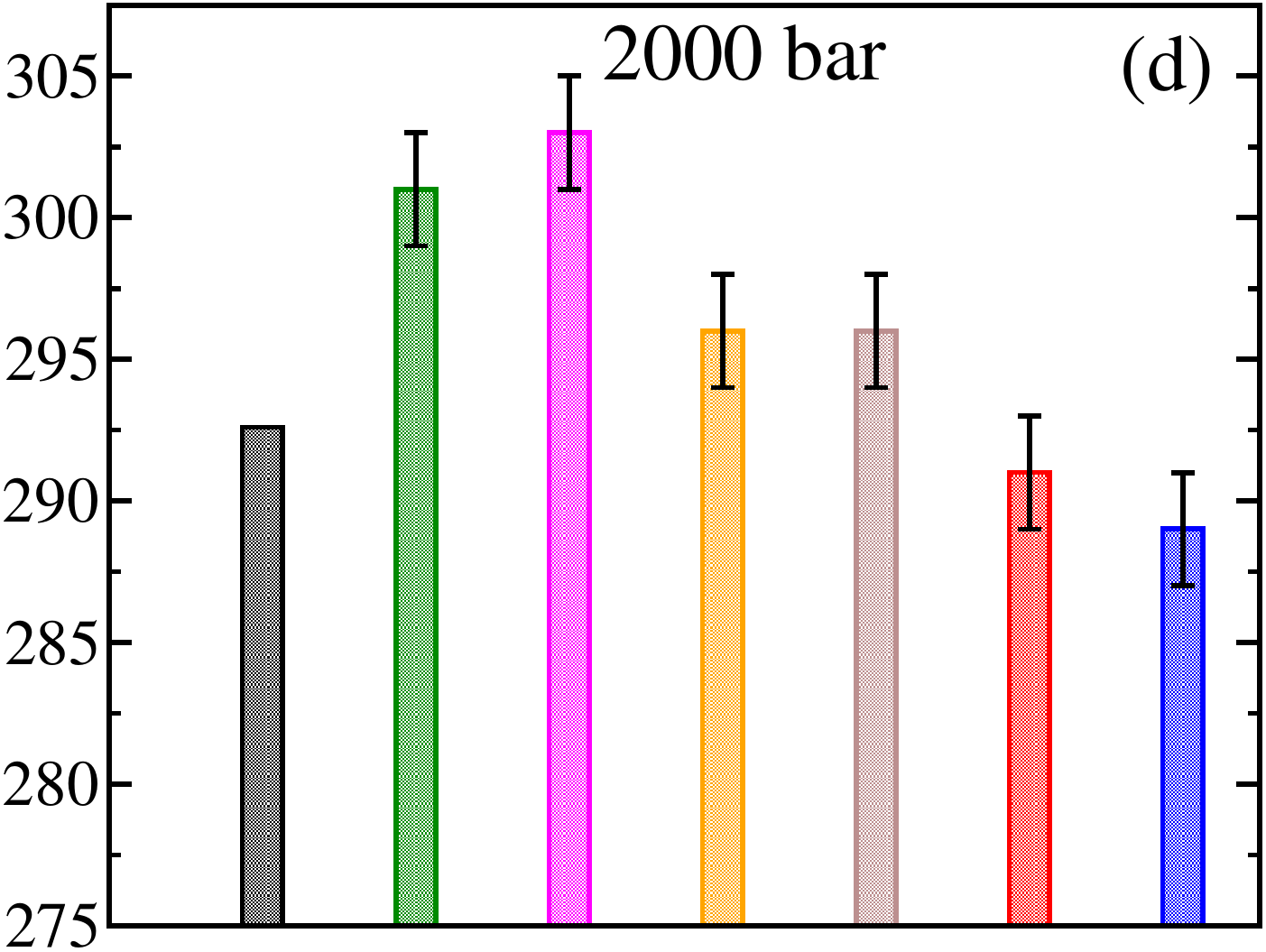}
     \end{subfigure}
     \begin{subfigure}[hbt]{0.25\textwidth}
         \centering         \includegraphics[width=1.0\textwidth]{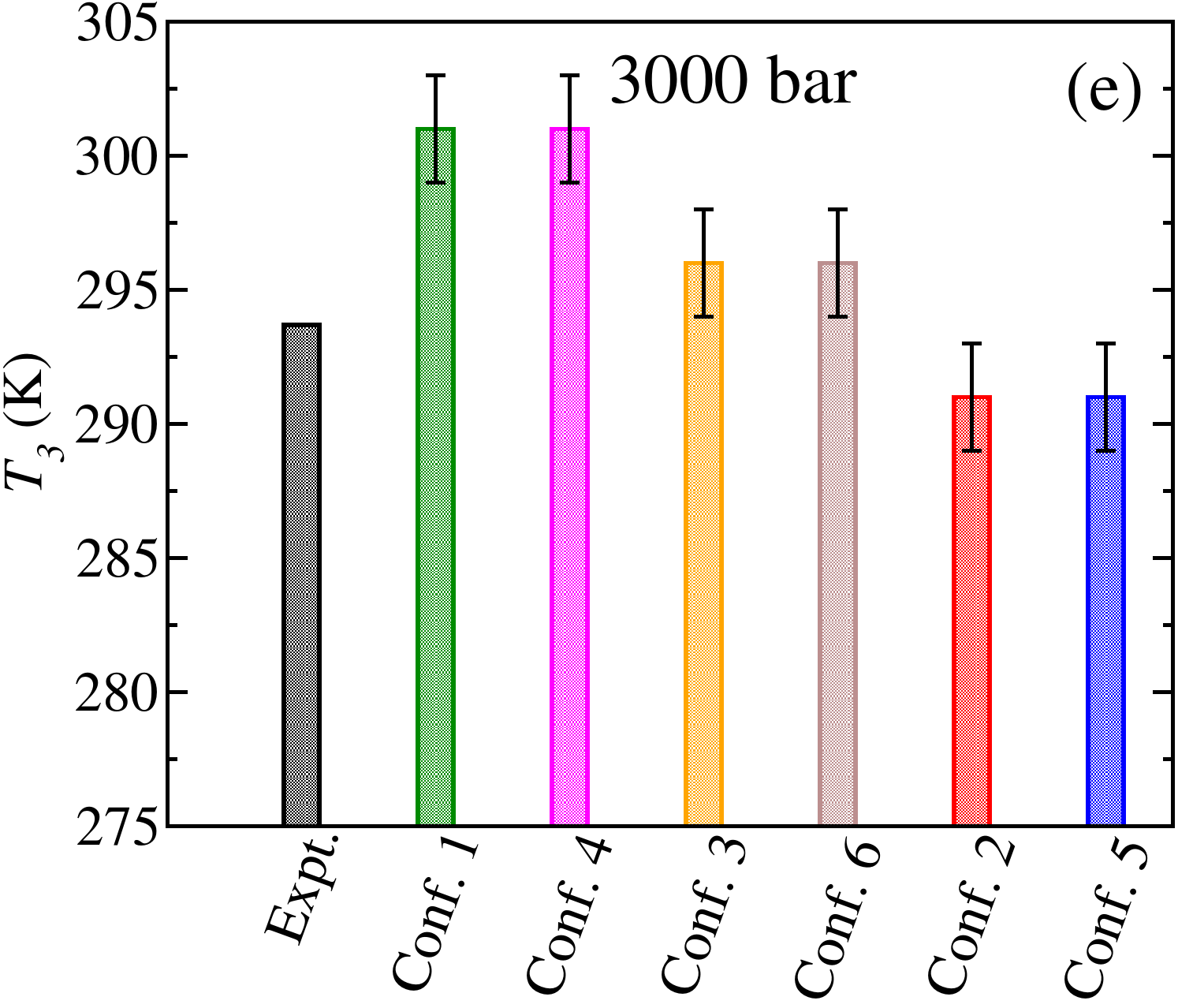}
     \end{subfigure}
     \hfill
     \begin{subfigure}[hbt]{0.23\textwidth}
         \centering         \includegraphics[width=1.0\textwidth]{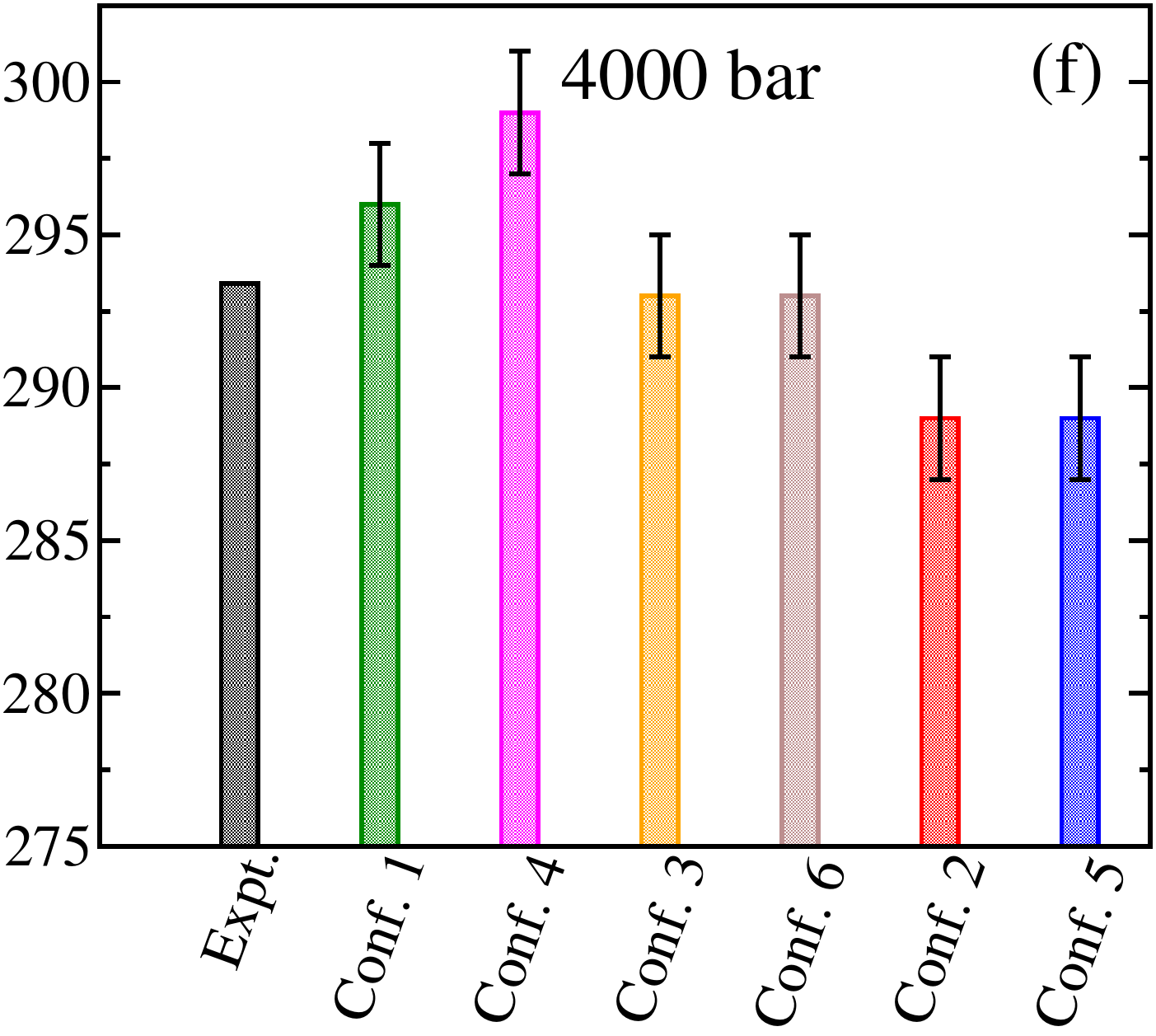}
     \end{subfigure}
     \hfill
     \begin{subfigure}[hbt]{0.23\textwidth}
         \centering         \includegraphics[width=1.0\textwidth]{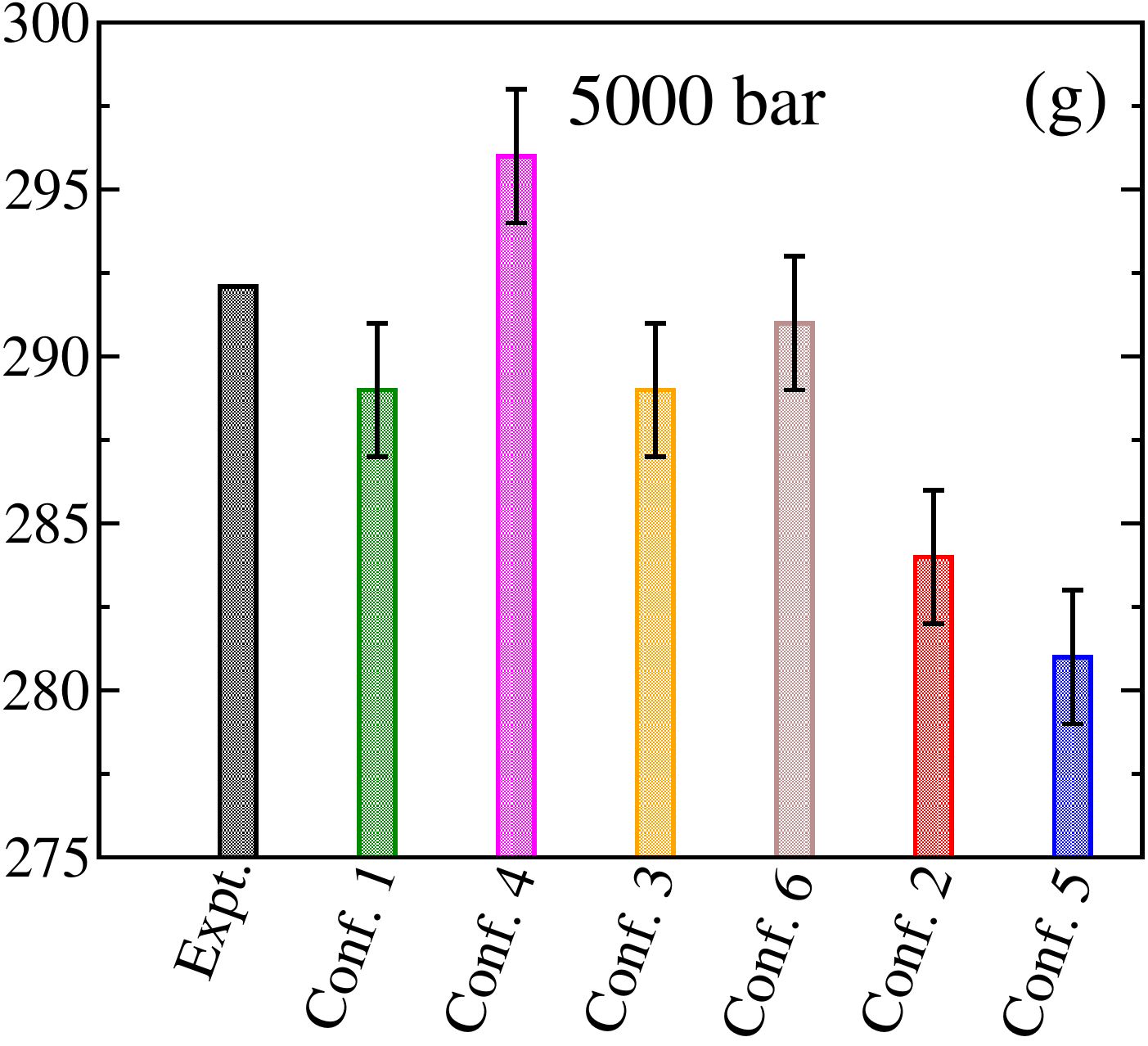}
     \end{subfigure}
     \hfill
     \begin{subfigure}[hbt]{0.23\textwidth}
         \centering         \includegraphics[width=1.0\textwidth]{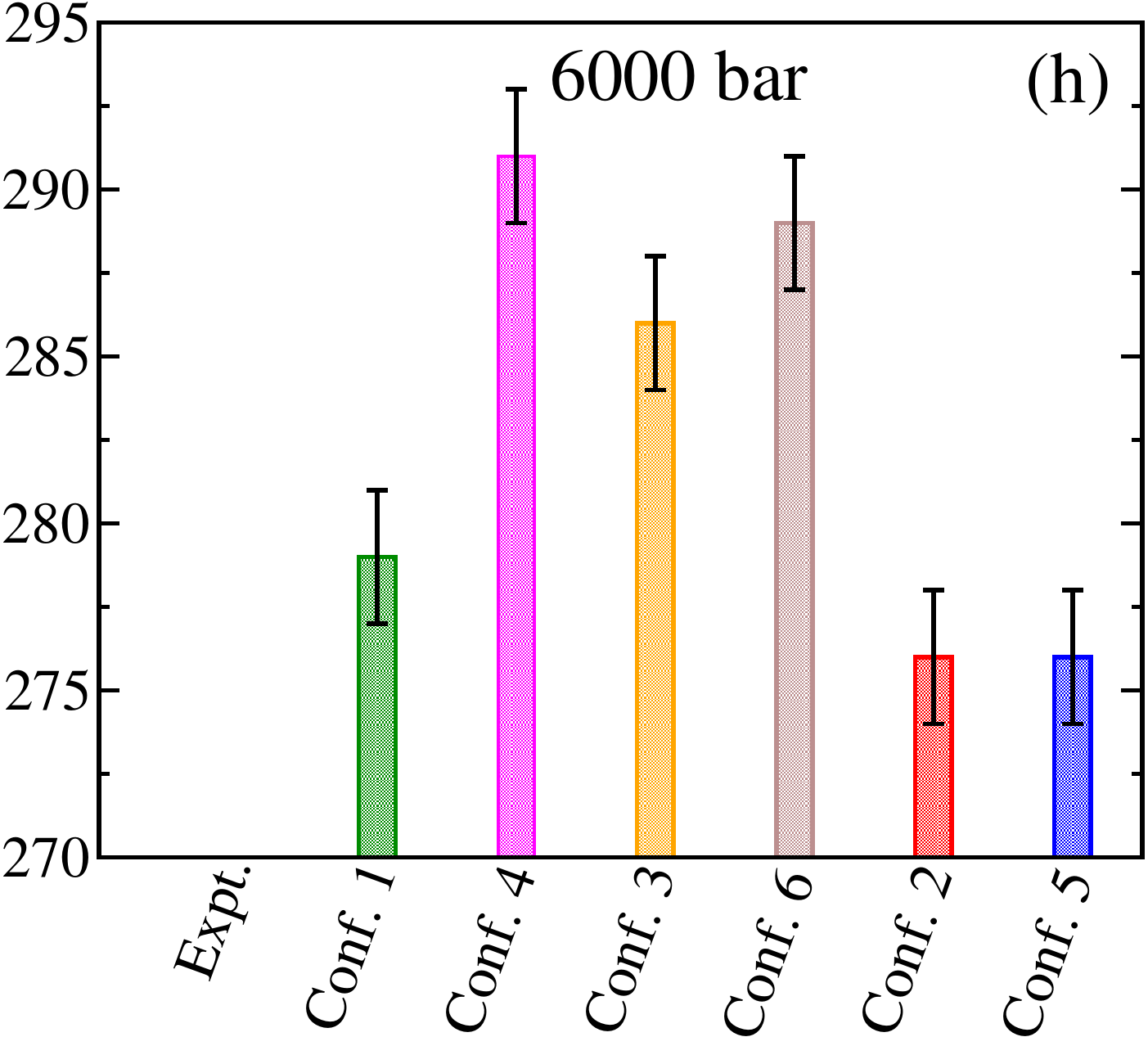}
     \end{subfigure}
\caption{\justifying{Comparison of the three-phase coexistence temperatures ($T_{3}$) for the CO$_{2}$ hydrate from the 6 size-dependent configurations analyzed in this work at several pressures. The experimental $T_{3}$ values taken from the literature,~\cite{Sloan2008a} at different pressures, are also included. In all cases, the composition
of the initial configuration is analogous, formed by a CO$_{2}$ hydrate phase in coexistence with a liquid water phase and a CO$_{2}$ liquid phase with the corresponding numbers of molecules in each configuration.}
}
\label{barras}
\end{figure*}

\subsection{Effect of hydrate thickness}

We have already discussed in Section~III.C the effect of passing from the configuration 0 ($2\times2\times2$) to systems with configurations 2 ($3\times3\times2$) and 5 ($4\times4\times2$). In both cases, we have analyzed the effect of increasing the interfacial area keeping invariant the thickness of the hydrate along the $z$-direction (perpendicular to the solid-fluid interface). But there is also an interesting comparison not made until now: the effect of increasing the thickness of the hydrate along the $z$-direction keeping constant the interfacial area. This can be done by comparing configurations 2 ($3\times3\times2$) and 3 ($3\times3\times3$), in which the thickness of the hydrate passes from $2$ to $3$ unit cells along the $z$-direction, and configuration 5 ($4\times4\times2$) and 6 ($4\times4\times4$). In the first case the interfacial area is $3\times3$ unit cells and in the second case $4\times4$. Note that in both comparisons the interfacial area remains unchanged.

We first consider the comparison between configurations 2 and 3 at $2000\,\text{bar}$. Following the same procedure used in previous cases (inspection of Figs.~\ref{conf2} and \ref{conf3}), the dissociation temperature predicted from both configurations are $291(2)$ and $296(2)\,\text{K}$, respectively. These results confirm that an increase in the thickness of the hydrate in the initial simulation box helps to stabilize the system. We have also investigated the same behavior at lower and higher temperatures. This information can be readily obtained from inspection of Figs.~\ref{pt_results}b and \ref{pt_results}c. The numerical results obtained from MD-NPT simulations are also presented in Table~\ref{t3_results}. As can be seen, the main effect of increasing the thickness of the hydrate phase is to shift the $T_{3}$ towards higher temperatures at all pressures. The displacement of the three-phase line is not uniform but varies as the pressure increases: $3-4\,\text{K}$ at low pressures ($100$ and $200\,\text{bar}$), $4-7\,\text{K}$ at intermediate pressures ($1000-4000\,\text{bar}$), and $5-10\,\text{K}$ at high pressures ($5000$ and $6000\,\text{bar}$).

We have also analyzed the changes observed when passing from configuration 5 ($4\times4\times2$) to configuration 6 ($4\times4\times4$). Note that in this case, the interfacial area is $4\times4$ instead of $3\times3$ units cells. In addition to that, the increase of the hydrate thickness varies from $2$ to $4$ hydrate unit cells along the $z$-direction (perpendicular to the interface). As can be seen in Figs.~\ref{pt_results}e and \ref{pt_results}f, the effect of increasing the hydrate thickness is also to shift the $T_{3}$ values toward higher temperatures. In this case, the increase of the stability of the hydrate is higher than in the previous case. Particularly, $5-7\,\text{K}$ at low pressures ($100$ and $200\,\text{bar}$), $4-8\,\text{K}$ at intermediate pressures ($1000-4000\,\text{bar}$), and $10-15\,\text{K}$ at high pressures ($5000$ and $6000\,\text{bar}$).

To recap, the main effect of increasing the hydrate thickness is to shift the dissociation line towards higher temperatures as the hydrate thickness increases, i.e., to increase the stability of the hydrate phase.

\subsection{Comparison with experimental data}

In the previous sections, we discuss the finite-size effects on the $T_{3}$ of the CO$_{2}$ hydrate in a wide range of pressures. This is performed using 6 different configurations and comparing the results with the configuration 0 obtained several years ago by M\'{\i}guez \emph{et al.}~\cite{Miguez2015a} In this Section, we focus on the comparison of the predictions obtained using these configurations with experimental data taken from the literature.

In the original work,~\cite{Miguez2015a} the authors proposed a modified Berthelot combining rule that allows to predict the dissociation line of the CO$_{2}$ hydrate in a wide range of pressures with confidence. Particularly, differences between predictions obtained from configuration 0 and experimental data taken from the literature~\cite{Sloan2008a} are below $1\,\text{K}$ from $100$ to $2000\,\text{K}$. At high pressures, at $3000\,\text{bar}$ and above, agreement between both results deteriorates. Although the model is able to capture the existence of the reentrant behavior of the dissociation line, the agreement is only qualitative. We recommend the reader to the original paper for a detailed account of this issue.~\cite{Miguez2015a}

Fig.~\ref{pt_results} shows the pressure-temperature composition of the dissociation line of the CO$_{2}$ hydrate as predicted from the model proposed by M\'{\i}guez \emph{et al.}~\cite{Miguez2015a}
using the 6 configurations considered in this work. The results obtained using configuration 0 are also presented, as well as the experimental data taken from the literature.~\cite{Sloan2008a} We only analyze the agreement between experimental data and predictions obtained using non-stoichiometric configurations (2, 3, 5, and 6) since stoichiometric systems do not provide the correct $T_{3}$ due to the presence of CO$_{2}$ liquid drops discussed in Section III. A. Let's concentrate on predictions obtained using configurations 3 and 6. As can be seen, predictions from configuration 3 overestimate the experimental $T_{3}$ in the whole range of pressures. Particularly, $T_{3}$ is overestimated $\sim 5\,\text{K}$ at low pressures, below $3000\,\text{bar}$. The agreement between predictions from configuration 6 and experimental data is similar.

The main conclusion is that finite-size effects also affect agreement with experimental data. What is the reason for this? The answer is simple. The unlike interaction parameter associated with the Berthelot rule, $\xi$, was established by M\'{\i}guez \emph{et al.}~\cite{Miguez2015a} using configuration 0, i.e., using a $2\times 2\times 2$ configuration and a certain cutoff distance for the dispersive interactions. We will not discuss here the effect of the cutoff distance on the $T_{3}$ of hydrates. We recommend to read the paper III of this series in which we analyze the effect of the dispersive interactions on the three-phase equilibria of CH$_{4}$ and CO$_{2}$ hydrates.~\cite{paperIII} According to this, the value $\xi=1.13$ obtained by M\'{\i}guez \emph{et al.}~\cite{Miguez2015a} allows to quantitatively predict the dissociation line of the CO$_{2}$ hydrate for the particular configuration ($2\times 2\times 2$) used to fit experimental data. As we have demonstrated in Section III. B, the use of larger configurations, i.e., configurations 3 and 6, produces a displacement of the dissociation line of the CO$_{2}$ towards higher temperatures, making the hydrate phase more stable. In fact, the same effect produces the use of a positive modification of the Berthelot rule ($\xi>1$). This suggests that the $\xi=1.13$ value is too high for describing in a quantitative way the dissociation line of the CO$_{2}$ hydrate using configurations not affected by finite-size effects, i.e., configurations 3 or 6. 

Finally, we have summarized the $T_{3}$ values obtained for each of the 6 size-dependent configurations, at all the pressure considered (from $100$ to $6000\,\text{bar}$), in the bar graphs of Fig.~\ref{barras}. In addition, we have also included the experimental $T_{3}$ for comparison reasons (first column). The order of the columns, from left to right, takes into account the effect of the stoichiometric configurations (1 and 4), of the overall size of the system (3 and 6), and of the size of the interfacial area (2 and 5). The first configurations in each of the three groups, 1, 2, and 3, correspond to the smaller system analyzed, and the second ones, 4, 5 and 6, to the larger systems. See Table~\ref{tabla-moleculas} for further details. As can be seen, the main conclusions stated in the previous sections can be observed in Fig.~\ref{barras} in the whole range of pressures analyed in this work: (1) overestimation of the T$_{3}$ values, with respect to the experimental data taken from the literature, in the case of the stoichiometric configurations (1 and 4); (2) similar $T_{3}$ values for configurations 3 ($3\times3\times3$) and 6 ($3\times3\times3$); and (3) configurations 2 and 5 exhibit the same $T_{3}$ values at all pressures (within the error bars) but below the experimental T$_{3}$ values.

\section{Concluding remarks}
In this second part of a three-paper series dedicated to investigating finite-size effects in methane and carbon dioxide, as well as the influence of dispersive interactions on both hydrates, our focus is on the CO$_{2}$ hydrate. Specifically, we delve into the finite-size effects affecting the determination of the three-phase coexistence temperature ($T_3$) for CO$_{2}$ hydrate, employing molecular dynamics simulations in conjunction with the direct coexistence technique. Adhering to the methodology outlined in the first paper~\cite{paperI}, we examine six size-dependent configurations using realistic water and CO$_{2}$ models (i.e., TIP4P/Ice and TraPPE) to assess the impact of size and composition on the estimation of $T_3$. Given the similarity of our findings to those in the first paper~\cite{paperI}, we provide a concise summary of the most pertinent results:

\begin{itemize}

\item {The simulation results obtained for the CO$_{2}$ hydrate confirm the sensitivity of $T_3$ depending on the size and composition of the system, explaining the discrepancies observed in the original work by M\'{\i}guez \emph{et al.}~\cite{Miguez2015a} in 2015. This is in agreement with the findings of paper I,~\cite{paperI} not only at a particular pressure but in a wide range of pressures considered.}

\item {Configurations with stoichiometric composition or less CO$_{2}$ molecules than the stoichiometric,
 at temperatures below $T_3$, evolve into a singular phase of CO$_{2}$ hydrate growth, as in the case of the methane hydrate.~\cite{paperI} This is confirmed in the whole dissociation line, from $100$ to $6000\,\text{bar}$. In this case, there is no a bubble of methane but a liquid drop of CO$_{2}$. The mechanism of growing is via the emergence of a liquid drop of CO$_{2}$ within the liquid and the subsequent formation of an oversaturated CO$_{2}$ solution in water. Conversely, an excess of molecules in the CO$_{2}$-rich liquid phase in the initial configuration leads to the coexistence of CO$_{2}$ hydrate and CO$_{2}$ liquid phases without the formation of drops.}

\item {Finite-size effects are pronounced in small systems with stoichiometric composition (e.g., configuration 1 with a unit cell of $2\times2\times2$), resulting in an overestimation of $T_3$ due to liquid drop formation during hydrate growth, causing a false stability of CO$_{2}$ hydrate by increasing CO$_{2}$ solubility. Note that this is a common conclusion in both methane and carbon dioxide hydrates.}

\item {Non-stoichiometric configurations with larger unit cells, like $3\times3\times3$ and $4\times4\times4$, show convergence of $T_3$ values, suggesting that finite-size effects for these system sizes, regardless of drop formation, can be safely neglected.}

\item {To study the T$_3$ of the CO$_{2}$ hydrate the best choice is configuration 3, which provides an accurate $T_3$ value, and affordable simulation times.}

\end{itemize}

The primary outcome of this study, entirely consistent with the findings in the initial paper of this series for CH$_{4}$ hydrate~\cite{paperI}, can be succinctly summarized as follows: when using the direct coexistence technique to estimate the $T_3$ of CO$_{2}$ hydrate, it is crucial to avoid small stoichiometric configurations such as configuration 1. These configurations tend to develop drops at the onset of the run, leading to an overestimation of the $T_3$ value. The recommended and optimal choice is configuration 3, which provides an accurate $T_3$ value, and the computational resources required for simulating this system are currently feasible.

This study, complementing the initial paper in the series~\cite{paperI}, presents valuable insights into the finite-size effects observed in simulations of CO$_{2}$ hydrate. The findings highlight the possibility of mitigating finite-size effects in estimating $T_3$ by thoughtfully selecting system configurations. We anticipate that these results will contribute to a better understanding of finite size effects in the determination of $T_3$ for methane hydrates, addressing discrepancies in the existing literature and assisting researchers in selecting appropriate system sizes for future investigations. Our future research will delve into examining the potential impact of cutoff values and guest types on $T_3$ values, exploring how these factors are influenced by finite-size effects.

\section*{Supplementary material}
See the supplementary material for the movie of the simulation trajectory at $285\,\text{K}$ and $2000\,\text{bar}$ for configuration $4$. The movie illustrates the  CO$_{2}$ molecules diffusion from the CO$_{2}$-rich liquid phase to the aqueous phase and the formation of the droplet.

\section*{Acknowledgments}

This work was funded by Ministerio de Ciencia e Innovaci\'on (Grant No.PID2019-105898GA-C22, PID2021-125081NB-I00 and PID2022-136919NB-C32), Junta de Andalucía (P20-00363), and Universidad de Huelva (P.O. FEDER UHU-1255522 and FEDER-UHU-202034), all four cofinanced by EU FEDER funds. This work was also funded by Project No.~ETSII-UPM20-PU01 from ``Ayudas Primeros Proyectos de la ETSII-UPM''. M.M.C. acknowledges CAM and UPM for financial support of this work through the CavItieS project No. APOYO-JOVENES-01HQ1S-129-B5E4MM from ``Accion financiada por la Comunidad de Madrid en el marco del Convenio Plurianual con la Universidad Politecnica de Madrid en la linea de actuacion estimulo a la investigacion de jovenes doctores'' and CAM under the Multiannual Agreement with UPM in the line Excellence Programme for University Professors, in the context of the V PRICIT (Regional Programme of Research and Technological Innovation). S.B. acknowledges Ayuntamiento de Madrid for a Residencia de Estudiantes grant. The authors also gratefully acknowledge the Universidad Politecnica de Madrid (www.upm.es) for providing computing resources on Magerit Supercomputer. We also acknowledge and additional computational resources from Centro de Supercomputaci\'on de Galicia (CESGA, Santiago de Compostela, Spain), at which some of the simulations were run.

\section*{AUTHORS DECLARATIONS}

\section*{Conflicts of interest}

The authors have no conflicts to disclose.

\section*{Data availability}

The data that support the findings of this study are available within the article.

\section*{REFERENCES}
\bibliography{bibfjblas}

\begin{thebibliography}{73}%
\makeatletter
\providecommand \@ifxundefined [1]{%
 \@ifx{#1\undefined}
}%
\providecommand \@ifnum [1]{%
 \ifnum #1\expandafter \@firstoftwo
 \else \expandafter \@secondoftwo
 \fi
}%
\providecommand \@ifx [1]{%
 \ifx #1\expandafter \@firstoftwo
 \else \expandafter \@secondoftwo
 \fi
}%
\providecommand \natexlab [1]{#1}%
\providecommand \enquote  [1]{``#1''}%
\providecommand \bibnamefont  [1]{#1}%
\providecommand \bibfnamefont [1]{#1}%
\providecommand \citenamefont [1]{#1}%
\providecommand \href@noop [0]{\@secondoftwo}%
\providecommand \href [0]{\begingroup \@sanitize@url \@href}%
\providecommand \@href[1]{\@@startlink{#1}\@@href}%
\providecommand \@@href[1]{\endgroup#1\@@endlink}%
\providecommand \@sanitize@url [0]{\catcode `\\12\catcode `\$12\catcode
  `\&12\catcode `\#12\catcode `\^12\catcode `\_12\catcode `\%12\relax}%
\providecommand \@@startlink[1]{}%
\providecommand \@@endlink[0]{}%
\providecommand \url  [0]{\begingroup\@sanitize@url \@url }%
\providecommand \@url [1]{\endgroup\@href {#1}{\urlprefix }}%
\providecommand \urlprefix  [0]{URL }%
\providecommand \Eprint [0]{\href }%
\providecommand \doibase [0]{http://dx.doi.org/}%
\providecommand \selectlanguage [0]{\@gobble}%
\providecommand \bibinfo  [0]{\@secondoftwo}%
\providecommand \bibfield  [0]{\@secondoftwo}%
\providecommand \translation [1]{[#1]}%
\providecommand \BibitemOpen [0]{}%
\providecommand \bibitemStop [0]{}%
\providecommand \bibitemNoStop [0]{.\EOS\space}%
\providecommand \EOS [0]{\spacefactor3000\relax}%
\providecommand \BibitemShut  [1]{\csname bibitem#1\endcsname}%
\let\auto@bib@innerbib\@empty
\bibitem [{\citenamefont {Sloan}(2003)}]{Sloan2003a}%
  \BibitemOpen
  \bibfield  {author} {\bibinfo {author} {\bibfnamefont {E.~D.}\ \bibnamefont
  {Sloan}},\ }\bibfield  {title} {\enquote {\bibinfo {title} {Fundamental
  principles and applications of natural gas hydrates},}\ }\href@noop {}
  {\bibfield  {journal} {\bibinfo  {journal} {Science}\ }\textbf {\bibinfo
  {volume} {426}},\ \bibinfo {pages} {353--359} (\bibinfo {year}
  {2003})}\BibitemShut {NoStop}%
\bibitem [{\citenamefont {Koh}, \citenamefont {Sum},\ and\ \citenamefont
  {Sloan}(2012)}]{Koh2012a}%
  \BibitemOpen
  \bibfield  {author} {\bibinfo {author} {\bibfnamefont {C.~A.}\ \bibnamefont
  {Koh}}, \bibinfo {author} {\bibfnamefont {A.~K.}\ \bibnamefont {Sum}}, \ and\
  \bibinfo {author} {\bibfnamefont {E.~D.}\ \bibnamefont {Sloan}},\ }\bibfield
  {title} {\enquote {\bibinfo {title} {State of the art: Natural gas hydrates
  as a natural resource},}\ }\href@noop {} {\bibfield  {journal} {\bibinfo
  {journal} {J. Nat. Gas Sci. Eng.}\ }\textbf {\bibinfo {volume} {8}},\
  \bibinfo {pages} {132--138} (\bibinfo {year} {2012})}\BibitemShut {NoStop}%
\bibitem [{\citenamefont {Sloan}\ and\ \citenamefont {Koh}(2008)}]{Sloan2008a}%
  \BibitemOpen
  \bibfield  {author} {\bibinfo {author} {\bibfnamefont {E.~D.}\ \bibnamefont
  {Sloan}}\ and\ \bibinfo {author} {\bibfnamefont {C.}~\bibnamefont {Koh}},\
  }\href@noop {} {\emph {\bibinfo {title} {{C}lathrate {H}ydrates of {N}atural
  {G}ases}}},\ \bibinfo {edition} {3rd}\ ed.\ (\bibinfo  {publisher} {CRC
  Press},\ \bibinfo {address} {New York},\ \bibinfo {year} {2008})\BibitemShut
  {NoStop}%
\bibitem [{\citenamefont {Ripmeester}\ and\ \citenamefont
  {Alavi}(2022)}]{Ripmeester2022a}%
  \BibitemOpen
  \bibfield  {author} {\bibinfo {author} {\bibfnamefont {J.~A.}\ \bibnamefont
  {Ripmeester}}\ and\ \bibinfo {author} {\bibfnamefont {S.}~\bibnamefont
  {Alavi}},\ }\href@noop {} {\emph {\bibinfo {title} {Clathrate Hydrates:
  Molecular Science and Characterization}}}\ (\bibinfo  {publisher} {Wiley-VCH:
  Weinheim, Germany},\ \bibinfo {year} {2022})\BibitemShut {NoStop}%
\bibitem [{\citenamefont {Barth{\'e}l{\'e}my}, \citenamefont {Weber},\ and\
  \citenamefont {Barbier}(2017)}]{barthelemy2017hydrogen}%
  \BibitemOpen
  \bibfield  {author} {\bibinfo {author} {\bibfnamefont {H.}~\bibnamefont
  {Barth{\'e}l{\'e}my}}, \bibinfo {author} {\bibfnamefont {M.}~\bibnamefont
  {Weber}}, \ and\ \bibinfo {author} {\bibfnamefont {F.}~\bibnamefont
  {Barbier}},\ }\bibfield  {title} {\enquote {\bibinfo {title} {Hydrogen
  storage: Recent improvements and industrial perspectives},}\ }\href@noop {}
  {\bibfield  {journal} {\bibinfo  {journal} {International Journal of Hydrogen
  Energy}\ }\textbf {\bibinfo {volume} {42}},\ \bibinfo {pages} {7254--7262}
  (\bibinfo {year} {2017})}\BibitemShut {NoStop}%
\bibitem [{\citenamefont {Chen}\ \emph {et~al.}(2023)\citenamefont {Chen},
  \citenamefont {Wang}, \citenamefont {Lang}, \citenamefont {Fan},\ and\
  \citenamefont {Li}}]{chen2023rapid}%
  \BibitemOpen
  \bibfield  {author} {\bibinfo {author} {\bibfnamefont {S.}~\bibnamefont
  {Chen}}, \bibinfo {author} {\bibfnamefont {Y.}~\bibnamefont {Wang}}, \bibinfo
  {author} {\bibfnamefont {X.}~\bibnamefont {Lang}}, \bibinfo {author}
  {\bibfnamefont {S.}~\bibnamefont {Fan}}, \ and\ \bibinfo {author}
  {\bibfnamefont {G.}~\bibnamefont {Li}},\ }\bibfield  {title} {\enquote
  {\bibinfo {title} {Rapid and high hydrogen storage in epoxycyclopentane
  hydrate at moderate pressure},}\ }\href@noop {} {\bibfield  {journal}
  {\bibinfo  {journal} {Energy}\ }\textbf {\bibinfo {volume} {268}},\ \bibinfo
  {pages} {126638} (\bibinfo {year} {2023})}\BibitemShut {NoStop}%
\bibitem [{\citenamefont {Zhang}\ \emph {et~al.}(2022)\citenamefont {Zhang},
  \citenamefont {Bhattacharjee}, \citenamefont {Zheng},\ and\ \citenamefont
  {Linga}}]{zhang2022hydrogen}%
  \BibitemOpen
  \bibfield  {author} {\bibinfo {author} {\bibfnamefont {Y.}~\bibnamefont
  {Zhang}}, \bibinfo {author} {\bibfnamefont {G.}~\bibnamefont
  {Bhattacharjee}}, \bibinfo {author} {\bibfnamefont {J.}~\bibnamefont
  {Zheng}}, \ and\ \bibinfo {author} {\bibfnamefont {P.}~\bibnamefont
  {Linga}},\ }\bibfield  {title} {\enquote {\bibinfo {title} {Hydrogen storage
  as clathrate hydrates in the presence of 1, 3-dioxolane as a dual-function
  promoter},}\ }\href@noop {} {\bibfield  {journal} {\bibinfo  {journal}
  {Chemical Engineering Journal}\ }\textbf {\bibinfo {volume} {427}},\ \bibinfo
  {pages} {131771} (\bibinfo {year} {2022})}\BibitemShut {NoStop}%
\bibitem [{\citenamefont {Kvenvolden}(1988)}]{CG_1988_71_41}%
  \BibitemOpen
  \bibfield  {author} {\bibinfo {author} {\bibfnamefont {K.~A.}\ \bibnamefont
  {Kvenvolden}},\ }\bibfield  {title} {\enquote {\bibinfo {title} {Methane
  hydrate - {A} major reservoir of carbon in the shallow geosphere},}\
  }\href@noop {} {\bibfield  {journal} {\bibinfo  {journal} {Chem. Geol.}\
  }\textbf {\bibinfo {volume} {71}},\ \bibinfo {pages} {41} (\bibinfo {year}
  {1988})}\BibitemShut {NoStop}%
\bibitem [{\citenamefont {MacDonald}(1990)}]{ARE_1990_15_53}%
  \BibitemOpen
  \bibfield  {author} {\bibinfo {author} {\bibfnamefont {G.~J.}\ \bibnamefont
  {MacDonald}},\ }\bibfield  {title} {\enquote {\bibinfo {title} {The future of
  methane as an energy resource},}\ }\href@noop {} {\bibfield  {journal}
  {\bibinfo  {journal} {Annu. Rev. Energy}\ }\textbf {\bibinfo {volume} {15}},\
  \bibinfo {pages} {53} (\bibinfo {year} {1990})}\BibitemShut {NoStop}%
\bibitem [{\citenamefont {Bourry}\ \emph {et~al.}(2007)\citenamefont {Bourry},
  \citenamefont {Charlou}, \citenamefont {Donval}, \citenamefont {Brunelli},
  \citenamefont {Focsa},\ and\ \citenamefont {Chazallon}}]{GRL_34_L22303_2007}%
  \BibitemOpen
  \bibfield  {author} {\bibinfo {author} {\bibfnamefont {C.}~\bibnamefont
  {Bourry}}, \bibinfo {author} {\bibfnamefont {J.~L.}\ \bibnamefont {Charlou}},
  \bibinfo {author} {\bibfnamefont {J.~P.}\ \bibnamefont {Donval}}, \bibinfo
  {author} {\bibfnamefont {M.}~\bibnamefont {Brunelli}}, \bibinfo {author}
  {\bibfnamefont {C.}~\bibnamefont {Focsa}}, \ and\ \bibinfo {author}
  {\bibfnamefont {B.}~\bibnamefont {Chazallon}},\ }\bibfield  {title} {\enquote
  {\bibinfo {title} {X-ray synchroton diffraction study of natural gas hydrates
  from african margin},}\ }\href@noop {} {\bibfield  {journal} {\bibinfo
  {journal} {Geophys. Res. Lett.}\ }\textbf {\bibinfo {volume} {34}},\ \bibinfo
  {pages} {L22303} (\bibinfo {year} {2007})}\BibitemShut {NoStop}%
\bibitem [{\citenamefont {Lu}\ \emph {et~al.}(2007)\citenamefont {Lu},
  \citenamefont {Seo}, \citenamefont {Lee}, \citenamefont {Moudrakovski},
  \citenamefont {Ripmeester}, \citenamefont {Chapman}, \citenamefont {Coffin},
  \citenamefont {Gardner},\ and\ \citenamefont {Pohlman}}]{N_2007_445_303}%
  \BibitemOpen
  \bibfield  {author} {\bibinfo {author} {\bibfnamefont {H.}~\bibnamefont
  {Lu}}, \bibinfo {author} {\bibfnamefont {Y.}~\bibnamefont {Seo}}, \bibinfo
  {author} {\bibfnamefont {J.}~\bibnamefont {Lee}}, \bibinfo {author}
  {\bibfnamefont {I.}~\bibnamefont {Moudrakovski}}, \bibinfo {author}
  {\bibfnamefont {J.~A.}\ \bibnamefont {Ripmeester}}, \bibinfo {author}
  {\bibfnamefont {N.~R.}\ \bibnamefont {Chapman}}, \bibinfo {author}
  {\bibfnamefont {R.~B.}\ \bibnamefont {Coffin}}, \bibinfo {author}
  {\bibfnamefont {G.}~\bibnamefont {Gardner}}, \ and\ \bibinfo {author}
  {\bibfnamefont {J.}~\bibnamefont {Pohlman}},\ }\bibfield  {title} {\enquote
  {\bibinfo {title} {Complex gas hydrate from the cascadia margin},}\
  }\href@noop {} {\bibfield  {journal} {\bibinfo  {journal} {Nature}\ }\textbf
  {\bibinfo {volume} {445}},\ \bibinfo {pages} {303} (\bibinfo {year}
  {2007})}\BibitemShut {NoStop}%
\bibitem [{\citenamefont {Lal}\ and\ \citenamefont
  {Nashed}(2020)}]{additives_book_hydrates}%
  \BibitemOpen
  \bibfield  {author} {\bibinfo {author} {\bibfnamefont {B.}~\bibnamefont
  {Lal}}\ and\ \bibinfo {author} {\bibfnamefont {O.}~\bibnamefont {Nashed}},\
  }\href@noop {} {\emph {\bibinfo {title} {Chemical Additives for Gas
  Hydrates}}}\ (\bibinfo  {publisher} {Springer},\ \bibinfo {year}
  {2020})\BibitemShut {NoStop}%
\bibitem [{\citenamefont {Ghiasi}, \citenamefont {Mohammadi},\ and\
  \citenamefont {Zendehboudi}(2021)}]{GHIASI2021114804}%
  \BibitemOpen
  \bibfield  {author} {\bibinfo {author} {\bibfnamefont {M.~M.}\ \bibnamefont
  {Ghiasi}}, \bibinfo {author} {\bibfnamefont {A.~H.}\ \bibnamefont
  {Mohammadi}}, \ and\ \bibinfo {author} {\bibfnamefont {S.}~\bibnamefont
  {Zendehboudi}},\ }\bibfield  {title} {\enquote {\bibinfo {title} {Modeling
  stability conditions of methane clathrate hydrate in ionic liquid aqueous
  solutions},}\ }\href@noop {} {\bibfield  {journal} {\bibinfo  {journal} {J.
  Mol. Liq.}\ }\textbf {\bibinfo {volume} {325}},\ \bibinfo {pages} {114804}
  (\bibinfo {year} {2021})}\BibitemShut {NoStop}%
\bibitem [{\citenamefont {Tanaka}, \citenamefont {Yagasaki},\ and\
  \citenamefont {Matsumoto}(2020)}]{JPCC_1_80_2020}%
  \BibitemOpen
  \bibfield  {author} {\bibinfo {author} {\bibfnamefont {H.}~\bibnamefont
  {Tanaka}}, \bibinfo {author} {\bibfnamefont {T.}~\bibnamefont {Yagasaki}}, \
  and\ \bibinfo {author} {\bibfnamefont {M.}~\bibnamefont {Matsumoto}},\
  }\bibfield  {title} {\enquote {\bibinfo {title} {On the occurrence of
  clathrate hydrates in extreme conditions: Dissociation pressures and
  occupancies at cryogenic temperatures with application to planetary
  systems},}\ }\href@noop {} {\bibfield  {journal} {\bibinfo  {journal}
  {Planet. Sci. J.}\ }\textbf {\bibinfo {volume} {1}},\ \bibinfo {pages} {80}
  (\bibinfo {year} {2020})}\BibitemShut {NoStop}%
\bibitem [{\citenamefont {Conde}, \citenamefont {Rovere},\ and\ \citenamefont
  {Gallo}(2017{\natexlab{a}})}]{PCCP_2017_19_9566}%
  \BibitemOpen
  \bibfield  {author} {\bibinfo {author} {\bibfnamefont {M.~M.}\ \bibnamefont
  {Conde}}, \bibinfo {author} {\bibfnamefont {M.}~\bibnamefont {Rovere}}, \
  and\ \bibinfo {author} {\bibfnamefont {P.}~\bibnamefont {Gallo}},\ }\bibfield
   {title} {\enquote {\bibinfo {title} {Spontaneous {N}a{C}l-doped ice at
  seawater conditions: focus on the mechanisms of ion inclusion},}\ }\href@noop
  {} {\bibfield  {journal} {\bibinfo  {journal} {Phys. Chem. Chem. Phys.}\
  }\textbf {\bibinfo {volume} {19}},\ \bibinfo {pages} {9566} (\bibinfo {year}
  {2017}{\natexlab{a}})}\BibitemShut {NoStop}%
\bibitem [{\citenamefont {Conde}, \citenamefont {Rovere},\ and\ \citenamefont
  {Gallo}(2021)}]{D1CP02638K}%
  \BibitemOpen
  \bibfield  {author} {\bibinfo {author} {\bibfnamefont {M.~M.}\ \bibnamefont
  {Conde}}, \bibinfo {author} {\bibfnamefont {M.}~\bibnamefont {Rovere}}, \
  and\ \bibinfo {author} {\bibfnamefont {P.}~\bibnamefont {Gallo}},\ }\bibfield
   {title} {\enquote {\bibinfo {title} {Spontaneous nacl-doped ices ih{,} ic{,}
  iii{,} v and vi. understanding the mechanism of ion inclusion and its
  dependence on the crystalline structure of ice},}\ }\href {\doibase
  10.1039/D1CP02638K} {\bibfield  {journal} {\bibinfo  {journal} {Phys. Chem.
  Chem. Phys.}\ }\textbf {\bibinfo {volume} {23}},\ \bibinfo {pages}
  {22897--22911} (\bibinfo {year} {2021})}\BibitemShut {NoStop}%
\bibitem [{\citenamefont {Prieto-Ballesteros}\ \emph
  {et~al.}(2005)\citenamefont {Prieto-Ballesteros}, \citenamefont {Kargel},
  \citenamefont {Fernández-Sampedro}, \citenamefont {Selsis}, \citenamefont
  {Martínez},\ and\ \citenamefont {Hogenboom}}]{PRIETOBALLESTEROS2005491}%
  \BibitemOpen
  \bibfield  {author} {\bibinfo {author} {\bibfnamefont {O.}~\bibnamefont
  {Prieto-Ballesteros}}, \bibinfo {author} {\bibfnamefont {J.~S.}\ \bibnamefont
  {Kargel}}, \bibinfo {author} {\bibfnamefont {M.}~\bibnamefont
  {Fernández-Sampedro}}, \bibinfo {author} {\bibfnamefont {F.}~\bibnamefont
  {Selsis}}, \bibinfo {author} {\bibfnamefont {E.~S.}\ \bibnamefont
  {Martínez}}, \ and\ \bibinfo {author} {\bibfnamefont {D.~L.}\ \bibnamefont
  {Hogenboom}},\ }\bibfield  {title} {\enquote {\bibinfo {title} {Evaluation of
  the possible presence of clathrate hydrates in europa's icy shell or
  seafloor},}\ }\href {\doibase https://doi.org/10.1016/j.icarus.2005.02.021}
  {\bibfield  {journal} {\bibinfo  {journal} {Icarus}\ }\textbf {\bibinfo
  {volume} {177}},\ \bibinfo {pages} {491--505} (\bibinfo {year} {2005})},\
  \bibinfo {note} {europa Icy Shell}\BibitemShut {NoStop}%
\bibitem [{\citenamefont {Kargel}\ \emph {et~al.}(2000)\citenamefont {Kargel},
  \citenamefont {Kaye}, \citenamefont {Head}, \citenamefont {Marion},
  \citenamefont {Sassen}, \citenamefont {Crowley}, \citenamefont {Ballesteros},
  \citenamefont {Grant},\ and\ \citenamefont {Hogenboom}}]{KARGEL2000226}%
  \BibitemOpen
  \bibfield  {author} {\bibinfo {author} {\bibfnamefont {J.~S.}\ \bibnamefont
  {Kargel}}, \bibinfo {author} {\bibfnamefont {J.~Z.}\ \bibnamefont {Kaye}},
  \bibinfo {author} {\bibfnamefont {J.~W.}\ \bibnamefont {Head}}, \bibinfo
  {author} {\bibfnamefont {G.~M.}\ \bibnamefont {Marion}}, \bibinfo {author}
  {\bibfnamefont {R.}~\bibnamefont {Sassen}}, \bibinfo {author} {\bibfnamefont
  {J.~K.}\ \bibnamefont {Crowley}}, \bibinfo {author} {\bibfnamefont {O.~P.}\
  \bibnamefont {Ballesteros}}, \bibinfo {author} {\bibfnamefont {S.~A.}\
  \bibnamefont {Grant}}, \ and\ \bibinfo {author} {\bibfnamefont {D.~L.}\
  \bibnamefont {Hogenboom}},\ }\bibfield  {title} {\enquote {\bibinfo {title}
  {Europa's crust and ocean: Origin, composition, and the prospects for
  life},}\ }\href {\doibase https://doi.org/10.1006/icar.2000.6471} {\bibfield
  {journal} {\bibinfo  {journal} {Icarus}\ }\textbf {\bibinfo {volume} {148}},\
  \bibinfo {pages} {226--265} (\bibinfo {year} {2000})}\BibitemShut {NoStop}%
\bibitem [{\citenamefont {Prieto-Ballesteros}\ \emph
  {et~al.}(2006)\citenamefont {Prieto-Ballesteros}, \citenamefont {Kargel},
  \citenamefont {Faire{\'e}n}, \citenamefont {Fern{\'a}ndez-Remolar},
  \citenamefont {Dohm},\ and\ \citenamefont {Amils}}]{10.1130/G22311.1}%
  \BibitemOpen
  \bibfield  {author} {\bibinfo {author} {\bibfnamefont {O.}~\bibnamefont
  {Prieto-Ballesteros}}, \bibinfo {author} {\bibfnamefont {J.~S.}\ \bibnamefont
  {Kargel}}, \bibinfo {author} {\bibfnamefont {A.~G.}\ \bibnamefont
  {Faire{\'e}n}}, \bibinfo {author} {\bibfnamefont {D.~C.}\ \bibnamefont
  {Fern{\'a}ndez-Remolar}}, \bibinfo {author} {\bibfnamefont {J.~M.}\
  \bibnamefont {Dohm}}, \ and\ \bibinfo {author} {\bibfnamefont
  {R.}~\bibnamefont {Amils}},\ }\bibfield  {title} {\enquote {\bibinfo {title}
  {{Interglacial clathrate destabilization on Mars: Possible contributing
  source of its atmospheric methane}},}\ }\href@noop {} {\bibfield  {journal}
  {\bibinfo  {journal} {Geology}\ }\textbf {\bibinfo {volume} {34}},\ \bibinfo
  {pages} {149--152} (\bibinfo {year} {2006})}\BibitemShut {NoStop}%
\bibitem [{\citenamefont {Pettinelli}\ \emph {et~al.}(2015)\citenamefont
  {Pettinelli}, \citenamefont {Cosciotti}, \citenamefont {Di~Paolo},
  \citenamefont {Lauro}, \citenamefont {Mattei}, \citenamefont {Orosei},\ and\
  \citenamefont {Vannaroni}}]{https://doi.org/10.1002/2014RG000463}%
  \BibitemOpen
  \bibfield  {author} {\bibinfo {author} {\bibfnamefont {E.}~\bibnamefont
  {Pettinelli}}, \bibinfo {author} {\bibfnamefont {B.}~\bibnamefont
  {Cosciotti}}, \bibinfo {author} {\bibfnamefont {F.}~\bibnamefont {Di~Paolo}},
  \bibinfo {author} {\bibfnamefont {S.~E.}\ \bibnamefont {Lauro}}, \bibinfo
  {author} {\bibfnamefont {E.}~\bibnamefont {Mattei}}, \bibinfo {author}
  {\bibfnamefont {R.}~\bibnamefont {Orosei}}, \ and\ \bibinfo {author}
  {\bibfnamefont {G.}~\bibnamefont {Vannaroni}},\ }\bibfield  {title} {\enquote
  {\bibinfo {title} {Dielectric properties of jovian satellite ice analogs for
  subsurface radar exploration: A review},}\ }\href {\doibase
  https://doi.org/10.1002/2014RG000463} {\bibfield  {journal} {\bibinfo
  {journal} {Reviews of Geophysics}\ }\textbf {\bibinfo {volume} {53}},\
  \bibinfo {pages} {593--641} (\bibinfo {year} {2015})}\BibitemShut {NoStop}%
\bibitem [{\citenamefont {Peter}(2018)}]{peter2018reduction}%
  \BibitemOpen
  \bibfield  {author} {\bibinfo {author} {\bibfnamefont {S.~C.}\ \bibnamefont
  {Peter}},\ }\bibfield  {title} {\enquote {\bibinfo {title} {Reduction of co2
  to chemicals and fuels: a solution to global warming and energy crisis},}\
  }\href@noop {} {\bibfield  {journal} {\bibinfo  {journal} {ACS Energy
  Letters}\ }\textbf {\bibinfo {volume} {3}},\ \bibinfo {pages} {1557--1561}
  (\bibinfo {year} {2018})}\BibitemShut {NoStop}%
\bibitem [{\citenamefont {Yoro}\ and\ \citenamefont
  {Daramola}(2020)}]{yoro2020co2}%
  \BibitemOpen
  \bibfield  {author} {\bibinfo {author} {\bibfnamefont {K.~O.}\ \bibnamefont
  {Yoro}}\ and\ \bibinfo {author} {\bibfnamefont {M.~O.}\ \bibnamefont
  {Daramola}},\ }\bibfield  {title} {\enquote {\bibinfo {title} {Co2 emission
  sources, greenhouse gases, and the global warming effect},}\ }in\ \href@noop
  {} {\emph {\bibinfo {booktitle} {Advances in carbon capture}}}\ (\bibinfo
  {publisher} {Elsevier},\ \bibinfo {year} {2020})\ pp.\ \bibinfo {pages}
  {3--28}\BibitemShut {NoStop}%
\bibitem [{\citenamefont {D'Alessandro}, \citenamefont {Smit},\ and\
  \citenamefont {Long}(2010)}]{d2010carbon}%
  \BibitemOpen
  \bibfield  {author} {\bibinfo {author} {\bibfnamefont {D.~M.}\ \bibnamefont
  {D'Alessandro}}, \bibinfo {author} {\bibfnamefont {B.}~\bibnamefont {Smit}},
  \ and\ \bibinfo {author} {\bibfnamefont {J.~R.}\ \bibnamefont {Long}},\
  }\bibfield  {title} {\enquote {\bibinfo {title} {Carbon dioxide capture:
  prospects for new materials},}\ }\href@noop {} {\bibfield  {journal}
  {\bibinfo  {journal} {Angewandte Chemie International Edition}\ }\textbf
  {\bibinfo {volume} {49}},\ \bibinfo {pages} {6058--6082} (\bibinfo {year}
  {2010})}\BibitemShut {NoStop}%
\bibitem [{\citenamefont {Gao}\ \emph {et~al.}(2020)\citenamefont {Gao},
  \citenamefont {Liang}, \citenamefont {Wang}, \citenamefont {Jiang},
  \citenamefont {Zhang}, \citenamefont {Zheng}, \citenamefont {Xie},
  \citenamefont {Toe}, \citenamefont {Zhu}, \citenamefont {Wang} \emph
  {et~al.}}]{gao2020industrial}%
  \BibitemOpen
  \bibfield  {author} {\bibinfo {author} {\bibfnamefont {W.}~\bibnamefont
  {Gao}}, \bibinfo {author} {\bibfnamefont {S.}~\bibnamefont {Liang}}, \bibinfo
  {author} {\bibfnamefont {R.}~\bibnamefont {Wang}}, \bibinfo {author}
  {\bibfnamefont {Q.}~\bibnamefont {Jiang}}, \bibinfo {author} {\bibfnamefont
  {Y.}~\bibnamefont {Zhang}}, \bibinfo {author} {\bibfnamefont
  {Q.}~\bibnamefont {Zheng}}, \bibinfo {author} {\bibfnamefont
  {B.}~\bibnamefont {Xie}}, \bibinfo {author} {\bibfnamefont {C.~Y.}\
  \bibnamefont {Toe}}, \bibinfo {author} {\bibfnamefont {X.}~\bibnamefont
  {Zhu}}, \bibinfo {author} {\bibfnamefont {J.}~\bibnamefont {Wang}},  \emph
  {et~al.},\ }\bibfield  {title} {\enquote {\bibinfo {title} {Industrial carbon
  dioxide capture and utilization: state of the art and future challenges},}\
  }\href@noop {} {\bibfield  {journal} {\bibinfo  {journal} {Chemical Society
  Reviews}\ }\textbf {\bibinfo {volume} {49}},\ \bibinfo {pages} {8584--8686}
  (\bibinfo {year} {2020})}\BibitemShut {NoStop}%
\bibitem [{\citenamefont {Wang}, \citenamefont {Zhang},\ and\ \citenamefont
  {Lipi{\'n}ski}(2020)}]{wang2020research}%
  \BibitemOpen
  \bibfield  {author} {\bibinfo {author} {\bibfnamefont {X.}~\bibnamefont
  {Wang}}, \bibinfo {author} {\bibfnamefont {F.}~\bibnamefont {Zhang}}, \ and\
  \bibinfo {author} {\bibfnamefont {W.}~\bibnamefont {Lipi{\'n}ski}},\
  }\bibfield  {title} {\enquote {\bibinfo {title} {Research progress and
  challenges in hydrate-based carbon dioxide capture applications},}\
  }\href@noop {} {\bibfield  {journal} {\bibinfo  {journal} {Applied Energy}\
  }\textbf {\bibinfo {volume} {269}},\ \bibinfo {pages} {114928} (\bibinfo
  {year} {2020})}\BibitemShut {NoStop}%
\bibitem [{\citenamefont {Nguyen}\ \emph {et~al.}(2022)\citenamefont {Nguyen},
  \citenamefont {La}, \citenamefont {Huynh},\ and\ \citenamefont
  {Nguyen}}]{nguyen2022technical}%
  \BibitemOpen
  \bibfield  {author} {\bibinfo {author} {\bibfnamefont {N.~N.}\ \bibnamefont
  {Nguyen}}, \bibinfo {author} {\bibfnamefont {V.~T.}\ \bibnamefont {La}},
  \bibinfo {author} {\bibfnamefont {C.~D.}\ \bibnamefont {Huynh}}, \ and\
  \bibinfo {author} {\bibfnamefont {A.~V.}\ \bibnamefont {Nguyen}},\ }\bibfield
   {title} {\enquote {\bibinfo {title} {Technical and economic perspectives of
  hydrate-based carbon dioxide capture},}\ }\href@noop {} {\bibfield  {journal}
  {\bibinfo  {journal} {Applied Energy}\ }\textbf {\bibinfo {volume} {307}},\
  \bibinfo {pages} {118237} (\bibinfo {year} {2022})}\BibitemShut {NoStop}%
\bibitem [{\citenamefont {Zheng}\ \emph {et~al.}(2020)\citenamefont {Zheng},
  \citenamefont {Chong}, \citenamefont {Qureshi},\ and\ \citenamefont
  {Linga}}]{zheng2020carbon}%
  \BibitemOpen
  \bibfield  {author} {\bibinfo {author} {\bibfnamefont {J.}~\bibnamefont
  {Zheng}}, \bibinfo {author} {\bibfnamefont {Z.~R.}\ \bibnamefont {Chong}},
  \bibinfo {author} {\bibfnamefont {M.~F.}\ \bibnamefont {Qureshi}}, \ and\
  \bibinfo {author} {\bibfnamefont {P.}~\bibnamefont {Linga}},\ }\bibfield
  {title} {\enquote {\bibinfo {title} {Carbon dioxide sequestration via gas
  hydrates: a potential pathway toward decarbonization},}\ }\href@noop {}
  {\bibfield  {journal} {\bibinfo  {journal} {Energy \& Fuels}\ }\textbf
  {\bibinfo {volume} {34}},\ \bibinfo {pages} {10529--10546} (\bibinfo {year}
  {2020})}\BibitemShut {NoStop}%
\bibitem [{\citenamefont {M{\'\i}guez}\ \emph {et~al.}(2015)\citenamefont
  {M{\'\i}guez}, \citenamefont {Conde}, \citenamefont {Torr{\'e}},
  \citenamefont {Blas}, \citenamefont {Pi{\~n}eiro},\ and\ \citenamefont
  {Vega}}]{Miguez2015a}%
  \BibitemOpen
  \bibfield  {author} {\bibinfo {author} {\bibfnamefont {J.~M.}\ \bibnamefont
  {M{\'\i}guez}}, \bibinfo {author} {\bibfnamefont {M.~M.}\ \bibnamefont
  {Conde}}, \bibinfo {author} {\bibfnamefont {J.-P.}\ \bibnamefont
  {Torr{\'e}}}, \bibinfo {author} {\bibfnamefont {F.~J.}\ \bibnamefont {Blas}},
  \bibinfo {author} {\bibfnamefont {M.~M.}\ \bibnamefont {Pi{\~n}eiro}}, \ and\
  \bibinfo {author} {\bibfnamefont {C.}~\bibnamefont {Vega}},\ }\bibfield
  {title} {\enquote {\bibinfo {title} {Molecular dynamics simulation of
  {CO$_2$} hydrates: Prediction of three phase coexistence line},}\ }\href
  {\doibase https://doi.org/10.1063/1.4916119} {\bibfield  {journal} {\bibinfo
  {journal} {J. Chem. Phys.}\ }\textbf {\bibinfo {volume} {142}},\ \bibinfo
  {pages} {124505} (\bibinfo {year} {2015})}\BibitemShut {NoStop}%
\bibitem [{\citenamefont {Tung}\ \emph {et~al.}(2011)\citenamefont {Tung},
  \citenamefont {Chen}, \citenamefont {Chen},\ and\ \citenamefont
  {Lin}}]{Tung2011a}%
  \BibitemOpen
  \bibfield  {author} {\bibinfo {author} {\bibfnamefont {Y.-T.}\ \bibnamefont
  {Tung}}, \bibinfo {author} {\bibfnamefont {L.-J.}\ \bibnamefont {Chen}},
  \bibinfo {author} {\bibfnamefont {Y.-P.}\ \bibnamefont {Chen}}, \ and\
  \bibinfo {author} {\bibfnamefont {S.-T.}\ \bibnamefont {Lin}},\ }\bibfield
  {title} {\enquote {\bibinfo {title} {Growth of structure i carbon dioxide
  hydrate from molecular dynamics simulations},}\ }\href@noop {} {\bibfield
  {journal} {\bibinfo  {journal} {The Journal of Physical Chemistry C}\
  }\textbf {\bibinfo {volume} {115}},\ \bibinfo {pages} {7504--7515} (\bibinfo
  {year} {2011})}\BibitemShut {NoStop}%
\bibitem [{\citenamefont {Horn}\ \emph {et~al.}(2004)\citenamefont {Horn},
  \citenamefont {Swope}, \citenamefont {Pitera}, \citenamefont {Madura},
  \citenamefont {Dick}, \citenamefont {Hura},\ and\ \citenamefont
  {Head-Gordon}}]{horn2004development}%
  \BibitemOpen
  \bibfield  {author} {\bibinfo {author} {\bibfnamefont {H.~W.}\ \bibnamefont
  {Horn}}, \bibinfo {author} {\bibfnamefont {W.~C.}\ \bibnamefont {Swope}},
  \bibinfo {author} {\bibfnamefont {J.~W.}\ \bibnamefont {Pitera}}, \bibinfo
  {author} {\bibfnamefont {J.~D.}\ \bibnamefont {Madura}}, \bibinfo {author}
  {\bibfnamefont {T.~J.}\ \bibnamefont {Dick}}, \bibinfo {author}
  {\bibfnamefont {G.~L.}\ \bibnamefont {Hura}}, \ and\ \bibinfo {author}
  {\bibfnamefont {T.}~\bibnamefont {Head-Gordon}},\ }\bibfield  {title}
  {\enquote {\bibinfo {title} {Development of an improved four-site water model
  for biomolecular simulations: Tip4p-ew},}\ }\href@noop {} {\bibfield
  {journal} {\bibinfo  {journal} {The Journal of chemical physics}\ }\textbf
  {\bibinfo {volume} {120}},\ \bibinfo {pages} {9665--9678} (\bibinfo {year}
  {2004})}\BibitemShut {NoStop}%
\bibitem [{\citenamefont {Harris}\ and\ \citenamefont
  {Yung}(1995)}]{harris1995carbon}%
  \BibitemOpen
  \bibfield  {author} {\bibinfo {author} {\bibfnamefont {J.~G.}\ \bibnamefont
  {Harris}}\ and\ \bibinfo {author} {\bibfnamefont {K.~H.}\ \bibnamefont
  {Yung}},\ }\bibfield  {title} {\enquote {\bibinfo {title} {Carbon dioxide's
  liquid-vapor coexistence curve and critical properties as predicted by a
  simple molecular model},}\ }\href@noop {} {\bibfield  {journal} {\bibinfo
  {journal} {The Journal of Physical Chemistry}\ }\textbf {\bibinfo {volume}
  {99}},\ \bibinfo {pages} {12021--12024} (\bibinfo {year} {1995})}\BibitemShut
  {NoStop}%
\bibitem [{\citenamefont {Costandy}\ \emph {et~al.}(2015)\citenamefont
  {Costandy}, \citenamefont {Michalisa}, \citenamefont {Tsimpanogiannis},
  \citenamefont {Stubos},\ and\ \citenamefont {Economou}}]{Costandy2015a}%
  \BibitemOpen
  \bibfield  {author} {\bibinfo {author} {\bibfnamefont {J.}~\bibnamefont
  {Costandy}}, \bibinfo {author} {\bibfnamefont {V.~K.}\ \bibnamefont
  {Michalisa}}, \bibinfo {author} {\bibfnamefont {I.~N.}\ \bibnamefont
  {Tsimpanogiannis}}, \bibinfo {author} {\bibfnamefont {A.~K.}\ \bibnamefont
  {Stubos}}, \ and\ \bibinfo {author} {\bibfnamefont {I.~G.}\ \bibnamefont
  {Economou}},\ }\bibfield  {title} {\enquote {\bibinfo {title} {The role of
  intermolecular interactions in the prediction of the phase equilibria of
  carbon dioxide hydrates},}\ }\href {\doibase
  https://doi.org/10.1063/1.4929805} {\bibfield  {journal} {\bibinfo  {journal}
  {J. Chem. Phys.}\ }\textbf {\bibinfo {volume} {143}},\ \bibinfo {pages}
  {094506} (\bibinfo {year} {2015})}\BibitemShut {NoStop}%
\bibitem [{\citenamefont {Waage}, \citenamefont {Vlugt},\ and\ \citenamefont
  {Kjelstrup}(2017)}]{Waage2017a}%
  \BibitemOpen
  \bibfield  {author} {\bibinfo {author} {\bibfnamefont {M.~H.}\ \bibnamefont
  {Waage}}, \bibinfo {author} {\bibfnamefont {T.~J.~H.}\ \bibnamefont {Vlugt}},
  \ and\ \bibinfo {author} {\bibfnamefont {S.}~\bibnamefont {Kjelstrup}},\
  }\bibfield  {title} {\enquote {\bibinfo {title} {Phase diagram of methane and
  carbon dioxide hydrates computed by {Monte Carlo} simulations},}\ }\href
  {\doibase https://doi.org/10.1021/acs.jpcb.7b03071} {\bibfield  {journal}
  {\bibinfo  {journal} {J. Phys. Chem. B}\ }\textbf {\bibinfo {volume} {121}},\
  \bibinfo {pages} {7336--7350} (\bibinfo {year} {2017})}\BibitemShut {NoStop}%
\bibitem [{\citenamefont {Jiao}\ \emph {et~al.}(2021)\citenamefont {Jiao},
  \citenamefont {Wang}, \citenamefont {Li}, \citenamefont {Zhao},\ and\
  \citenamefont {Wan}}]{Jiao2021a}%
  \BibitemOpen
  \bibfield  {author} {\bibinfo {author} {\bibfnamefont {L.}~\bibnamefont
  {Jiao}}, \bibinfo {author} {\bibfnamefont {Z.}~\bibnamefont {Wang}}, \bibinfo
  {author} {\bibfnamefont {J.}~\bibnamefont {Li}}, \bibinfo {author}
  {\bibfnamefont {P.}~\bibnamefont {Zhao}}, \ and\ \bibinfo {author}
  {\bibfnamefont {R.}~\bibnamefont {Wan}},\ }\bibfield  {title} {\enquote
  {\bibinfo {title} {Stability and dissociation studies of co2 hydrate under
  different systems using molecular dynamic simulations},}\ }\href@noop {}
  {\bibfield  {journal} {\bibinfo  {journal} {Journal of Molecular Liquids}\
  }\textbf {\bibinfo {volume} {338}},\ \bibinfo {pages} {116788} (\bibinfo
  {year} {2021})}\BibitemShut {NoStop}%
\bibitem [{\citenamefont {Hao}\ \emph {et~al.}(2023)\citenamefont {Hao},
  \citenamefont {Li}, \citenamefont {Meng}, \citenamefont {Sun}, \citenamefont
  {Huang}, \citenamefont {Bu},\ and\ \citenamefont {Li}}]{Hao2023a}%
  \BibitemOpen
  \bibfield  {author} {\bibinfo {author} {\bibfnamefont {X.}~\bibnamefont
  {Hao}}, \bibinfo {author} {\bibfnamefont {C.}~\bibnamefont {Li}}, \bibinfo
  {author} {\bibfnamefont {Q.}~\bibnamefont {Meng}}, \bibinfo {author}
  {\bibfnamefont {J.}~\bibnamefont {Sun}}, \bibinfo {author} {\bibfnamefont
  {L.}~\bibnamefont {Huang}}, \bibinfo {author} {\bibfnamefont
  {Q.}~\bibnamefont {Bu}}, \ and\ \bibinfo {author} {\bibfnamefont
  {C.}~\bibnamefont {Li}},\ }\bibfield  {title} {\enquote {\bibinfo {title}
  {Molecular dynamics simulation of the three-phase equilibrium line of co2
  hydrate with opc water model},}\ }\href@noop {} {\bibfield  {journal}
  {\bibinfo  {journal} {ACS omega}\ }\textbf {\bibinfo {volume} {8}},\ \bibinfo
  {pages} {39847--39854} (\bibinfo {year} {2023})}\BibitemShut {NoStop}%
\bibitem [{\citenamefont {Qiu}\ \emph {et~al.}(2018)\citenamefont {Qiu},
  \citenamefont {Bai}, \citenamefont {Sun}, \citenamefont {Yu}, \citenamefont
  {Yang}, \citenamefont {Li}, \citenamefont {Yang}, \citenamefont {Huang},\
  and\ \citenamefont {Du}}]{Qiu2018a}%
  \BibitemOpen
  \bibfield  {author} {\bibinfo {author} {\bibfnamefont {N.}~\bibnamefont
  {Qiu}}, \bibinfo {author} {\bibfnamefont {X.}~\bibnamefont {Bai}}, \bibinfo
  {author} {\bibfnamefont {N.}~\bibnamefont {Sun}}, \bibinfo {author}
  {\bibfnamefont {X.}~\bibnamefont {Yu}}, \bibinfo {author} {\bibfnamefont
  {L.}~\bibnamefont {Yang}}, \bibinfo {author} {\bibfnamefont {Y.}~\bibnamefont
  {Li}}, \bibinfo {author} {\bibfnamefont {M.}~\bibnamefont {Yang}}, \bibinfo
  {author} {\bibfnamefont {Q.}~\bibnamefont {Huang}}, \ and\ \bibinfo {author}
  {\bibfnamefont {S.}~\bibnamefont {Du}},\ }\bibfield  {title} {\enquote
  {\bibinfo {title} {Grand canonical monte carlo simulations on phase
  equilibria of methane, carbon dioxide, and their mixture hydrates},}\
  }\href@noop {} {\bibfield  {journal} {\bibinfo  {journal} {The Journal of
  Physical Chemistry B}\ }\textbf {\bibinfo {volume} {122}},\ \bibinfo {pages}
  {9724--9737} (\bibinfo {year} {2018})}\BibitemShut {NoStop}%
\bibitem [{\citenamefont {El~Meragawi}\ \emph {et~al.}(2016)\citenamefont
  {El~Meragawi}, \citenamefont {Diamantonis}, \citenamefont {Tsimpanogiannis},\
  and\ \citenamefont {Economou}}]{El2016a}%
  \BibitemOpen
  \bibfield  {author} {\bibinfo {author} {\bibfnamefont {S.}~\bibnamefont
  {El~Meragawi}}, \bibinfo {author} {\bibfnamefont {N.~I.}\ \bibnamefont
  {Diamantonis}}, \bibinfo {author} {\bibfnamefont {I.~N.}\ \bibnamefont
  {Tsimpanogiannis}}, \ and\ \bibinfo {author} {\bibfnamefont {I.~G.}\
  \bibnamefont {Economou}},\ }\bibfield  {title} {\enquote {\bibinfo {title}
  {Hydrate--fluid phase equilibria modeling using pc-saft and peng--robinson
  equations of state},}\ }\href@noop {} {\bibfield  {journal} {\bibinfo
  {journal} {Fluid Phase Equilibria}\ }\textbf {\bibinfo {volume} {413}},\
  \bibinfo {pages} {209--219} (\bibinfo {year} {2016})}\BibitemShut {NoStop}%
\bibitem [{\citenamefont {J{\"a}ger}\ \emph {et~al.}(2013)\citenamefont
  {J{\"a}ger}, \citenamefont {Vin{\v{s}}}, \citenamefont {Gernert},
  \citenamefont {Span},\ and\ \citenamefont {Hrub{\`y}}}]{jager2013a}%
  \BibitemOpen
  \bibfield  {author} {\bibinfo {author} {\bibfnamefont {A.}~\bibnamefont
  {J{\"a}ger}}, \bibinfo {author} {\bibfnamefont {V.}~\bibnamefont
  {Vin{\v{s}}}}, \bibinfo {author} {\bibfnamefont {J.}~\bibnamefont {Gernert}},
  \bibinfo {author} {\bibfnamefont {R.}~\bibnamefont {Span}}, \ and\ \bibinfo
  {author} {\bibfnamefont {J.}~\bibnamefont {Hrub{\`y}}},\ }\bibfield  {title}
  {\enquote {\bibinfo {title} {Phase equilibria with hydrate formation in h2o+
  co2 mixtures modeled with reference equations of state},}\ }\href@noop {}
  {\bibfield  {journal} {\bibinfo  {journal} {Fluid Phase Equilibria}\ }\textbf
  {\bibinfo {volume} {338}},\ \bibinfo {pages} {100--113} (\bibinfo {year}
  {2013})}\BibitemShut {NoStop}%
\bibitem [{\citenamefont {Sun}\ \emph {et~al.}(2005)\citenamefont {Sun},
  \citenamefont {Zhao}, \citenamefont {Kiselev},\ and\ \citenamefont
  {McCabe}}]{Sun2005b}%
  \BibitemOpen
  \bibfield  {author} {\bibinfo {author} {\bibfnamefont {L.}~\bibnamefont
  {Sun}}, \bibinfo {author} {\bibfnamefont {H.}~\bibnamefont {Zhao}}, \bibinfo
  {author} {\bibfnamefont {S.~B.}\ \bibnamefont {Kiselev}}, \ and\ \bibinfo
  {author} {\bibfnamefont {C.}~\bibnamefont {McCabe}},\ }\bibfield  {title}
  {\enquote {\bibinfo {title} {{Predicting Mixture Phase Equilibria and
  Critical Behavior Using the SAFT-VRX Approach}},}\ }\href@noop {} {\bibfield
  {journal} {\bibinfo  {journal} {Fluid Phase Equil.}\ }\textbf {\bibinfo
  {volume} {228-229}},\ \bibinfo {pages} {275--282} (\bibinfo {year} {2005})},\
  \bibinfo {note} {\textbf{Sun2005b}}\BibitemShut {NoStop}%
\bibitem [{\citenamefont {Blazquez}\ \emph {et~al.}(2023)\citenamefont
  {Blazquez}, \citenamefont {M~Conde}, \citenamefont {Vega},\ and\
  \citenamefont {Sanz}}]{blazquez2023growth}%
  \BibitemOpen
  \bibfield  {author} {\bibinfo {author} {\bibfnamefont {S.}~\bibnamefont
  {Blazquez}}, \bibinfo {author} {\bibfnamefont {M.}~\bibnamefont {M~Conde}},
  \bibinfo {author} {\bibfnamefont {C.}~\bibnamefont {Vega}}, \ and\ \bibinfo
  {author} {\bibfnamefont {E.}~\bibnamefont {Sanz}},\ }\bibfield  {title}
  {\enquote {\bibinfo {title} {Growth rate of co2 and ch4 hydrates by means of
  molecular dynamics simulations},}\ }\href@noop {} {\bibfield  {journal}
  {\bibinfo  {journal} {The Journal of Chemical Physics}\ }\textbf {\bibinfo
  {volume} {159}} (\bibinfo {year} {2023})}\BibitemShut {NoStop}%
\bibitem [{\citenamefont {Wang}\ \emph {et~al.}(2023)\citenamefont {Wang},
  \citenamefont {Lu}, \citenamefont {Zhang}, \citenamefont {Fan}, \citenamefont
  {Li}, \citenamefont {Zhang}, \citenamefont {Zhao}, \citenamefont {Yang},\
  and\ \citenamefont {Song}}]{Wang2023a}%
  \BibitemOpen
  \bibfield  {author} {\bibinfo {author} {\bibfnamefont {H.}~\bibnamefont
  {Wang}}, \bibinfo {author} {\bibfnamefont {Y.}~\bibnamefont {Lu}}, \bibinfo
  {author} {\bibfnamefont {X.}~\bibnamefont {Zhang}}, \bibinfo {author}
  {\bibfnamefont {Q.}~\bibnamefont {Fan}}, \bibinfo {author} {\bibfnamefont
  {Q.}~\bibnamefont {Li}}, \bibinfo {author} {\bibfnamefont {L.}~\bibnamefont
  {Zhang}}, \bibinfo {author} {\bibfnamefont {J.}~\bibnamefont {Zhao}},
  \bibinfo {author} {\bibfnamefont {L.}~\bibnamefont {Yang}}, \ and\ \bibinfo
  {author} {\bibfnamefont {Y.}~\bibnamefont {Song}},\ }\bibfield  {title}
  {\enquote {\bibinfo {title} {Molecular dynamics of carbon sequestration via
  forming co2 hydrate in a marine environment},}\ }\href@noop {} {\bibfield
  {journal} {\bibinfo  {journal} {Energy \& Fuels}\ } (\bibinfo {year}
  {2023})}\BibitemShut {NoStop}%
\bibitem [{\citenamefont {Fern{\'a}ndez-Fern{\'a}ndez}\ \emph
  {et~al.}(2019)\citenamefont {Fern{\'a}ndez-Fern{\'a}ndez}, \citenamefont
  {P{\'e}rez-Rodr{\'\i}guez}, \citenamefont {Comesa{\~n}a},\ and\ \citenamefont
  {Pi{\~n}eiro}}]{Fernandez-Fernandez2019a}%
  \BibitemOpen
  \bibfield  {author} {\bibinfo {author} {\bibfnamefont {A.~M.}\ \bibnamefont
  {Fern{\'a}ndez-Fern{\'a}ndez}}, \bibinfo {author} {\bibfnamefont
  {M.}~\bibnamefont {P{\'e}rez-Rodr{\'\i}guez}}, \bibinfo {author}
  {\bibfnamefont {A.}~\bibnamefont {Comesa{\~n}a}}, \ and\ \bibinfo {author}
  {\bibfnamefont {M.~M.}\ \bibnamefont {Pi{\~n}eiro}},\ }\bibfield  {title}
  {\enquote {\bibinfo {title} {Three-phase equilibrium curve shift for methane
  hydrate in oceanic conditions calculated from molecular dynamics
  simulations},}\ }\href {\doibase
  https://doi.org/10.1016/j.molliq.2018.10.146} {\bibfield  {journal} {\bibinfo
   {journal} {J. Mol. Liq.}\ }\textbf {\bibinfo {volume} {274}},\ \bibinfo
  {pages} {426--433} (\bibinfo {year} {2019})}\BibitemShut {NoStop}%
\bibitem [{\citenamefont {Blazquez}, \citenamefont {Vega},\ and\ \citenamefont
  {Conde}(2023)}]{Blazquez2023b}%
  \BibitemOpen
  \bibfield  {author} {\bibinfo {author} {\bibfnamefont {S.}~\bibnamefont
  {Blazquez}}, \bibinfo {author} {\bibfnamefont {C.}~\bibnamefont {Vega}}, \
  and\ \bibinfo {author} {\bibfnamefont {M.~M.}\ \bibnamefont {Conde}},\
  }\bibfield  {title} {\enquote {\bibinfo {title} {Three phase equilibria of
  the methane hydrate in nacl solutions: A simulation study},}\ }\href@noop {}
  {\bibfield  {journal} {\bibinfo  {journal} {J. Mol. Liq.}\ }\textbf {\bibinfo
  {volume} {383}},\ \bibinfo {pages} {122031} (\bibinfo {year}
  {2023})}\BibitemShut {NoStop}%
\bibitem [{\citenamefont {Grabowska}\ \emph {et~al.}(2023)\citenamefont
  {Grabowska}, \citenamefont {Bl{\'a}zquez}, \citenamefont {Sanz},
  \citenamefont {Noya}, \citenamefont {Zer{\'o}n}, \citenamefont {Algaba},
  \citenamefont {M{\'{\i}}guez}, \citenamefont {Blas},\ and\ \citenamefont
  {Vega}}]{Grabowska2022b}%
  \BibitemOpen
  \bibfield  {author} {\bibinfo {author} {\bibfnamefont {J.}~\bibnamefont
  {Grabowska}}, \bibinfo {author} {\bibfnamefont {S.}~\bibnamefont
  {Bl{\'a}zquez}}, \bibinfo {author} {\bibfnamefont {E.}~\bibnamefont {Sanz}},
  \bibinfo {author} {\bibfnamefont {E.~G.}\ \bibnamefont {Noya}}, \bibinfo
  {author} {\bibfnamefont {I.~M.}\ \bibnamefont {Zer{\'o}n}}, \bibinfo {author}
  {\bibfnamefont {J.}~\bibnamefont {Algaba}}, \bibinfo {author} {\bibfnamefont
  {J.~M.}\ \bibnamefont {M{\'{\i}}guez}}, \bibinfo {author} {\bibfnamefont
  {F.~J.}\ \bibnamefont {Blas}}, \ and\ \bibinfo {author} {\bibfnamefont
  {C.}~\bibnamefont {Vega}},\ }\bibfield  {title} {\enquote {\bibinfo {title}
  {Homogeneous nucleation rate of methane hydrate formation under experimental
  conditions from seeding simulations},}\ }\href@noop {} {\bibfield  {journal}
  {\bibinfo  {journal} {J. Chem. Phys.}\ }\textbf {\bibinfo {volume} {158}}
  (\bibinfo {year} {2023})}\BibitemShut {NoStop}%
\bibitem [{\citenamefont {Algaba}\ \emph
  {et~al.}(2023{\natexlab{a}})\citenamefont {Algaba}, \citenamefont
  {Zer{\'o}n}, \citenamefont {M{\'\i}guez}, \citenamefont {Grabowska},
  \citenamefont {Blazquez}, \citenamefont {Sanz}, \citenamefont {Vega},\ and\
  \citenamefont {Blas}}]{Algaba2023a}%
  \BibitemOpen
  \bibfield  {author} {\bibinfo {author} {\bibfnamefont {J.}~\bibnamefont
  {Algaba}}, \bibinfo {author} {\bibfnamefont {I.~M.}\ \bibnamefont
  {Zer{\'o}n}}, \bibinfo {author} {\bibfnamefont {J.~M.}\ \bibnamefont
  {M{\'\i}guez}}, \bibinfo {author} {\bibfnamefont {J.}~\bibnamefont
  {Grabowska}}, \bibinfo {author} {\bibfnamefont {S.}~\bibnamefont {Blazquez}},
  \bibinfo {author} {\bibfnamefont {E.}~\bibnamefont {Sanz}}, \bibinfo {author}
  {\bibfnamefont {C.}~\bibnamefont {Vega}}, \ and\ \bibinfo {author}
  {\bibfnamefont {F.~J.}\ \bibnamefont {Blas}},\ }\bibfield  {title} {\enquote
  {\bibinfo {title} {Solubility of carbon dioxide in water: Some useful results
  for hydrate nucleation},}\ }\href@noop {} {\bibfield  {journal} {\bibinfo
  {journal} {The Journal of Chemical Physics}\ }\textbf {\bibinfo {volume}
  {158}} (\bibinfo {year} {2023}{\natexlab{a}})}\BibitemShut {NoStop}%
\bibitem [{\citenamefont
  {Panagiotopoulos}(1994)}]{panagiotopoulos1994molecular}%
  \BibitemOpen
  \bibfield  {author} {\bibinfo {author} {\bibfnamefont {A.~Z.}\ \bibnamefont
  {Panagiotopoulos}},\ }\bibfield  {title} {\enquote {\bibinfo {title}
  {Molecular simulation of phase coexistence: Finite-size effects and
  determination of critical parameters for two-and three-dimensional
  lennard-jones fluids},}\ }\href@noop {} {\bibfield  {journal} {\bibinfo
  {journal} {International journal of thermophysics}\ }\textbf {\bibinfo
  {volume} {15}},\ \bibinfo {pages} {1057--1072} (\bibinfo {year}
  {1994})}\BibitemShut {NoStop}%
\bibitem [{\citenamefont {Orea}, \citenamefont {L{\'o}pez-Lemus},\ and\
  \citenamefont {Alejandre}(2005)}]{orea2005oscillatory}%
  \BibitemOpen
  \bibfield  {author} {\bibinfo {author} {\bibfnamefont {P.}~\bibnamefont
  {Orea}}, \bibinfo {author} {\bibfnamefont {J.}~\bibnamefont
  {L{\'o}pez-Lemus}}, \ and\ \bibinfo {author} {\bibfnamefont {J.}~\bibnamefont
  {Alejandre}},\ }\bibfield  {title} {\enquote {\bibinfo {title} {Oscillatory
  surface tension due to finite-size effects},}\ }\href@noop {} {\bibfield
  {journal} {\bibinfo  {journal} {The Journal of chemical physics}\ }\textbf
  {\bibinfo {volume} {123}} (\bibinfo {year} {2005})}\BibitemShut {NoStop}%
\bibitem [{\citenamefont {Binder}\ and\ \citenamefont
  {M{\"u}ller}(2000)}]{binder2000computer}%
  \BibitemOpen
  \bibfield  {author} {\bibinfo {author} {\bibfnamefont {K.}~\bibnamefont
  {Binder}}\ and\ \bibinfo {author} {\bibfnamefont {M.}~\bibnamefont
  {M{\"u}ller}},\ }\bibfield  {title} {\enquote {\bibinfo {title} {Computer
  simulation of profiles of interfaces between coexisting phases: Do we
  understand their finite size effects?}}\ }\href@noop {} {\bibfield  {journal}
  {\bibinfo  {journal} {International Journal of Modern Physics C}\ }\textbf
  {\bibinfo {volume} {11}},\ \bibinfo {pages} {1093--1113} (\bibinfo {year}
  {2000})}\BibitemShut {NoStop}%
\bibitem [{\citenamefont {V{\"o}rtler}, \citenamefont {Sch{\"a}fer},\ and\
  \citenamefont {Smith}(2008)}]{vortler2008simulation}%
  \BibitemOpen
  \bibfield  {author} {\bibinfo {author} {\bibfnamefont {H.~L.}\ \bibnamefont
  {V{\"o}rtler}}, \bibinfo {author} {\bibfnamefont {K.}~\bibnamefont
  {Sch{\"a}fer}}, \ and\ \bibinfo {author} {\bibfnamefont {W.~R.}\ \bibnamefont
  {Smith}},\ }\bibfield  {title} {\enquote {\bibinfo {title} {Simulation of
  chemical potentials and phase equilibria in two-and three-dimensional
  square-well fluids: finite size effects},}\ }\href@noop {} {\bibfield
  {journal} {\bibinfo  {journal} {The Journal of Physical Chemistry B}\
  }\textbf {\bibinfo {volume} {112}},\ \bibinfo {pages} {4656--4661} (\bibinfo
  {year} {2008})}\BibitemShut {NoStop}%
\bibitem [{\citenamefont {Conde}, \citenamefont {Rovere},\ and\ \citenamefont
  {Gallo}(2017{\natexlab{b}})}]{Conde2017a}%
  \BibitemOpen
  \bibfield  {author} {\bibinfo {author} {\bibfnamefont {M.~M.}\ \bibnamefont
  {Conde}}, \bibinfo {author} {\bibfnamefont {M.}~\bibnamefont {Rovere}}, \
  and\ \bibinfo {author} {\bibfnamefont {P.}~\bibnamefont {Gallo}},\ }\bibfield
   {title} {\enquote {\bibinfo {title} {High precision determination of the
  melting points of water {TIP4P/2005} and water {TIP4P/Ice} models by the
  direct coexistence technique},}\ }\href {\doibase
  https://doi.org/10.1063/1.4790647} {\bibfield  {journal} {\bibinfo  {journal}
  {J. Chem. Phys.}\ }\textbf {\bibinfo {volume} {147}},\ \bibinfo {pages}
  {244506} (\bibinfo {year} {2017}{\natexlab{b}})}\BibitemShut {NoStop}%
\bibitem [{\citenamefont {Conde}\ and\ \citenamefont
  {Vega}(2010)}]{Conde2010a}%
  \BibitemOpen
  \bibfield  {author} {\bibinfo {author} {\bibfnamefont {M.~M.}\ \bibnamefont
  {Conde}}\ and\ \bibinfo {author} {\bibfnamefont {C.}~\bibnamefont {Vega}},\
  }\bibfield  {title} {\enquote {\bibinfo {title} {Determining the three-phase
  coexistence line in methane hydrates using computer simulations},}\ }\href
  {\doibase https://doi.org/10.1063/1.3466751} {\bibfield  {journal} {\bibinfo
  {journal} {J. Chem. Phys.}\ }\textbf {\bibinfo {volume} {133}},\ \bibinfo
  {pages} {064507} (\bibinfo {year} {2010})}\BibitemShut {NoStop}%
\bibitem [{\citenamefont {Buch}, \citenamefont {Sandler},\ and\ \citenamefont
  {Sadlej}(1998)}]{Buch1998a}%
  \BibitemOpen
  \bibfield  {author} {\bibinfo {author} {\bibfnamefont {V.}~\bibnamefont
  {Buch}}, \bibinfo {author} {\bibfnamefont {P.}~\bibnamefont {Sandler}}, \
  and\ \bibinfo {author} {\bibfnamefont {J.}~\bibnamefont {Sadlej}},\
  }\bibfield  {title} {\enquote {\bibinfo {title} {Simulations of h2o solid,
  liquid, and clusters, with an emphasis on ferroelectric ordering transition
  in hexagonal ice},}\ }\href@noop {} {\bibfield  {journal} {\bibinfo
  {journal} {J. Phys. Chem. B}\ }\textbf {\bibinfo {volume} {102}},\ \bibinfo
  {pages} {8641--8653} (\bibinfo {year} {1998})}\BibitemShut {NoStop}%
\bibitem [{\citenamefont {Bernal}\ and\ \citenamefont
  {Fowler}(1933)}]{Bernal1933a}%
  \BibitemOpen
  \bibfield  {author} {\bibinfo {author} {\bibfnamefont {J.~D.}\ \bibnamefont
  {Bernal}}\ and\ \bibinfo {author} {\bibfnamefont {R.~H.}\ \bibnamefont
  {Fowler}},\ }\bibfield  {title} {\enquote {\bibinfo {title} {Simulations of
  h2o solid, liquid, and clusters, with an emphasis on ferroelectric ordering
  transition in hexagonal ice},}\ }\href@noop {} {\bibfield  {journal}
  {\bibinfo  {journal} {J. Chem. Phys.}\ }\textbf {\bibinfo {volume} {1}},\
  \bibinfo {pages} {515--548} (\bibinfo {year} {1933})}\BibitemShut {NoStop}%
\bibitem [{\citenamefont {{van der Spoel}}\ \emph {et~al.}(2005)\citenamefont
  {{van der Spoel}}, \citenamefont {Lindahl}, \citenamefont {Hess},
  \citenamefont {Groenhof}, \citenamefont {Mark},\ and\ \citenamefont
  {Berendsen}}]{VanDerSpoel2005a}%
  \BibitemOpen
  \bibfield  {author} {\bibinfo {author} {\bibfnamefont {D.}~\bibnamefont {{van
  der Spoel}}}, \bibinfo {author} {\bibfnamefont {E.}~\bibnamefont {Lindahl}},
  \bibinfo {author} {\bibfnamefont {B.}~\bibnamefont {Hess}}, \bibinfo {author}
  {\bibfnamefont {G.}~\bibnamefont {Groenhof}}, \bibinfo {author}
  {\bibfnamefont {A.~E.}\ \bibnamefont {Mark}}, \ and\ \bibinfo {author}
  {\bibfnamefont {H.~J.}\ \bibnamefont {Berendsen}},\ }\bibfield  {title}
  {\enquote {\bibinfo {title} {Gromacs: Fast, flexible, and free},}\
  }\href@noop {} {\bibfield  {journal} {\bibinfo  {journal} {J. Comput. Chem.}\
  }\textbf {\bibinfo {volume} {26}},\ \bibinfo {pages} {1701--1718} (\bibinfo
  {year} {2005})}\BibitemShut {NoStop}%
\bibitem [{\citenamefont {Potoff}\ and\ \citenamefont
  {Siepmann}(2001)}]{Potoff2001a}%
  \BibitemOpen
  \bibfield  {author} {\bibinfo {author} {\bibfnamefont {J.~J.}\ \bibnamefont
  {Potoff}}\ and\ \bibinfo {author} {\bibfnamefont {J.~I.}\ \bibnamefont
  {Siepmann}},\ }\bibfield  {title} {\enquote {\bibinfo {title} {Vapor-liquid
  equilibria of mixtures containing alkanes, carbon dioxide, and nitrogen},}\
  }\href@noop {} {\bibfield  {journal} {\bibinfo  {journal} {AIChE Journal.}\
  }\textbf {\bibinfo {volume} {47}},\ \bibinfo {pages} {1676--1682} (\bibinfo
  {year} {2001})}\BibitemShut {NoStop}%
\bibitem [{\citenamefont {Abascal}\ \emph {et~al.}(2005)\citenamefont
  {Abascal}, \citenamefont {Sanz}, \citenamefont {Fern\'andez},\ and\
  \citenamefont {Vega}}]{Abascal2005b}%
  \BibitemOpen
  \bibfield  {author} {\bibinfo {author} {\bibfnamefont {J.~L.~F.}\
  \bibnamefont {Abascal}}, \bibinfo {author} {\bibfnamefont {E.}~\bibnamefont
  {Sanz}}, \bibinfo {author} {\bibfnamefont {R.~G.}\ \bibnamefont
  {Fern\'andez}}, \ and\ \bibinfo {author} {\bibfnamefont {C.}~\bibnamefont
  {Vega}},\ }\bibfield  {title} {\enquote {\bibinfo {title} {A potential model
  for the study of ices and amorphous water: {TIP4P/Ice}},}\ }\href@noop {}
  {\bibfield  {journal} {\bibinfo  {journal} {J. Chem. Phys.}\ }\textbf
  {\bibinfo {volume} {122}},\ \bibinfo {pages} {234511--1--234511--9} (\bibinfo
  {year} {2005})}\BibitemShut {NoStop}%
\bibitem [{\citenamefont {Cuendet}\ and\ \citenamefont
  {Gunsteren}(2007)}]{Cuendet2007a}%
  \BibitemOpen
  \bibfield  {author} {\bibinfo {author} {\bibfnamefont {M.~A.}\ \bibnamefont
  {Cuendet}}\ and\ \bibinfo {author} {\bibfnamefont {W.~F.~V.}\ \bibnamefont
  {Gunsteren}},\ }\bibfield  {title} {\enquote {\bibinfo {title} {On the
  calculation of velocity-dependent properties in molecular dynamics
  simulations using the leapfrog integration algorithm},}\ }\href@noop {}
  {\bibfield  {journal} {\bibinfo  {journal} {J. Chem. Phys.}\ }\textbf
  {\bibinfo {volume} {127}},\ \bibinfo {pages} {184102/1--9} (\bibinfo {year}
  {2007})}\BibitemShut {NoStop}%
\bibitem [{\citenamefont {Nos{\'e}}(1984)}]{Nose1984a}%
  \BibitemOpen
  \bibfield  {author} {\bibinfo {author} {\bibfnamefont {S.}~\bibnamefont
  {Nos{\'e}}},\ }\bibfield  {title} {\enquote {\bibinfo {title} {A molecular
  dynamics method for simulations in the canonical ensemble},}\ }\href@noop {}
  {\bibfield  {journal} {\bibinfo  {journal} {Mol. Phys.}\ }\textbf {\bibinfo
  {volume} {52}},\ \bibinfo {pages} {255--268} (\bibinfo {year}
  {1984})}\BibitemShut {NoStop}%
\bibitem [{\citenamefont {Parrinello}\ and\ \citenamefont
  {Rahman}(1981)}]{Parrinello1981a}%
  \BibitemOpen
  \bibfield  {author} {\bibinfo {author} {\bibfnamefont {M.}~\bibnamefont
  {Parrinello}}\ and\ \bibinfo {author} {\bibfnamefont {A.}~\bibnamefont
  {Rahman}},\ }\bibfield  {title} {\enquote {\bibinfo {title} {Polymorphic
  transitions in single crystals: A new molecular dynamics method},}\
  }\href@noop {} {\bibfield  {journal} {\bibinfo  {journal} {J. Appl. Phys.}\
  }\textbf {\bibinfo {volume} {52}},\ \bibinfo {pages} {7182--7190} (\bibinfo
  {year} {1981})}\BibitemShut {NoStop}%
\bibitem [{\citenamefont {Essmann}\ \emph {et~al.}(1995)\citenamefont
  {Essmann}, \citenamefont {Perera}, \citenamefont {Berkowitz}, \citenamefont
  {Darden}, \citenamefont {Lee},\ and\ \citenamefont
  {Pedersen}}]{Essmann1995a}%
  \BibitemOpen
  \bibfield  {author} {\bibinfo {author} {\bibfnamefont {U.}~\bibnamefont
  {Essmann}}, \bibinfo {author} {\bibfnamefont {L.}~\bibnamefont {Perera}},
  \bibinfo {author} {\bibfnamefont {M.~L.}\ \bibnamefont {Berkowitz}}, \bibinfo
  {author} {\bibfnamefont {T.}~\bibnamefont {Darden}}, \bibinfo {author}
  {\bibfnamefont {H.}~\bibnamefont {Lee}}, \ and\ \bibinfo {author}
  {\bibfnamefont {L.~G.}\ \bibnamefont {Pedersen}},\ }\bibfield  {title}
  {\enquote {\bibinfo {title} {A smooth particle mesh {Ewald} method},}\
  }\href@noop {} {\bibfield  {journal} {\bibinfo  {journal} {J. Chem. Phys.}\
  }\textbf {\bibinfo {volume} {103}},\ \bibinfo {pages} {8577--8593} (\bibinfo
  {year} {1995})}\BibitemShut {NoStop}%
\bibitem [{\citenamefont {Blazquez}\ \emph {et~al.}(2024)\citenamefont
  {Blazquez}, \citenamefont {Algaba}, \citenamefont {M{\'{\i}}guez},
  \citenamefont {Vega}, \citenamefont {Blas},\ and\ \citenamefont
  {Conde}}]{paperI}%
  \BibitemOpen
  \bibfield  {author} {\bibinfo {author} {\bibfnamefont {S.}~\bibnamefont
  {Blazquez}}, \bibinfo {author} {\bibfnamefont {J.}~\bibnamefont {Algaba}},
  \bibinfo {author} {\bibfnamefont {J.~M.}\ \bibnamefont {M{\'{\i}}guez}},
  \bibinfo {author} {\bibfnamefont {C.}~\bibnamefont {Vega}}, \bibinfo {author}
  {\bibfnamefont {F.~J.}\ \bibnamefont {Blas}}, \ and\ \bibinfo {author}
  {\bibfnamefont {M.~M.}\ \bibnamefont {Conde}},\ }\bibfield  {title} {\enquote
  {\bibinfo {title} {Three-phase equilibria of hydrates from computer
  simulation. {I}. {F}inite- size effects in the methane hydrate},}\
  }\href@noop {} {\bibfield  {journal} {\bibinfo  {journal} {The Journal of
  Chemical Physics, submitted}\ } (\bibinfo {year} {2024})}\BibitemShut
  {NoStop}%
\bibitem [{\citenamefont {Grabowska}\ \emph {et~al.}(2022)\citenamefont
  {Grabowska}, \citenamefont {Bl{\'a}zquez}, \citenamefont {Sanz},
  \citenamefont {Zer{\'o}n}, \citenamefont {Algaba}, \citenamefont
  {M{\'{\i}}guez}, \citenamefont {Blas},\ and\ \citenamefont
  {Vega}}]{Grabowska2022a}%
  \BibitemOpen
  \bibfield  {author} {\bibinfo {author} {\bibfnamefont {J.}~\bibnamefont
  {Grabowska}}, \bibinfo {author} {\bibfnamefont {S.}~\bibnamefont
  {Bl{\'a}zquez}}, \bibinfo {author} {\bibfnamefont {E.}~\bibnamefont {Sanz}},
  \bibinfo {author} {\bibfnamefont {I.~M.}\ \bibnamefont {Zer{\'o}n}}, \bibinfo
  {author} {\bibfnamefont {J.}~\bibnamefont {Algaba}}, \bibinfo {author}
  {\bibfnamefont {J.~M.}\ \bibnamefont {M{\'{\i}}guez}}, \bibinfo {author}
  {\bibfnamefont {F.~J.}\ \bibnamefont {Blas}}, \ and\ \bibinfo {author}
  {\bibfnamefont {C.}~\bibnamefont {Vega}},\ }\bibfield  {title} {\enquote
  {\bibinfo {title} {Solubility of methane in water: some useful results for
  hydrate nucleation},}\ }\href@noop {} {\bibfield  {journal} {\bibinfo
  {journal} {J. Phys. Chem. B}\ }\textbf {\bibinfo {volume} {126}},\ \bibinfo
  {pages} {8553--8570} (\bibinfo {year} {2022})}\BibitemShut {NoStop}%
\bibitem [{\citenamefont {Walsh}\ \emph {et~al.}(2009)\citenamefont {Walsh},
  \citenamefont {Koh}, \citenamefont {Sloan}, \citenamefont {Sum},\ and\
  \citenamefont {Wu}}]{Walsh2009a}%
  \BibitemOpen
  \bibfield  {author} {\bibinfo {author} {\bibfnamefont {M.~R.}\ \bibnamefont
  {Walsh}}, \bibinfo {author} {\bibfnamefont {C.~A.}\ \bibnamefont {Koh}},
  \bibinfo {author} {\bibfnamefont {E.~D.}\ \bibnamefont {Sloan}}, \bibinfo
  {author} {\bibfnamefont {A.~K.}\ \bibnamefont {Sum}}, \ and\ \bibinfo
  {author} {\bibfnamefont {D.~T.}\ \bibnamefont {Wu}},\ }\bibfield  {title}
  {\enquote {\bibinfo {title} {Microsecond simulations of spontaneous methane
  hydrate nucleation and growth},}\ }\href@noop {} {\bibfield  {journal}
  {\bibinfo  {journal} {Science}\ }\textbf {\bibinfo {volume} {326}},\ \bibinfo
  {pages} {1095--1098} (\bibinfo {year} {2009})}\BibitemShut {NoStop}%
\bibitem [{\citenamefont {Walsh}\ \emph {et~al.}(2011)\citenamefont {Walsh},
  \citenamefont {Beckham}, \citenamefont {Koh}, \citenamefont {Sloan},
  \citenamefont {Wu},\ and\ \citenamefont {Sum}}]{Walsh2011a}%
  \BibitemOpen
  \bibfield  {author} {\bibinfo {author} {\bibfnamefont {M.~R.}\ \bibnamefont
  {Walsh}}, \bibinfo {author} {\bibfnamefont {G.~T.}\ \bibnamefont {Beckham}},
  \bibinfo {author} {\bibfnamefont {C.~A.}\ \bibnamefont {Koh}}, \bibinfo
  {author} {\bibfnamefont {E.~D.}\ \bibnamefont {Sloan}}, \bibinfo {author}
  {\bibfnamefont {D.~T.}\ \bibnamefont {Wu}}, \ and\ \bibinfo {author}
  {\bibfnamefont {A.~K.}\ \bibnamefont {Sum}},\ }\bibfield  {title} {\enquote
  {\bibinfo {title} {Methane hydrate nucleation rates from molecular dynamics
  simulations: Effects of aqueous methane concentration, interfacial curvature,
  and system size},}\ }\href {\doibase https://doi.org/10.1021/jp206483q}
  {\bibfield  {journal} {\bibinfo  {journal} {J. Phys. Chem. C}\ }\textbf
  {\bibinfo {volume} {115}},\ \bibinfo {pages} {21241--21248} (\bibinfo {year}
  {2011})}\BibitemShut {NoStop}%
\bibitem [{\citenamefont {Liang}\ and\ \citenamefont
  {Kusalik}(2011)}]{Liang2011a}%
  \BibitemOpen
  \bibfield  {author} {\bibinfo {author} {\bibfnamefont {S.}~\bibnamefont
  {Liang}}\ and\ \bibinfo {author} {\bibfnamefont {P.~G.}\ \bibnamefont
  {Kusalik}},\ }\bibfield  {title} {\enquote {\bibinfo {title} {Exploring
  nucleation of h 2 s hydrates},}\ }\href@noop {} {\bibfield  {journal}
  {\bibinfo  {journal} {Chemical Science}\ }\textbf {\bibinfo {volume} {2}},\
  \bibinfo {pages} {1286--1292} (\bibinfo {year} {2011})}\BibitemShut {NoStop}%
\bibitem [{\citenamefont {Yagasaki}\ \emph {et~al.}(2014)\citenamefont
  {Yagasaki}, \citenamefont {Matsumoto}, \citenamefont {Andoh}, \citenamefont
  {Okazaki},\ and\ \citenamefont {Tanaka}}]{Yagasaki2014a}%
  \BibitemOpen
  \bibfield  {author} {\bibinfo {author} {\bibfnamefont {T.}~\bibnamefont
  {Yagasaki}}, \bibinfo {author} {\bibfnamefont {M.}~\bibnamefont {Matsumoto}},
  \bibinfo {author} {\bibfnamefont {Y.}~\bibnamefont {Andoh}}, \bibinfo
  {author} {\bibfnamefont {S.}~\bibnamefont {Okazaki}}, \ and\ \bibinfo
  {author} {\bibfnamefont {H.}~\bibnamefont {Tanaka}},\ }\bibfield  {title}
  {\enquote {\bibinfo {title} {Effect of bubble formation on the dissociation
  of methane hydrate in water: A molecular dynamics study},}\ }\href@noop {}
  {\bibfield  {journal} {\bibinfo  {journal} {J. Phys. Chem. B}\ }\textbf
  {\bibinfo {volume} {118}},\ \bibinfo {pages} {1900} (\bibinfo {year}
  {2014})}\BibitemShut {NoStop}%
\bibitem [{\citenamefont {Fang}\ \emph {et~al.}(2023)\citenamefont {Fang},
  \citenamefont {Moultos}, \citenamefont {L{\"u}}, \citenamefont {Sun},
  \citenamefont {Liu}, \citenamefont {Ning},\ and\ \citenamefont
  {Vlugt}}]{Fang2023a}%
  \BibitemOpen
  \bibfield  {author} {\bibinfo {author} {\bibfnamefont {B.}~\bibnamefont
  {Fang}}, \bibinfo {author} {\bibfnamefont {O.~A.}\ \bibnamefont {Moultos}},
  \bibinfo {author} {\bibfnamefont {T.}~\bibnamefont {L{\"u}}}, \bibinfo
  {author} {\bibfnamefont {J.}~\bibnamefont {Sun}}, \bibinfo {author}
  {\bibfnamefont {Z.}~\bibnamefont {Liu}}, \bibinfo {author} {\bibfnamefont
  {F.}~\bibnamefont {Ning}}, \ and\ \bibinfo {author} {\bibfnamefont {T.~J.}\
  \bibnamefont {Vlugt}},\ }\bibfield  {title} {\enquote {\bibinfo {title}
  {Effects of nanobubbles on methane hydrate dissociation: A molecular
  simulation study},}\ }\href@noop {} {\bibfield  {journal} {\bibinfo
  {journal} {Fuel}\ }\textbf {\bibinfo {volume} {345}},\ \bibinfo {pages}
  {128230} (\bibinfo {year} {2023})}\BibitemShut {NoStop}%
\bibitem [{\citenamefont {Bagherzadeh}\ \emph {et~al.}(2015)\citenamefont
  {Bagherzadeh}, \citenamefont {Alavi}, \citenamefont {Ripmeester},\ and\
  \citenamefont {Englezos}}]{Bagherzadeh2015a}%
  \BibitemOpen
  \bibfield  {author} {\bibinfo {author} {\bibfnamefont {S.~A.}\ \bibnamefont
  {Bagherzadeh}}, \bibinfo {author} {\bibfnamefont {S.}~\bibnamefont {Alavi}},
  \bibinfo {author} {\bibfnamefont {J.}~\bibnamefont {Ripmeester}}, \ and\
  \bibinfo {author} {\bibfnamefont {P.}~\bibnamefont {Englezos}},\ }\bibfield
  {title} {\enquote {\bibinfo {title} {Formation of methane nano-bubbles during
  hydrate decomposition and their effect on hydrate growth},}\ }\href@noop {}
  {\bibfield  {journal} {\bibinfo  {journal} {J. Chem. Phys.}\ }\textbf
  {\bibinfo {volume} {142}},\ \bibinfo {pages} {214701} (\bibinfo {year}
  {2015})}\BibitemShut {NoStop}%
\bibitem [{\citenamefont {Hall}, \citenamefont {Zhang},\ and\ \citenamefont
  {Kusalik}(2016)}]{Hall2016a}%
  \BibitemOpen
  \bibfield  {author} {\bibinfo {author} {\bibfnamefont {K.~W.}\ \bibnamefont
  {Hall}}, \bibinfo {author} {\bibfnamefont {Z.}~\bibnamefont {Zhang}}, \ and\
  \bibinfo {author} {\bibfnamefont {P.~G.}\ \bibnamefont {Kusalik}},\
  }\bibfield  {title} {\enquote {\bibinfo {title} {Unraveling mixed hydrate
  formation: Microscopic insights into early stage behavior},}\ }\href@noop {}
  {\bibfield  {journal} {\bibinfo  {journal} {The Journal of Physical Chemistry
  B}\ }\textbf {\bibinfo {volume} {120}},\ \bibinfo {pages} {13218--13223}
  (\bibinfo {year} {2016})}\BibitemShut {NoStop}%
\bibitem [{\citenamefont {MacDowell}, \citenamefont {Shen},\ and\ \citenamefont
  {Errington}(2006)}]{MacDowell2006a}%
  \BibitemOpen
  \bibfield  {author} {\bibinfo {author} {\bibfnamefont {L.~G.}\ \bibnamefont
  {MacDowell}}, \bibinfo {author} {\bibfnamefont {V.~K.}\ \bibnamefont {Shen}},
  \ and\ \bibinfo {author} {\bibfnamefont {J.~R.}\ \bibnamefont {Errington}},\
  }\bibfield  {title} {\enquote {\bibinfo {title} {{Nucleation and cavitation
  of spherical, cylindrical, and slablike droplets and bubbles in small
  systems}},}\ }\href@noop {} {\bibfield  {journal} {\bibinfo  {journal} {The
  Journal of Chemical Physics}\ }\textbf {\bibinfo {volume} {125}},\ \bibinfo
  {pages} {034705} (\bibinfo {year} {2006})}\BibitemShut {NoStop}%
\bibitem [{\citenamefont {Singh}\ \emph {et~al.}(2019)\citenamefont {Singh},
  \citenamefont {Palmer}, \citenamefont {Panagiotopoulos},\ and\ \citenamefont
  {Debenedetti}}]{Singh2019a}%
  \BibitemOpen
  \bibfield  {author} {\bibinfo {author} {\bibfnamefont {R.~S.}\ \bibnamefont
  {Singh}}, \bibinfo {author} {\bibfnamefont {J.~C.}\ \bibnamefont {Palmer}},
  \bibinfo {author} {\bibfnamefont {A.~Z.}\ \bibnamefont {Panagiotopoulos}}, \
  and\ \bibinfo {author} {\bibfnamefont {P.~G.}\ \bibnamefont {Debenedetti}},\
  }\bibfield  {title} {\enquote {\bibinfo {title} {{Thermodynamic analysis of
  the stability of planar interfaces between coexisting phases and its
  application to supercooled water}},}\ }\href@noop {} {\bibfield  {journal}
  {\bibinfo  {journal} {The Journal of Chemical Physics}\ }\textbf {\bibinfo
  {volume} {150}},\ \bibinfo {pages} {224503} (\bibinfo {year}
  {2019})}\BibitemShut {NoStop}%
\bibitem [{\citenamefont {Montero~de Hijes}\ and\ \citenamefont
  {Vega}(2022)}]{Montero2022a}%
  \BibitemOpen
  \bibfield  {author} {\bibinfo {author} {\bibfnamefont {P.}~\bibnamefont
  {Montero~de Hijes}}\ and\ \bibinfo {author} {\bibfnamefont {C.}~\bibnamefont
  {Vega}},\ }\bibfield  {title} {\enquote {\bibinfo {title} {On the
  thermodynamics of curved interfaces and the nucleation of hard spheres in a
  finite system},}\ }\href@noop {} {\bibfield  {journal} {\bibinfo  {journal}
  {The Journal of Chemical Physics}\ }\textbf {\bibinfo {volume} {156}},\
  \bibinfo {pages} {014505} (\bibinfo {year} {2022})}\BibitemShut {NoStop}%
\bibitem [{\citenamefont {Algaba}\ \emph
  {et~al.}(2023{\natexlab{b}})\citenamefont {Algaba}, \citenamefont {Blazquez},
  \citenamefont {Míguez}, \citenamefont {Conde},\ and\ \citenamefont
  {Blas}}]{paperIII}%
  \BibitemOpen
  \bibfield  {author} {\bibinfo {author} {\bibfnamefont {J.}~\bibnamefont
  {Algaba}}, \bibinfo {author} {\bibfnamefont {S.}~\bibnamefont {Blazquez}},
  \bibinfo {author} {\bibfnamefont {J.~M.}\ \bibnamefont {Míguez}}, \bibinfo
  {author} {\bibfnamefont {M.~M.}\ \bibnamefont {Conde}}, \ and\ \bibinfo
  {author} {\bibfnamefont {F.~J.}\ \bibnamefont {Blas}},\ }\bibfield  {title}
  {\enquote {\bibinfo {title} {Three-phase equilibria of hydrates from computer
  simulation. {III}. {E}ffect of dispersive interactions in methane and carbon
  dioxide hydrates},}\ }\href@noop {} {\bibfield  {journal} {\bibinfo
  {journal} {The Journal of Chemical Physics, submitted}\ } (\bibinfo {year}
  {2023}{\natexlab{b}})}\BibitemShut {NoStop}%
\end{thebibliography}%
\end{document}